\documentclass[11pt, reqno, harvard]{amsart}

\usepackage{graphics,amsmath,amssymb,epsfig,stmaryrd,color,subfigure}
\usepackage{amsfonts}

\usepackage{hyperref}

\usepackage[
backend=biber,
style=apa,
sorting=nyt
]{biblatex}
\DeclareSourcemap{
  \maps[datatype=bibtex]{
    \map{
      \step[fieldset=issn, null]
      \step[fieldset=doi, null]
      \step[fieldset=url, null]
      \step[fieldset=urldate, null]
    }
  }
}

\hypersetup{
    colorlinks=true,
    urlcolor=blue,
    linkcolor=blue,
    citecolor=blue
}

\newtheorem{theorem}{Theorem}
\theoremstyle{plain}

\newtheorem{assumption}{Assumption}

\newtheorem{corollary}{Corollary}

\newtheorem{definition}{Definition}
\newtheorem{example}{Example}

\newtheorem{lemma}{Lemma}

\newtheorem{proposition}{Proposition}
\newtheorem{remark}{Remark}

\newenvironment{continued}[1][continued]{\begin{trivlist}
\item[\hskip \labelsep {\bfseries #1}]}{\end{trivlist}}

\newcommand\independent{\protect\mathpalette{\protect\independenT}{\perp}}
\def\independenT#1#2{\mathrel{\rlap{$#1#2$}\mkern2mu{#1#2}}}
\numberwithin{equation}{section}

\begin{document}

\title[Set identification in models with multiple equilibria]{Set identification in models with multiple equilibria}
\author[Galichon and Henry]{Alfred Galichon$^\dag$ and Marc Henry$^\S$     }
\noindent \date{ The present version is of 11 November 2009. This is a preprint of an article that has been published in \textit{The Review of Economic Studies}, Vol. 78, No. 4 (October 2011), pp. 1264-1298, by Oxford University Press, \url{https://doi.org/10.1093/restud/rdr008}. This paper is a
substantially revised version of the working paper
\cite{GH:2006a}, which was widely circulated since May 2006.
Support from NSF grant SES 0350770 to Princeton University and from
NSF grant SES 0532398 is gratefully acknowledged by both authors.
Galichon gratefully acknowledges support from Chaire EDF-Calyon ´
``Finance et DÈveloppement Durable'', Chaire Axa ``Assurance des
Risques Majeurs'' and Chaire Soci\'et\'e G\'en\'erale ``Risques
Financiers''. We are grateful to Enrique Sentana and three anonymous
referees, whose comments led to considerable improvements in the
paper. We thank Romuald Meango for outstanding research assistance. We
also thank Fr\'ed\'eric Bonnans, Pierre-Andr\'e Chiappori, Ivar
Ekeland, Guido Imbens, Francesca Molinari, Bernard Salani\'e, Elie
Tamer and especially Victor Chernozhukov for many helpful
discussions and seminar participants at Berkeley, Cambridge, the
Canadian Econometrics Study Group, Chicago, the CIRANO-CIREQ conference on the econometrics of interactions, Columbia, \'Ecole
polytechnique, Harvard-MIT, MIT Sloane OR, Northwestern, NYU,
Princeton, SAMSI, Stanford, Tokyo University, University College
London, the Weierstrass Institut and Yale for helpful comments (with
the usual disclaimer). Correspondence address: D\'epartement
d'\'economie, \'Ecole polytechnique, 91128 Palaiseau, France and
D\'epartement de sciences \'economiques, Universit\'e de Montréal,
C.P. 6128, succursale Centre-ville, Montréal QC H3C 3J7, Canada.
E-mail: alfred.galichon@polytechnique.edu and
marc.henry@umontreal.ca}

\begin{abstract}
We propose a computationally feasible way of deriving the identified
 features of models with multiple equilibria in pure or mixed strategies.
 It is shown that in the case of Shapley regular normal form games, the identified set is
 characterized by the inclusion of the true data distribution within the
 core of a Choquet capacity, which is interpreted as the generalized
 likelihood of the model. In turn, this inclusion is characterized by a
 finite set of inequalities and efficient and easily implementable
 combinatorial methods are described to check them.
 In all normal form games, the identified set is
 characterized in terms of the value of a submodular or convex
 optimization program. Efficient algorithms are then given and
 compared to check inclusion of a parameter in this identified set.
 The latter are illustrated with family bargaining games and oligopoly
 entry games.
\end{abstract}

\maketitle

{\footnotesize
 \textbf{Keywords}: multiple equilibria, optimal transportation,
 identification regions, core determining classes.

\textbf{JEL subject classification}: C13, C72}

\newpage
\section*{Introduction}
The empirical study of game theoretic models is complicated by the
presence of multiple equilibria. As noted in
\cite{Jovanovic:89}, the existence of multiple equilibria
generally leads to a failure of identification of the structural
parameters governing the model. \cite{BT:2006} and
\cite{ABBP:2007} give an account of the various ways this
identification issue was approached in the literature, where
identification of structural parameters is achieved through
equilibrium refinements, shape restrictions, informational
assumptions or the specification of equilibrium selection
mechanisms. An alternative approach is to eschew identification
strategies and base inference purely on the identified features of
the models with multiple equilibria, which are sets of values rather
than a single value of the structural parameter vector. This
approach is taken in the context of imperfectly competitive markets
by \cite{ABJ:2003} and \cite{CT:2006}. The inferential
method they use, however, relies on a set of restrictions which is
not guaranteed to exhaust all the restrictions embodied in the
model, and hence leads to more conservative inference than could be
achieved. This paper proposes a computationally feasible way of
recovering the identified feature of a model with multiple
equilibria, with specific applications to inference in participation
games such as oligopoly entry models and group bargaining models.

We first note that the likelihood implied by a model with multiple
equilibria can be represented by a non additive set function called
a \emph{Choquet capacity}. Seminal work on coherency conditions and nonadditive likelihoods
can be found in \cite{Heckman:78}, \cite{GLM:80}, \cite{Manski:90} and \cite{HSC:97} among
others. In cases where identification holds, \cite{DJ:94} and \cite{BMT:2002}  propose likelihood-based estimation methods that are robust to the lack of coherency. Otherwise, the nonadditive likelihood can be refined to a likelihood
represented by a probability measure if there exists a mechanism
that picks outcomes among the admissible equilibria in the region of
multiplicity. We give a formal definition of an \emph{equilibrium
selection mechanism}, and call such a mechanism \emph{compatible}
with the data if the likelihood of the model augmented with such a
mechanism is equal to the probabilities observed in the data. The
identified feature of the model therefore is the set of parameter
values such that there exists an equilibrium selection mechanism
compatible with the data. The main result of this paper is the
equivalence between the latter condition and the actual probability
of observed outcomes belonging to the core of the likelihood
predicted by the model, where the \emph{core} is a well known and well
studied notion in economics since the word was coined in
\cite{Gillies:53}. This results allows the computation of the
identified feature of models with multiple equilibria and a finite
number of observable outcomes, as it reduces the problem to that of
checking a finite number of moment inequalities.


The computational burden remains high in situations with a large
number of observable outcomes, since the number of inequalities to
be checked is equal to the number of subsets of the set of
observable outcomes. When the set of observable outcomes is
infinite, the problem remains infinite dimensional.
\cite{GH:2006a} and \cite{EGH:2008} include results
pertaining to that case. To lift the remaining computational burden,
we propose several alternative strategies, and discuss their relative
merits.

First, in cases with only pure strategy equilibria, we show that the model likelihood is a submodular function (the set function analogue of a convex function), and that checking that the data distribution belongs to the core of the likelihood is equivalent to minimizing a submodular function, a well studied problem (analogue of minimizing a convex function) with efficient algorithms and easily available off the shelf implementations. Second, we show that if only pure strategy equilibria are considered, a special case of submodular function minimization applies, which relies on optimal transportation methods, and provides more efficient algorithms for the problem of constructing the identified set. Finally, we introduce the notion of \emph{core determining
classes}, which are suitably low cardinality classes of sets that
are sufficient to characterize the core, and we provide some results
to exhibit such core determining classes when the model satisfies
some monotonicity properties. The method is illustrated on the family bargaining game of \cite{ES:2002} and the oligopoly entry
game of \cite{BT:2006}.

In cases where mixed strategies are included in the equilibrium
concept, the fundamental work by \cite{BMM:2008} was the first
to address the problem of constructing the identified set (whereas,
in the case of pure strategies, to the best our knowledge, our work
had been the first to address this problem, in
\cite{GH:2006a}). Our incremental contribution in the case of
mixed strategy equilibria is two-fold. When the game satisfies a
regularity condition defined in \cite{Shapley:71}, we show
that the identified set is still characterized by the core of the
model likelihood, hence by a finite number of inequalities and a
submodular optimization problem as before. In all other cases, we
give a simple and efficient algorithm to construct the identified
set based on convex optimization. This goes beyond the results in
BMM that characterize the identified set only by a continuum of
inequalities. Thus our results are complementary to the results of
BMM.

The methodology proposed here applies to the empirical study of any normal form game, with both pure and mixed strategy equilibria. In economics, coordination games and participation games are the most natural area of application, including models of labour force participation (including \cite{BV:85}), models of discrete choice with social interactions (including \cite{BD:2001}), oligopoly entry models (including \cite{ABJ:2003}, \cite{CT:2006}), auction models (including \cite{BHR:2005}), bargaining models (including \cite{ES:2002}),
network effects (including \cite{Sweeting:2004}).

Beyond the identification issue of computing the identified set
given the knowledge of the true distribution of observable
variables, the inference issue of constructing confidence regions
for structural parameters in models with multiple equilibria is
taken up in \cite{GH:2006d} and \cite{GH:2006c} to
complement the seminal contribution of \cite{CHT:2007}.
Related work on inference in partially identified models include
\cite{MT:2002}, \cite{IM:2004}, \cite{BM:2007},
\cite{RS:2006a}, \cite{Rosen:2006},
\cite{AS:2007}, \cite{Canay:2007}
among many others.

The remainder of the paper is organized as follows.
Section~\ref{subsection:general} describes the framework and general
results in the case where only equilibria in pure strategies are
considered, while section~\ref{subsection:leading examples}
specializes and illustrates them on leading examples of
participation games. Section~\ref{section:efficient computation}
describes three related methods to efficiently compute the
identified feature of the model based on the characterization from
section~\ref{subsection:general} and discusses their relative
merits. Section~\ref{section:BR44} illustrates the three methods
on an oligopoly entry game with two types of players,
section~\ref{section:mixed strategies} shows how the results extend
to the case where mixed strategy equilibria are also considered. Section~\ref{section:mixed strategies} also proposes extensive simulation results and section~\ref{section:application} applies the methodology to the study of the determinants of long term care option choices for elderly parents in American families. The
last section concludes, and proofs of the results are collected in
an appendix.

\section{Identified features of models with multiple equilibria}
\label{section: incomplete model specifications}
\subsection{Identified parameter sets in general models with multiple equilibria}
\label{subsection:general} The general framework is that of
\cite{KR:50} generalized by \cite{Jovanovic:89}. It
applies primarily to the empirical analysis of normal form games,
where only equilibria in pure strategies are considered. We defer
the extension of our results to mixed strategies to
section~\ref{section:mixed strategies}.

We consider three types of economic
variables. Outcome variables $Y$, exogenous explanatory variables
$X$, and latent variables, or random shocks, $\epsilon$. Outcome
variables and latent variables are assumed to belong to complete and
separable metric spaces, so that both outcomes and latent variables
could be discrete, continuous, they could be probability distributions
or stochastic processes. The economic model consists in
a set of restrictions on the joint behaviour of the variables listed
above. These restrictions may be induced by assumptions of
rationality of agents, and they generally depend on a set of unknown
structural parameters $\theta$. Without loss of generality, the
model may be formalized as a measurable correspondence (defined in
assumption~\ref{assumption:model} below) between the latent
variables $\epsilon$ and the outcome variables $Y$ indexed by the
exogenous variables $X$ and the vector of parameters $\theta$. We
call this correspondence $G$, and write $Y\in G(\epsilon\vert
X;\theta)$ to indicate admissible values of $Y$ given values of
$\epsilon$, $X$ and $\theta$. The econometrician will be assumed to
have access to a sample of independent and identically distributed
vectors $(Y,X)$, and the problem considered is that of estimating
the vector of parameters $\theta$. The latent variables $\epsilon$
is supposed to be distributed according to a parametric distribution
$\nu(.\vert X;\theta)$, where the indexing is meant to indicate that the
unknown parameters that enter in the distribution of latent
variables are contained in the vector $\theta$ of parameters to the
estimated. We collect these assumptions next.

\begin{assumption}\label{assumption:model}
An independent and identically distributed sample of copies of the
random vector $(Y,X)$ is available. The observable outcomes $Y$
conditionally distributed according to the probability distribution
$P(\cdot\vert X)$ on $\mathcal{Y}$, a  Polish space (i.e. a complete
and separable metric space) endowed with its Borel $\sigma$-algebra
of subsets $\mathcal{B}$) are related to unobservable variables
$\epsilon$ according to the model $Y\in G(\epsilon\vert X;\theta)$,
where $\theta$ belongs to an open subset $\Theta$ of
$\mathbb{R}^{d_\theta}$, $\epsilon$ is distributed according to the
probability measure $\nu(.\vert X;\theta)$ on $\mathcal{U}$ (also
Polish endowed with its Borel $\sigma$-algebra of subsets), and $G$
is a measurable correspondence\footnote{ In the first version
circulated, \cite{GH:2006a}, we used the notation $U$ for
$\epsilon$ and $\Gamma$ for~$G^{-1}$.}, i.e. such that for all open
subsets $A$ of $\mathcal{Y}$, $G^{-1}(A\vert X;\theta):=
\{\epsilon\in\mathcal{U}:\;G(\epsilon\vert X;\theta)\cap
A\ne\varnothing\}$ is measurable (a measurable correspondence is
also called \emph{random correspondence} or \emph{random set}, and
the requirement is very mild) for almost all $X$ and for all
$\theta\in\Theta$. Finally, the variables $(Y,X,\epsilon)$ are
defined on the same underlying probability space
$(\Omega,\mathcal{F},\mathbb{P})$.
\end{assumption}

\begin{example}
To illustrate assumption~\ref{assumption:model}, we consider a simple game
proposed by \cite{Jovanovic:89}. There are two firms with profit functions
$\pi_1(Y_1,Y_2,\epsilon_1,\epsilon_2;\theta)=(\theta Y_2-\epsilon_2)Y_1$ and
$\pi_2(Y_1,Y_2,\epsilon_1,\epsilon_2;\theta)=(\theta Y_1-\epsilon_1)Y_2$,
where $Y_i\in\{0,1\}$ is firm i's action, and $\epsilon=(\epsilon_1,\epsilon_2)'$ are
exogenous costs. The firms know their costs; the analyst, however,
knows only that $\epsilon$ is uniformly distributed on $[0,1]^2$, and that
the structural parameter $\theta$ is in $(0,1]$. There are two pure
strategy Nash equilibria. The first is $Y_1=Y_2=0$ for all
$\epsilon\in[0,1]^2$. The second is $Y_1=Y_2=1$ for all $\epsilon\in[0,\theta]^2$
and $Y_1=Y_2=0$ otherwise. Hence the model is described by the
correspondence: $G(\epsilon;\theta)=\{(0,0),(1,1)\}$ for all $\epsilon\in[0,\theta]^2$,
and $G(\epsilon;\theta)=\{(0,0)\}$ otherwise.
\label{example:Jovanovic}\end{example}

To conduct inference on the parameter vector $\theta$, one first
needs to determine the identified features of the model. Because the
correspondence $G$ may be multi-valued due to the presence of
multiple equilibria, the outcomes may not be uniquely determined by
the latent variable. In such cases, the likelihood of an outcome
falling in the subset $A$ of $\mathcal{Y}$ predicted by the model is
$\mathcal{L}(A\vert X;\theta)=\nu(G^{-1}(A\vert X;\theta)\vert X;\theta)$. Because of multiple equilibria, this likelihood may
sum to more than one, as we may have $A\cap B=\varnothing$, and
yet $G^{-1}(A\vert X;\theta)\cap G^{-1}(B\vert
X;\theta)\ne\varnothing$, so that we may have $\mathcal{L}(A\cup
B\vert X;\theta)<\mathcal{L}(A\vert X;\theta)+\mathcal{L} (B\vert
X;\theta)$. The set function $A\mapsto\mathcal{L}(A\vert X;\theta)=\nu(G^{-1}(A\vert
X;\theta)\vert X;\theta)$ is generally not additive, and is called a \emph{Choquet
capacity} (see \cite{Choquet:53}).
This non additivity of the model likelihood is referred to as ``lack of coherency'' in \cite{GLM:80} and \cite{Tamer:2003}.

\begin{definition}[Choquet capacity]\label{definition:Choquet}
A Choquet capacity $\mathcal{L}$ on a finite set $\mathcal{Y}$ is a
set function $\mathcal{L}:A\subseteq\mathcal{Y}\mapsto[0,1]$ which
is \begin{itemize} \item normalized, i.e.
$\mathcal{L}(\varnothing)=0$ and $\mathcal{L}(\mathcal{Y})=1$, \item
monotone, i.e. $\mathcal{L}(A)\leq\mathcal{L}(B)$, for any
$A\subseteq B\subseteq\mathcal{Y}$.\end{itemize}
\end{definition}

\begin{continued}[Example \protect\ref{example:Jovanovic} continued]
In example~\ref{example:Jovanovic}, $\nu(.\vert X;\theta)$ is the uniform
distribution on $[0,1]^2$ and the Choquet capacity $\nu(G^{-1})$
gives value $\nu G^{-1}(\{(0,0)\})=\nu([0,1]^2)=1$ to the set $\{(0,0)\}$ and value
$\nu G^{-1}(\{(1,1)\})=\nu([0,\theta]^2)=\theta^2$ to the set $\{(1,1)\}$.
Hence it is immediately apparent that
the Choquet capacity  $\nu G^{-1}$
is a set function that is not additive, as it sums to more than $1$.
\end{continued}

As discussed in \cite{Jovanovic:89}, \cite{BT:2006} and
\cite{CT:2006}, the model with multiple equilibria can be
completed with an equilibrium selection mechanism. Following
\cite{Jovanovic:89} and \cite{BT:2006} (See for instance
the formulation (2.20) page 66 of \cite{BT:2006}), we define
an equilibrium selection mechanism as a conditional distribution
$\pi_{Y\vert\epsilon,X}$ over equilibrium outcomes $Y$ in the
regions of multiplicity. By construction, an equilibrium selection
is allowed to depend on the latent variables $\epsilon$ even after
conditioning on $X$. This is summarized in the following definition.

\begin{definition}[Equilibrium selection mechanism]
An equilibrium selection mechanism is a conditional probability
$\pi(.\vert\epsilon,X)$ for $Y$ conditionally on $\epsilon$ and $X$, such
that the selected value of the outcome variable is actually an
equilibrium, or more formally, such that $\pi(.\vert\epsilon,X)$ has
support contained in $G(\epsilon\vert X;\theta)$.\end{definition}

The identified feature of the model is the smallest set of
parameters that cannot be rejected by the data. Hence, it is the set
of parameters for which one can find an equilibrium selection
mechanism which completes the model and equates probabilities of
outcomes predicted by the model with the probabilities obtained from
the data.

\begin{definition}[Compatible equilibrium selection mechanism]
The equilibrium selection mechanism $\pi(.\vert\epsilon,X)$ is
compatible with the data if the probabilities observed in the data
are equal to the probabilities predicted by the equilibrium
selection mechanism, or more formally (see for instance the
formulation (3.24) page 72 of \cite{BT:2006}) if for all $A$
measurable subset of $\mathcal{Y}$, $P(A\vert
X)=\int_\mathcal{U}\pi(A\vert\epsilon,X)\nu(d\epsilon\vert X;\theta).$
\end{definition}

Hence the identified set is the set of parameters $\theta$
such that there exists an \emph{equilibrium selection mechanism
compatible with the data}.

\begin{definition}[Identified set] We call \emph{identified set}
(sometimes somewhat redundantly called \emph{sharp identified set})
the set $\Theta_I$ of $\theta\in\Theta$ such that there exists an
equilibrium selection mechanism compatible with the data.
\end{definition}

The definition above is not an operational definition, in the sense
that it does not alow the computation of the identified set based on
the knowledge of the probabilities in the data because the
conditional distribution $\pi$ is an infinite dimensional nuisance
parameter. We now set out to show how to reduce the dimensionality
of the problem. Our equivalent formulation of the identified set is
based on an appeal to the notion of \emph{core} of the Choquet
capacity introduced in
definition~\ref{definition:Choquet}.

\begin{definition}[Core of a Choquet capacity]
The \emph{core} of a Choquet capacity $\mathcal{L}$ on $\mathcal{Y}$
is the collection of probability distributions $Q$ on $\mathcal{Y}$
such that for all $A\subseteq\mathcal{Y}$, $Q(A)\leq
\mathcal{L}(A)$.\end{definition}

In cooperative game theory, a Choquet capacity on a set
$\mathcal{Y}$ is interpreted as a game, where $\mathcal{Y}$ is the
set of players and $\mathcal{L}$ is the utility value or worth of
coalition $A\subseteq\mathcal{Y}$ and the core of the game
$\mathcal{L}$ is the collection of allocations that cannot be
improved upon by any coalition of players (see \cite{Moulin:95}).

\begin{continued}[Example \protect\ref{example:Jovanovic} continued]
In example~\ref{example:Jovanovic}, the core of the Choquet capacity
$\nu G^{-1}$ is the set of probabilities $P$ for the observed outcomes $(0,0)$
and $(1,1)$ such that $P(\{(0,0)\})\leq\nu G^{-1}(\{(0,0)\})=\nu([0,1]^2)=1$
and $P(\{(1,1)\})\leq\nu G^{-1}(\{(1,1)\})=\nu([0,\theta]^2)=\theta^2$. \end{continued}

The result we propose next\footnote{This result appeared as
equivalence between (ii') and (iv') in theorem~1' of the first
version circulated \cite{GH:2006a}.} shows the equivalence
between the existence of a compatible equilibrium selection
mechanism and the fact that the true distribution of the data
belongs to the core of the Choquet capacity that characterizes the
likelihood predicted by the model (which we shall call \emph{core of
the likelihood predicted by the model}).

\begin{theorem}\label{theorem:sharp}
The identified set $\Theta_I$ is equal to the set of parameters such
that the true distribution of the observable variables lies in the
core of the likelihood predicted by the model. Hence
\begin{eqnarray*}\Theta_I&=&\{\theta\in\Theta:\;\left(\forall
A\in\mathcal{B},\, P(A\vert X)\leq\mathcal{L}(A\vert
X;\theta)\right);\;X-\mathrm{a.s.}\}\end{eqnarray*}
\end{theorem}

\begin{continued}[Example \protect\ref{example:Jovanovic} continued]
In example~\ref{example:Jovanovic}, the identified set is the set of
values for $\theta$ such that $p\leq\theta^2$ and $1-p\leq1$ where
$p=P((Y_1,Y_2)=(1,1))$ is the true probability that the observable
variable takes the value $(1,1)$, i.e. that both firms enter the
market. Hence, $\Theta_I=[\sqrt{p},1]$.
\end{continued}

The first thing to note from this theorem is that the problem of
computing the identified set has been transformed into a finite
dimensional problem in the special case where ${\mathcal{Y}}$ is a
finite set (or equivalently, the support of the distribution $P$ of
observable outcomes has finite cardinality). Indeed, in the latter
case, the problem of computing the identified set is reduced to the
problem of computing a finite number of inequalities, i.e. $P(A\vert
X)\leq\mathcal{L}(A\vert X;\theta)=\nu(\epsilon:\;G(\epsilon\vert
X;\theta)\cap A\ne\varnothing\vert X;\theta)$ for each subset $A$ of
$\mathcal{Y}$. However, in cases where the cardinality of
$\mathcal{Y}$ is large, then the number of inequalities to be
checked is $2^{\mathrm{Card}(\mathcal{Y})}-2$, and the computational
burden is only partially lifted, and the second section of the paper
is devoted to efficient methods of computation of the identified set
based on the characterization of theorem~\ref{theorem:sharp}. First
we turn to the specialization of our results and concepts to some
leading examples, and illustrate them with a simplified version of a
family bargaining game studied in the literature.

\subsection{Some illustrative examples}
\label{subsection:leading examples}

\subsubsection{Models of market entry}\label{subsubsection: entry game}

A leading special example of the framework above is that of
empirical models of oligopoly entry, proposed in \cite{BR:90}
and \cite{Berry:92}, and considered in the framework of
partial identification by \cite{Tamer:2003},
\cite{ABJ:2003}, \cite{BT:2006}, \cite{CT:2006}
and \cite{PPHI:2004} among others. In this setup, economic
agents are firms who decide whether of not to enter a market.
Markets are indexed by $m$, $m=1,\ldots,M$ and firms that could
potentially enter the market are indexed by $i$, $i=1,\ldots,J$.
$Y_{im}$ is firm $i$'s strategy in market $m$, and it is equal to
$1$ if firm $i$ enters market $m$, and zero otherwise. $Y_m$ denotes
the vector $(Y_{1m},\ldots,Y_{Jm})^t$ of strategies of all the
firms. In standard notation, $Y_{-im}$ denotes the vector of
strategies of all firms except firm $i$. In models of oligopoly
entry, the profit $\pi_{im}$ of firm $i$ in market $m$ is allowed to
depend on strategies $Y_{-im}$ of other firms, as well as on a set
of profit shifters $X_{im}$ that are observed by all firms and the
econometrician, a profit shifter $\epsilon_{im}$ that is observed by
all the firms but not by the econometrician, and a vector of unknown
structural parameters, so that it can be written
$\pi_{im}(Y_m,X_{im},\epsilon_{im};\theta)=\pi_{im}(Y_{im},Y_{-im},X_{im},\epsilon_{im};\theta)$. If, for instance, firms are assumed to play Nash equilibria in pure
strategies in market $m$, their strategies $Y_{im}$ are such that
they yield higher profits than $1-Y_{im}$ given other firms'
strategies $Y_{-im}$. So the restrictions induced on the strategies
and latent profit shifters are
$\pi_{im}(Y_{im},Y_{-im},X_{im},\epsilon_{im};\theta)\geq\pi_{im}(1-Y_{im},Y_{-im},X_{im},\epsilon_{im};\theta)$
for all $i=1,\ldots,J$. Hence the model can be written $Y_m\in
G(\epsilon_m\vert X_m;\theta)$, where $X_m$ denotes the matrix of
observed profit shifters for firms $i=1,\ldots,J$, $\epsilon_m$
denotes the vector of latent profit shifters for firms
$i=1,\ldots,J$, and the correspondence $G$ is defined by
$G(\epsilon\vert
X;\theta)=\{Y:\;\pi_{i}(Y_{i},Y_{-i},X_{i},\epsilon_{i};\theta)
\geq\pi_{i}(1-Y_{i},Y_{-i},X_{i},\epsilon_{i};\theta);\mathrm{ all }
\;i=1,\ldots,J\}$, where the index $m$ is dropped when considering a
generic market.

\subsubsection{Family bargaining}\label{subsubsection:family bargaining}
For illustration purposes, we consider a simplified version of the
bargaining model of decision regarding the long term care of an
elderly parent in \cite{ES:2002}.

Consider a family with two children. The issue is which of the
children will become the primary care giver of an elderly and
disabled parent or whether the parent moves to a nursing home.

The payoff to family member $i$, $i=1,2$ is represented as the sum
of three terms. The first term $V_{ij}$ represents the value to
child $i$ of care option $j$, where $j>0$ means child $j$ becomes
the primary care giver and $j=0$ means the parent is moved to a
nursing home. The matrix $V=(V_{ij})_{ij}$ is known to both
children. We suppose it takes the form
\[V=\left(\begin{array}{ccc}0 &2\theta& 4\theta\\0&
4\theta &2\theta\end{array}\right)\] where $\theta$ is a nonnegative
parameter, unknown to the analyst.

Both children simultaneously decide whether or not to take part in
the long term care decision. Suppose $M$ is the set of children who
participate. The option chosen is option $j$ which maximizes the sum
$\sum_{i\in M}V_{ij}$ among the available options (only
participating children can become primary care givers). It is
assumed that participants abide with the decision and that benefits
are then shared equally among children participating in the decision
through a monetary transfer $s_i$, which is the second term in the
children's payoff.

The third term $\epsilon_i$ in the payoff is a random benefit from
participation, which is $0$ for children who decide not to
participate and distributed according to $\nu(.\vert \theta)$ for children
who participate. All children observe the realizations of
$\epsilon$, whereas the analyst only knows its distribution.

The matrix of payoffs of the participation game is given in
table~\ref{table:payoffs}.

\begin{table}\label{table:payoffs}
\begin{center}
\begin{tabular}{c|c|c}Child 1$\backslash$ Child 2&N&P\\\hline
N&$0,0$&$4\theta,2\theta+\epsilon_2$\\\hline
P&$2\theta+\epsilon_1,4\theta$&$3\theta+\epsilon_1,3\theta+\epsilon_2$\\\hline
\end{tabular}
\end{center}
\caption{}\end{table}

We derive the equilibrium correspondence, restricting the analysis to pure strategy Nash equilibria for now. We shall later add equilibria in mixed strategies to illustrate results of section~\ref{section:mixed strategies}.

\begin{itemize}
\item[] $\{(0,0)\}$ is a Nash equilibrium in pure strategies if and only
if $\epsilon_2<-2\theta$ and $\epsilon_1<-2\theta$.
\item[] $\{(1,1)\}$ is a Nash equilibrium in pure strategies if and only
if $\epsilon_2>\theta$ and $\epsilon_1>\theta$.
\item[] $\{(0,1)\}$ is a Nash equilibrium in pure strategies if and only
if $\epsilon_2>-2\theta$ and $\epsilon_1<\theta$.
\item[] $\{(1,0)\}$ is a Nash equilibrium in pure strategies if and only
if $\epsilon_2<\theta$ and $\epsilon_1>-2\theta$.
\end{itemize}

\begin{figure}[htbp!]
\centering
\subfigure[{\scriptsize Model Correspondence}]{
\includegraphics[width=8cm]{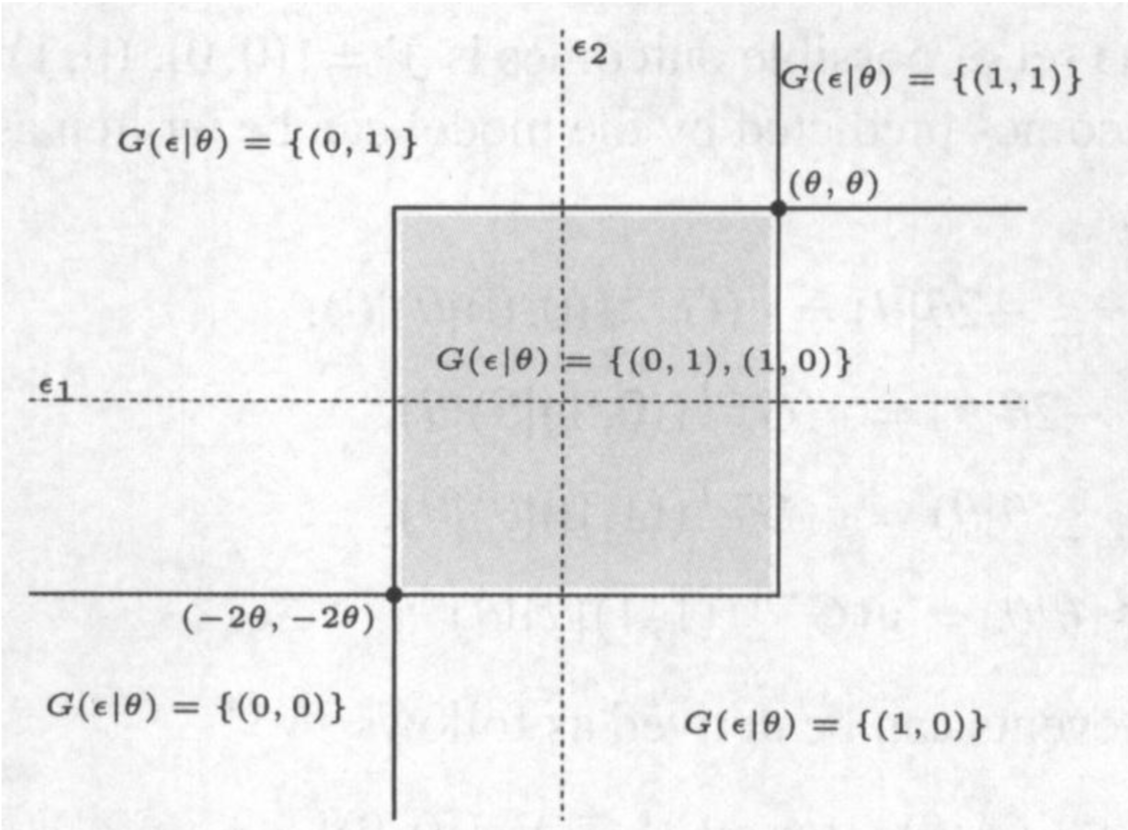}
\label{figure:FB22}
}
\subfigure[{\scriptsize Inequality to check}]{
\includegraphics[width=8cm]{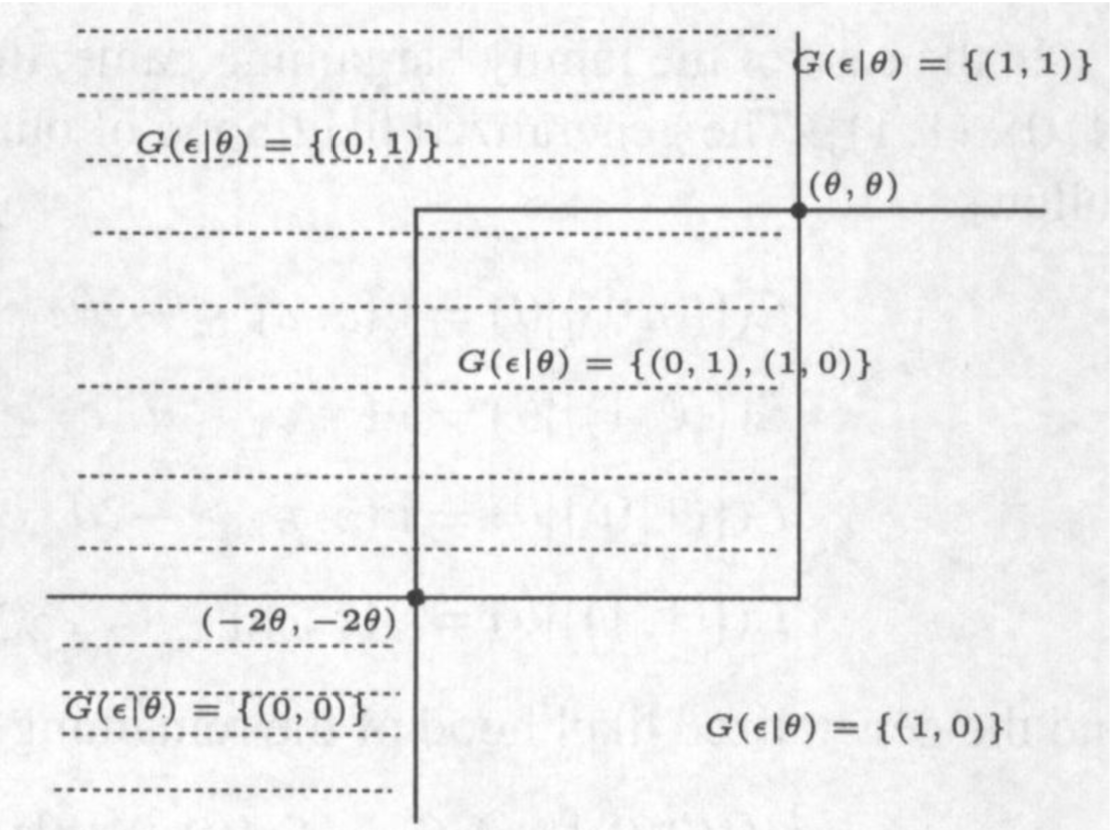}
\label{figure:FB22inequalities}
}
\caption[]{Family Game}
\label{figure:FB}
\end{figure}

The equilibrium correspondence $G(\epsilon\vert \theta)$ is
represented in figure~\ref{figure:FB}(a) as a function of
$\epsilon$. The dotted lines represents the axes in the $\epsilon$
space, and the full lines represent the frontiers of the regions
defining the correspondence $G$. The shaded area is the area of
multiplicity, where $G(\epsilon\vert X;\theta)$ contains two values
$(0,1)$ and $(1,0)$.

The model thus described is incomplete in the sense that more
information is required in the regions of multiplicity to determine
which equilibrium will be selected. Without knowledge of such an
equilibrium selection mechanism, the likelihood predicted by the
model can be written as follows. As before $\mathcal{Y}$ denotes the
set of possible outcomes. The likelihood of observation $y$ is
$\mathcal{L}(y\vert \theta)=\nu(\epsilon:\;y\in G(\epsilon\vert
\theta)\vert \theta) =\nu(G^{-1}(y\vert \theta)\vert
\theta)$, for all $y\in\mathcal{Y}$ and
$\sum_{y\in\mathcal{Y}}\mathcal{L}(y\vert \theta)\geq1$, where the
inequality may be strict if there are regions of multiplicity.

\begin{continued}[Example \protect\ref{subsubsection:family bargaining} continued]
In the case of the family bargaining game, the set of possible
outcomes is $\mathcal{Y} = \{(0,0)$, $(0,1)$, $(1,0)$, $(1,1)\}$.
The likelihood of outcomes predicted by the model can be written as
follows.
\begin{eqnarray*}
&&\mathcal{L}(\{(0,0)\}\vert
\theta)=\nu(\epsilon:\;\epsilon_1\leq-2\theta,\;
\epsilon_2\leq-2\theta\vert \theta)
=\nu(G^{-1}((0,0)\vert \theta)\vert \theta)\\
&&\mathcal{L}(\{(0,1)\}\vert \theta)=
\nu(\epsilon:\;\epsilon_1\leq\theta,\;\epsilon_2\geq-2\theta\vert \theta)
=\nu(G^{-1}((0,1)\vert \theta)\vert \theta)\\
&&\mathcal{L}(\{(1,0)\}\vert \theta)=
\nu(\epsilon:\;\epsilon_1\geq-2\theta,\;\epsilon_2\leq\theta\vert \theta)
=\nu(G^{-1}((1,0)\vert \theta)\vert \theta)\\
&&\mathcal{L}(\{(1,1)\}\vert \theta)=
\nu(\epsilon:\;\epsilon_1\geq\theta,\;\epsilon_2\geq\theta\vert \theta) =\nu(G^{-1}((1,1)\vert \theta)\vert\theta)
\end{eqnarray*}
and the likelihood of the remaining events can be derived as follows
\begin{eqnarray*} &&\mathcal{L}(\{(0,0)\}\cup A\vert \theta)=
\mathcal{L}(\{(0,0)\}\vert \theta)+\mathcal{L}(A\vert
\theta),\;\mbox{ for all }\;A\subseteq\mathcal{Y}\backslash\{(0,0)\}\\
&&\mathcal{L}(\{(1,1)\}\cup A\vert \theta)=
\mathcal{L}(\{(1,1)\}\vert \theta)+\mathcal{L}(A\vert
\theta),\;\mbox{ for all }\;A\subseteq\mathcal{Y}\backslash\{(1,1)\}\\
&&\mathcal{L}(\{(0,1),(1,0)\}\vert \theta)=
1-\mathcal{L}(\{(0,0),(1,1)\}\vert \theta)
\end{eqnarray*}
The likelihood predicted by the model is the set function
$A\mapsto\mathcal{L}(A\vert \theta)=\nu (G^{-1}(A\vert
\theta)\vert \theta)$ for $A$ subset of $\mathcal{Y}= \{(0,1)$,
$(0,1)$, $(1,0)$, $(1,1)\}$. This set function is a Choquet
capacity and if the support of $\nu$ is sufficiently large, the
likelihood sums to more than one, because the region of multiple
equilibria is ``counted twice''. This is related to the non-additive
feature of Choquet capacities, as seen here with the inequality
$\nu(G^{-1}(\{(0,1)\}\cup\{(1,0)\}\vert \theta)\vert \theta)<
\nu(G^{-1}(\{(0,1)\}\vert \theta)\vert
\theta)+\nu(G^{-1}(\{(1,0)\}\vert \theta)\vert \theta)$, since
the latter is equal to the former plus
$\nu(\epsilon:\;-2\theta\leq\epsilon_1\leq\theta,
-2\theta\leq\epsilon_2\leq\theta\vert \theta)$.
\end{continued}

The model can be completed by adding an equilibrium selection
mechanism which will pick out a single equilibrium for each value of
the latent variable $\epsilon$ in the region of multiplicity. As
formally defined in the previous section, an equilibrium selection
mechanism is a conditional probability $\pi(.\vert\epsilon,X)$ with
support included in $G(\epsilon\vert X;\theta)$. It is compatible
with the data if the probabilities it predicts are equal to the true
probabilities of the observable variables.

\begin{continued}[Example \protect\ref{subsubsection:family bargaining} continued]
In the family bargaining example example, we have for $i,j=0,1$:
\begin{eqnarray*}
P((i,j)\vert
X)=\int_\mathcal{U}\pi((i,j)\vert\epsilon,X)\nu(d\epsilon\vert X;\theta)
\end{eqnarray*}
\end{continued}

As noted above, since the model contains no prior information about
which outcome is selected in the regions of multiplicity, the
identified set $\Theta_I$ for the parameter vector $\theta$ is the
set of parameters for which one can find an equilibrium selection
mechanism which completes the model and equates probabilities of
outcomes predicted by the model with true outcome frequencies. The
definition of the identification region using a semiparametric
likelihood representation, where the equilibrium selection mechanism
is included as the infinite dimensional nuisance parameter $\pi$ is
impractical, so we use theorem~\ref{theorem:sharp} to provide an
operational method to compute $\Theta_I$. The existence of a
compatible selection mechanism is equivalent to the fact that the
true distribution $P$ of observed outcomes lies in the core of the
Choquet capacity $\mathcal{L}=\nu G^{-1}$ defined by the model.
Hence, we have $\Theta_I=\{\theta\in\Theta:\;\left( \forall
A\in2^\mathcal{Y};\;P(A\vert X)\leq\mathcal{L}(A\vert
X;\theta);\right) \;X\,\mathrm{ a.s.}\}$ where $2^S$ denotes the set
of all subsets of a set $S$.

\begin{continued}[Example \protect\ref{subsubsection:family bargaining} continued]
In the case of the family bargaining game, the identified region is
the set of parameter vectors that satisfy the 16 inequalities
$P(A)\leq \nu(G^{-1}(A\vert \theta)\vert \theta)$ for all sets $A$
in $2^\mathcal{Y}$. Figure~\ref{figure:FB}(b) gives a representation
of one of the inequalities to be checked. The probability of the
outcome being either $(0,0)$ or $(0,1)$ needs to be no larger than
the probability of the latent variable lying in the set covered with
horizontal dashed lines.
\end{continued}

\section{Efficient computation of the identified set: which inequalities to check, and how to check them?}
\label{section:efficient computation} We now describe three
approaches to the effective computation of the identified set based
on our characterization of theorem~\ref{theorem:sharp}.

The first approach, described in
section~\ref{subsection:submodular} is based on the observation that the likelihood is a submodular set function (the set function analogue of a convex function), and that checking the condition of theorem~\ref{theorem:sharp} is equivalent to minimizing a submodular function, which is a well studied problem (analogous to minimizing a convex function) and easily and efficiently implementable. It extends readily to
the more general case when equilibria in mixed strategies are also
allowed, as will be discussed in the computational part of
section~\ref{section:mixed strategies}.

The second approach, described in section~\ref{subsection:transportation}, is a special case of the first, which relies on the highly efficient algorithms (and easily
available packaged implementations) for optimal transportation
problems. As a combinatorial optimization method, it is a well known special case of the second approach, and it is computationally more efficient, and therefore recommended to practitioners who restrict attention to equilibria in pure strategies.

The third approach, based on the notion of \emph{core determining} classes of
sets and providing a dramatic reduction in the computational complexity
under specific assumptions on the game under study, is described in
section~\ref{subsection:cd}, where it is shown how the exponential problem of checking
$2^{|\mathcal{Y}|}-2$ inequalities in theorem~\ref{theorem:sharp}
can be replaced by the problem polynomial problem of checking
$2|\mathcal{Y}|-2$ inequalities instead.

\subsection{Submodular optimization}\label{subsection:submodular}
The first proposal to deal with the complexity of the problem of checking inequalities in theorem~\ref{theorem:sharp} is a method of general validity based on the minimization of a submodular function, the discrete equivalent of a convex function, which is a well known problem in combinatorial optimization, and for which efficient algorithms and easily available off the shelf implementations exist.

\begin{definition}[Submodular function]
A set function $\mathcal{L}:\mathcal{Y}\rightarrow\mathbb{R}$ is called submodular if for each $A,B\subseteq\mathcal{Y}$, we have $\mathcal{L}(A\cup B)+\mathcal{L}(A\cap B)\leq\mathcal{L}(A)+\mathcal{L}(B)$. Note that in the special case where $\mathcal{L}$ is a probability measure, the latter inequality holds as an equality.\label{definition:submodular}\end{definition}

Submodularity for set functions is the analogue of convexity, and the problem
of minimizing a submodular function is well studied (see for instance \cite{Topkis:98} chapter~2). A complete account of the theory can be found in \cite{Fujishige:2005} and off the shelf matlab implementation of the most efficient known algorithms can be found in Andreas Krause's SFO toolbox.

We now show that checking inequalities involved in the characterization of the identified set in theorem~\ref{theorem:sharp} is indeed equivalent to the minimization of a submodular function. Theorem~\ref{theorem:sharp} shows that the identified set is the set of values of $\theta$ such that $X$-almost surely, we have the domination $\forall A\subseteq\mathcal{Y}$, $P(A\vert X)\leq\mathcal{L}(A\vert X;\theta)$, or equivalently $\min_{A\subseteq\mathcal{Y}}\left(\mathcal{L}(A\vert X;\theta)-P(A\vert X)\right)\geq0$. We first note that the function in the minimization above is indeed submodular.

\begin{lemma}[Submodularity of the likelihood] For all $\theta\in\Theta$, $X$-almost surely, the set function on $\mathcal{Y}$ defined for all $A\subseteq\mathcal{Y}$ by $A\mapsto\mathcal{L}(A\vert X;\theta)-P(A\vert X)$ is submodular.\label{lemma:submodular}\end{lemma}

The most efficient generic way to check that a convex function is everywhere non negative is to minimize it and to verify that the minimum is non negative. In the same way, we propose to check inequalities of theorem~\ref{theorem:sharp} by minimizing the submodular function defined for all $A\subseteq\mathcal{Y}$ by $A\mapsto\mathcal{L}(A\vert X;\theta)-P(A\vert X)$, and checking that the minimum is indeed non negative. Of course, we can speed up the algorithm by stopping short of the minimum as soon as a negative value is found.

\begin{theorem}[Computation of the identified set]\label{theorem:submodular function}
The identified set is obtained by minimization of a submodular set function: $\Theta_I=\left\{\theta\in\Theta:\;\min_{B\subseteq\mathcal{Y}}
\left(\mathcal{L}(B\vert X;\theta)-P(B\vert X)\right)=0,\;X-\mbox{a.s.}\right\}$.
\end{theorem}

More details of the procedure are given in
section~\ref{section:mixed strategies}, as this method is one of the
recommended methods of construction of the identified set in the
case where equilibria in mixed strategies are considered. Below is a
description of a special case of submodular optimization, which is
more efficient, and applies to the case where only equilibria in
pure strategies are considered.

\subsection{Optimal transportation approach}
\label{subsection:transportation}

In the special case with only pure strategy equilibria that we are considering until section~\ref{section:mixed strategies}, the model likelihood $\mathcal{L}$ is a very special case of submodular function, since it is derived as the distribution function of a random set, or random correspondence $\mathcal{L}(A\vert X;\theta)=\nu(\epsilon:\;G(\epsilon\vert X;\theta)\cap A\ne\varnothing\vert X;\theta)$. It is that property that allows the following refinement of the method, and the simpler and more efficient technique proposed below. When mixed equilibria are considered, this improvement in efficiency is no longer available, precisely because in general the model likelihood is no longer the distribution of a random set.

To describe the method, we need the following notations and
definitions. The terminology used is intended to be
self-explanatory. Otherwise, the reader is referred to
\cite{PS:98} for standard definitions in graph theory. Call
$\mathcal{U}^\ast$ the set of predicted combinations of equilibria,
 formally $\mathcal{U}^\ast=\{G(\epsilon\vert
X;\theta);\epsilon\in\mathcal{U}\}$ (we suppress reference to the
dependence of $\mathcal{U}^\ast$ on $\theta$ for notational
convenience). Hence $\mathcal{U}^\ast$ contains subsets of
$\mathcal{Y}$, but is typically of much lower cardinality than the
power set $2^\mathcal{Y}$. Further consider the bi-partite graph
$\mathcal{G}(\theta,X)$ in $\mathcal{Y}\times\mathcal{U}^\ast$. The
latter is defined as the set of pairs
$(y,u)\in\mathcal{Y}\times\mathcal{U}^\ast$ such that $y\in u$. Each
vertex $y$ in $\mathcal{Y}$ has weight $\mathbb{P}(Y=y\vert X)$ and
each vertex $u\in\mathcal{U}^\ast$ has weight
$\mathbb{P}(G(\epsilon\vert X;\theta)=u\vert X)$. The graph contains
edges $(y,u)$ linking an element $y\in\mathcal{Y}$ to an element
$u\in\mathcal{U}^\ast$ if the former is an element of the latter
(i.e. $y\in u$). Finally, call $P(y\vert X)=\mathbb{P}(Y=y\vert X)$
the actual probabilities of observable variables $y\in\mathcal{Y}$,
and call $Q(.\vert X;\theta)$ the probabilities $Q(u\vert
X;\theta)=\mathbb{P}(G(\epsilon\vert X;\theta)=u\vert X)$. If we
consider $G$ (keeping the same notation for simplicity) as a
correspondence from $\mathcal{U}^\ast$ to $\mathcal{Y}$, then,
formally $G(u)=u$, and we have shown in theorem~\ref{theorem:sharp}
that $\theta$ belongs to the identified set if and only if for any
subset $A$ of $\mathcal{Y}$, $P(A\vert X)\leq Q(G^{-1}(A)\vert
X;\theta)$. \cite{GH:2006d} show that it is equivalent to the
existence of a joint probability $\pi$ on $\mathcal{G}(\theta,X)$
with marginal distributions $P(.\vert X)$ and $Q(.\vert X;\theta)$.
This is summarized in the following proposition\footnote{ This
result is a special case of equivalence between (ii') and (iii') in
theorem~1' of the previous version circulated
\cite{GH:2006a}.}.

\begin{theorem}\label{theorem:matching}
The parameter value $\theta$ belongs to the identified set if and only if
there exists a probability on $\mathcal{Y}\times\mathcal{U}^\ast$ with domain $\mathcal{G}(X;\theta)$
and with marginal probabilities $P(.\vert X)$ and $Q(.\vert X;\theta)$.
\end{theorem}

Note that one implication in theorem~\ref{theorem:matching} is very easy to prove.
Call $U$ the random element with distribution $Q$. If a joint probability
exists with the required properties, then ${Y\in A}\Rightarrow U\in G^{-1}(A)$, so
that $1_{\{Y\in A\}}\leq1_{\{U\in G^{-1}(A)\}}$, $\pi$-almost surely.
Taking expectation, we have $\mathbb{E}_\pi(1_{\{Y\in A\}})\leq
\mathbb{E}_\pi(1_{\{U\in G^{-1}(A)\}})$, or equivalently $P(A\vert X)\leq Q(G^{-1}(A))\vert X;\theta)$.
The converse is much more involved and relies on optimal transportation theory.
A similar result is proved in theorem~3 of \cite{Artstein:83}, based on an extension of the marriage lemma.

We illustrate this requirement on our
example~\ref{subsubsection:family bargaining}.

\begin{figure}[htbp]
\centering
\includegraphics[width=15cm]{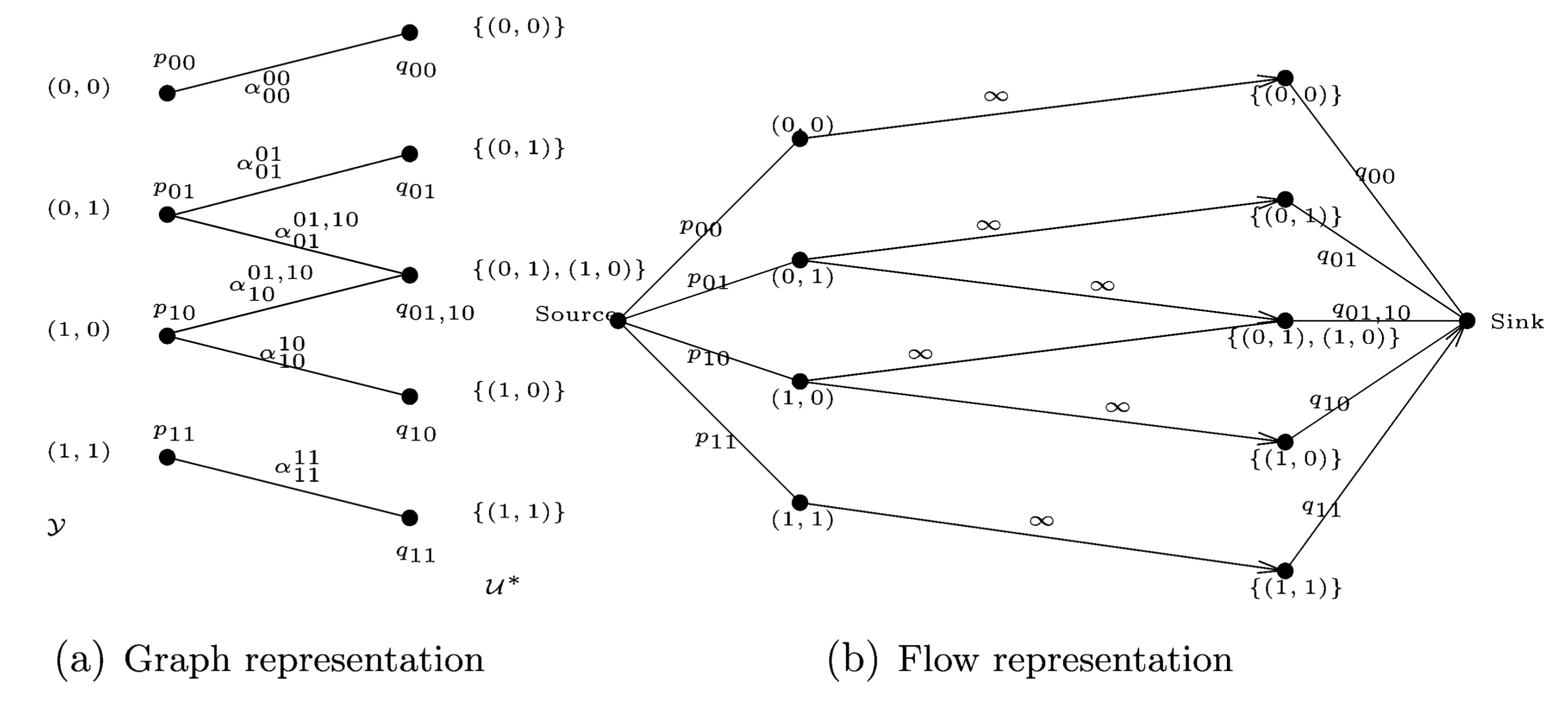}
\caption[]{Family game (continued)}
\label{figure:FBgraph}
\end{figure}

\begin{continued}[Example \protect\ref{subsubsection:family bargaining} continued]
For the case of the family bargaining game,
$\mathcal{U}^\ast=\large\{\{(0,0)\}$, $\{(0,1)\}$, $\{(1,0)\}$,
$\{(1,1)\}$, $\{(0,1),(1,0)\}\large\}$. The bi-partite graph
representing the admissible connections between observable outcomes
and combinations of equilibria is shown in
figure~\ref{figure:FBgraph}(a), where $p_y$ denotes
$\mathbb{P}(Y=y\vert X)$ and $q_u=\mathbb{P}(G(\epsilon\vert
X;\theta)=u\vert X)$. The existence of a joint probability on
$\mathcal{Y}\times\mathcal{U}^\ast$ supported on
$\mathcal{G}(X;\theta)$ with marginal probabilities $p_y$,
$y\in\mathcal{Y}$ and $q_u$, $u\in\mathcal{U}^\ast$ can be
represented graphically by a set of non negative numbers attached to
each edge of the graph, that sum to 1, and such that the weight of
each vertex is equal to the sum of the weights on the edges that
reach it. For instance, a joint probability is denoted $\alpha_y^u$,
$(y,u)\in\mathcal{Y}\times\mathcal{U}^\ast$ and must satisfy
$\alpha_y^u\geq0$ for all
$(y,u)\in\mathcal{Y}\times\mathcal{U}^\ast$, $\alpha_y^u=0$ if
$y\notin u$, $\sum_{y\in u}\alpha_y^u=1$ and equalities such as
$p_{01}=\alpha_{01}^{01}+\alpha_{01}^{01,10}$ and
$q_{01,10}=\alpha_{01}^{01,10}+\alpha_{10}^{01,10}$.
\end{continued}

Since we have now formulated the problem of computing the identified
set as a problem involving the existence of a probability measure
with given marginal distributions, we can appeal to efficient
computational methods in the optimal transportation literature. The
problem of sending $p_y$, $y\in\mathcal{Y}$ \emph{units} of a
\emph{good} from sources $y\in\mathcal{Y}$ to $q_u$,
$u\in\mathcal{U}^\ast$ \emph{units} in \emph{terminals}
$u\in\mathcal{U}^\ast$ at minimum cost of transportation, where
costs are attached to each pair
 $(y,u)\in\mathcal{Y}\times\mathcal{U}^\ast$ is called an optimal transportation problem, and first appears in the economics literature with \cite{Koopmans:49}.
 Our problem can be reduced to an optimal transportation problem
 with 0-1 cost of transportation, where a pair $(y,u)$ is assigned cost zero if it belongs to $\mathcal{G}(X;\theta)$,
 and 1 otherwise, and there exists a joint law on $\mathcal{G}(X;\theta)$ with marginals $P$ and $Q$ (in other words,
 $\theta$ is in the identified set) if and only if there is a zero cost solution to the optimization problem thus defined.

As explained in \cite{FF:57} (see also \cite{PS:98}
section~7.4 page~143), there is an equivalent dual formulation of
this minimum cost of transportation problem as a maximum flow
problem described in figure~\ref{figure:FBgraph}(b). The edges in
the graph with zero cost in the minimum cost of transportation
problem have infinite \emph{carrying capacity} (not to be confused
with Choquet capacity) in the dual maximal flow problem. Hence
efficient maximum flow programs (such as maxflow.m in the Matlab BGL
library) can be applied directly to the network described in
figure~\ref{figure:FBgraph}(b): Mass flows from the source to the
sink through the network. The number on each edge is the maximum
mass that can flow through that edge. The source sends $p_y$ mass
exactly to each node corresponding to elements of $\mathcal{Y}$, and
the sink receives $q_u$ mass from each node corresponding to an
element of $\mathcal{U}^\ast$. Between edges in $\mathcal{Y}$ and
$\mathcal{U}^\ast$, mass can flow freely through pairs $(y,u)$ such
that $y\in u$ (full lines with infinite carrying capacity), and not
at all through pairs $(y,u)$ such that $y\notin u$. The parameter
value $\theta$ is in the identified set if and only if the maximum
flow program returns a maximum flow of exactly $1$ (note that the
network carrying capacities depend on $\theta$ through the
probabilities $q_u$). The full procedure is described in detail on
the oligopoly entry example of section~\ref{section:BR44}.

\subsection{Core determining classes}
\label{subsection:cd} As we have seen in the first section,
theorem~\ref{theorem:sharp} allows to reduce the problem of
computing the identified set to that of checking a set of
inequalities. However, the computational burden is only partially
lifted, as the number of inequalities to check can be very large if
the cardinality of the outcome space is large. In this section, we
shall analyze ways of reducing this remaining computational burden,
by eliminating redundant inequalities in the computation of the
identified set. This is formalized with the concept of \emph{core
determining classes}, which was first introduced in section~3.2.2
page 27 of \cite{GH:2006a}.

\begin{definition}
A class $\mathcal{A}$ of measurable subsets of $\mathcal{Y}$ is
called \emph{core determining} for the Choquet capacity
$\mathcal{L}$ on $\mathcal{Y}$ if it is sufficient to characterize
the core of $\mathcal{L}$, i.e. if a probability $Q$ is in
$\mathrm{core}(\mathcal{L})$ when $Q(A)\leq\mathcal{L}(A)$ for all
$A\in\mathcal{A}$. In other words, $Q(A) \leq\mathcal{L}(A)$ for all
$A\in\mathcal{A}$ implies $Q(A)\leq\mathcal{L}(A)$ for all $A$ measurable.
\end{definition}

A core determining class $\mathcal{A}$ allows the elimination of all
the inequalities $Q(A)\leq\mathcal{L}(A)$ for $A\notin\mathcal{A}$
when checking whether a probability $Q$ belongs to the core of a
Choquet capacity $\mathcal{L}$. Since the likelihood predicted by
the model $Y\in G(\epsilon\vert X;\theta)$ was characterized by the
Choquet capacity $A\mapsto\mathcal{L}(A\vert
X;\theta)=\nu(\epsilon:G(\epsilon\vert X;\theta)\cap
A\ne\varnothing\vert X;\theta)$, a core determining class of sets is
sufficient to characterize the identified region $\Theta_I$ as
summarized in the following proposition.

\begin{proposition}
If $\mathcal{A}$ is core determining for the Choquet
capacity $A\mapsto\mathcal{L}(A\vert
X;\theta)=\nu(\epsilon:G(\epsilon\vert X;\theta)\cap
A\ne\varnothing\vert X;\theta)$ $X$-almost surely, and for all $\theta$, then $\Theta_I=\{\theta\in\Theta:
\forall A\in\mathcal{A}(\theta),\,P(A\vert X)\leq \mathcal{L}(A\vert
X),\,X-\mathrm{a.s.}\}$.
\end{proposition}

The challenge therefore becomes that of finding a core determining
class $\mathcal{A}$ in order to reduce the number of inequalities to
be checked to the cardinality of $\mathcal{A}$. We first consider
the case of our example~\ref{subsubsection:family bargaining},
before turning to a criterion that will prove useful in exhibiting
core determining classes in many important cases.

\begin{continued}[Example \protect\ref{subsubsection:family bargaining} continued]
We return to the family bargaining game and consider some proposals
for the computation of the identified set proposed in the
literature. We call \emph{Singleton class} the class of singleton
sets $(\{(0,0)\}$, $\{(0,1)\}$, $\{(1,0)\}$, $\{(1,1)\})$, since it
corresponds to the class of sets proposed in \cite{ABJ:2003}
specialized to this simple case. It is immediate to see that the
\emph{Singleton} class is not core determining in general. Indeed,
if $\epsilon$ has large enough support, the two equalities
$P(\{(0,1)\})=\nu(G(\epsilon\vert
\theta)\cap\{(0,1)\}\ne\varnothing\vert \theta)=
\nu(\epsilon:\;\epsilon_1\leq\theta,\;\epsilon_2\geq-2\theta\vert
\theta)$ and $P(\{(1,0)\})=\nu(G(\epsilon\vert
\theta)\cap\{(1,0)\}\ne\varnothing\vert \theta)=
\nu(\epsilon:\;\epsilon_1\geq-2\theta,\;\epsilon_2\leq\theta\vert
\theta)$ jointly imply $P(\{(0,1),(1,0)\})>\nu(G(\epsilon\vert
\theta)\cap\{(0,1),(1,0)\}\ne\varnothing\vert \theta)
=\nu(\epsilon:\;[\epsilon_1\geq-2\theta\,\mathrm{ or
}\,\epsilon_2\geq-2\theta] \,\mathrm{ and
}\;[\epsilon_1\leq\theta\,\mathrm{ or }\,\epsilon_2\leq\theta]\vert
\theta)$.
\end{continued}

We now show how to identify core determining classes more generally
to avoid painstaking case-by-case elimination of redundant
inequalities. To that end, we give general conditions under which
one can find a core determining class of low cardinality. Recall
that a subset $A$ of an ordered set (with ordering $\preceq$) is
said to be \emph{connected} if any $a$ such that $\inf A\preceq
a\preceq \sup A$ belongs to $A$.

\begin{assumption}[Monotonicity]\label{assumption:monotonicity}
There exists an ordering $\precsim_{\mathcal{Y}}$ of the set of outcomes $\mathcal{Y}$ and an ordering
$\precsim_{\mathcal{U}}$ of the set of
latent variables $\mathcal{U}$ such that
$G(\epsilon\vert X;\theta)$ is connected for all $\epsilon\in\mathcal{U}$, $X$-a.s., all $\theta$,
and $\sup G(\epsilon_2\vert X;\theta)\succsim_\mathcal{Y}\sup G(\epsilon_1\vert X;\theta)$
and $\inf G(\epsilon_2\vert X;\theta)\succsim_\mathcal{Y}\inf G(\epsilon_1\vert X;\theta)$
when $\epsilon_1\precsim_\mathcal{U}\epsilon_2$.
Both ordering are allowed to depend on the exogenous variables $X$,
but the dependence is suppressed in the notation for clarity.
\end{assumption}

\begin{remark}
This assumption is related to monotone comparative statics in
supermodular games (see \cite{Topkis:98},
\cite{Vives:90} and \cite{MR:90}).\end{remark}

We illustrate this assumption on our family bargaining game before
stating the theorem and applying it to the more sophisticated case
of an oligopoly entry game with two types of players presented in
\cite{BT:2006}.

\begin{figure}[htbp]
\begin{center}
\includegraphics[width=15cm]{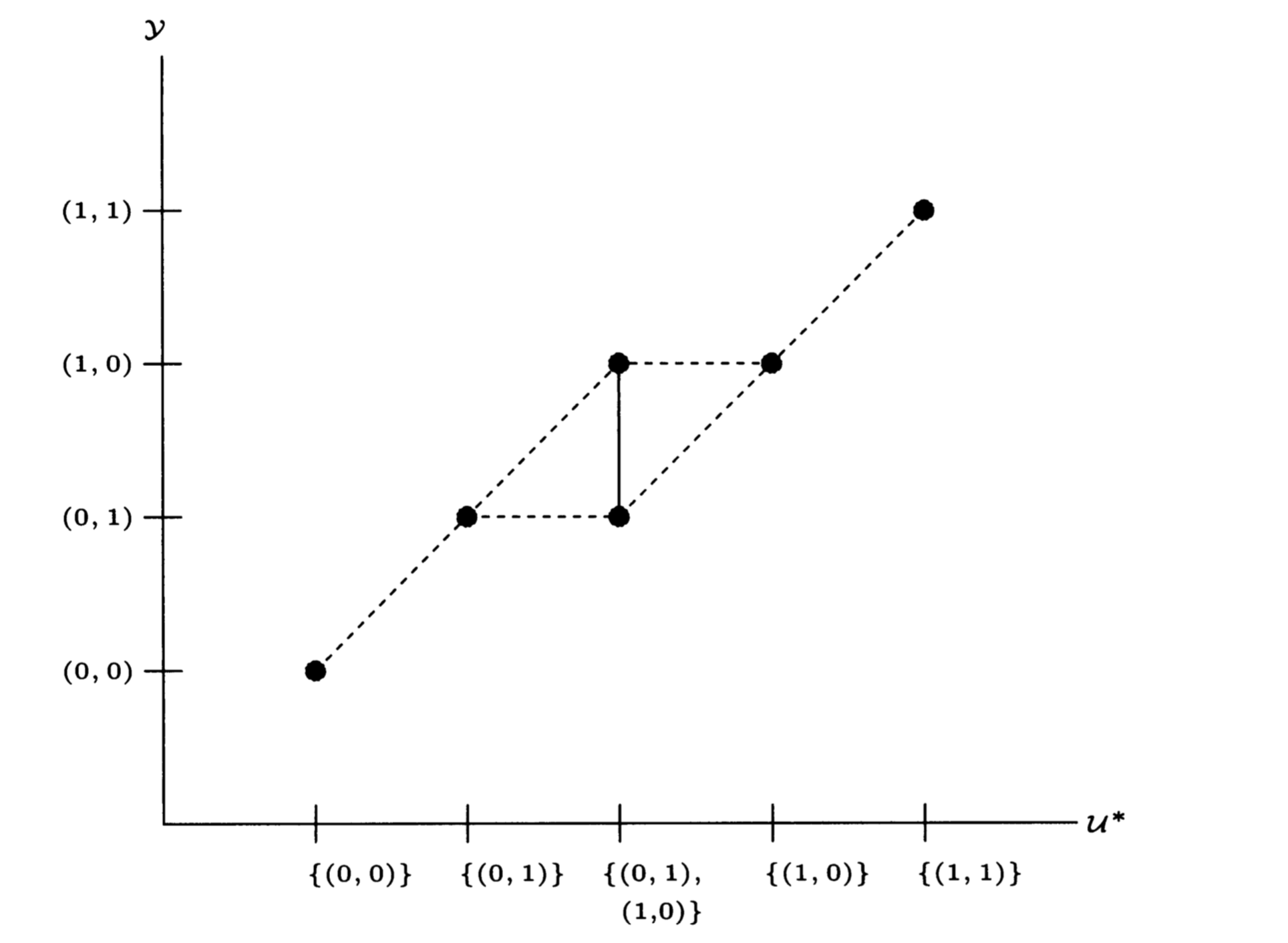}
\caption{{\scriptsize The monotonicity requirement in
assumption~\ref{assumption:monotonicity} is satisfied for this
choice of orderings in the family bargaining example. (The thick dots
represent the correspondence $G(.\vert \theta)$).
$\mathcal{U}^\ast$ denotes the ordered set of combinations of
equilibria.}} \label{figure:FB22mon}
\end{center}
\end{figure}

\begin{continued}[Example \protect\ref{subsubsection:family bargaining} continued]
In the family bargaining game, the orderings are very simple to
construct. A lexicographic order on $\mathcal{Y}$ is suitable, with
$(0,0)\precsim_\mathcal{Y}(0,1)\precsim_\mathcal{Y}(1,0)\precsim_\mathcal{Y}(1,1)$.
On $\mathcal{U}$ the ordering is related to predicted outcomes. All
$\epsilon$ producing the same predicted outcomes will be equivalent,
and the ordering on predicted outcomes is
$\{(0,0)\}\precsim_\mathcal{U}\{(0,1)\}\precsim_\mathcal{U}\{(0,1),(1,0)\}\precsim_\mathcal{U}\{1,0\}
\precsim_\mathcal{U}\{1,1\}$, where $A_1\precsim_\mathcal{U}A_2$ is
a short-hand notation for $\epsilon_1\precsim_\mathcal{U}\epsilon_2$
if $G(\epsilon_1\vert X;\theta)=A_1$ and $G(\epsilon_2\vert
X;\theta)=A_2$. It is straightforward to check
assumption~\ref{assumption:monotonicity}. For instance, taking
$\epsilon_1$ such that $G(\epsilon_1\vert \theta)=\{(0,1)\}$ and
$\epsilon_2$ such that $G(\epsilon_2\vert \theta)=\{(0,1),(1,0)\}$
we have $\sup G(\epsilon_1\vert \theta)
=(0,1)\precsim_\mathcal{Y}(1,0)=\sup G(\epsilon_2\vert \theta)$ and
$\inf G(\epsilon_1\vert \theta) =(0,1)=\inf G(\epsilon_2\vert
\theta)$. This is illustrated in figure~\ref{figure:FB22mon}.
\end{continued}

We are now in a position to state the theorem\footnote{ This result
is a reformulation of theorem~3d of the previous version circulated
\cite{GH:2006a}.}, which is the main tool in the construction
of core determining classes, and hence in the computation of the
identified set.

\begin{theorem}\label{theorem:cd}
Suppose assumption~\ref{assumption:monotonicity} is satisfied with
orderings $\precsim_\mathcal{Y}$ and $\precsim_\mathcal{U}$. Call
$I$ the cardinality of $\mathcal{Y}$, and list outcomes (elements of
$\mathcal{Y}$) in increasing order (with respect to ordering
$\precsim_\mathcal{Y}$) as $y_1,\ldots,y_I$. Then
$\mathcal{A}=(\{y_1,\ldots,y_i\},\{y_i,\ldots,y_I\}:\;i=1\ldots,I)$
is core determining.
\end{theorem}

Theorem~\ref{theorem:cd} allows to reduce the cardinality of the
power set $2^\mathcal{Y}$ to twice the cardinality of $\mathcal{Y}$
minus $2$ (since the inequality needn't be checked on the whole set
$\mathcal{Y}$), as we illustrate in our
example~\ref{subsubsection:family bargaining}.

\begin{continued}[Example \protect\ref{subsubsection:family bargaining} continued]
In the family bargaining example,
assumption~\ref{assumption:monotonicity} is satisfied, as seen on
figure~\ref{figure:FB22mon}, with the ordering described above.
Hence the class $\left(\{(0,0)\}\right.$, $\{(0,0),(0,1)\}$,
$\{(0,0),(0,1),(1,0)\}$, $\{(0,1),(1,0),(1,1)\}$, $\{(1,0),(1,1)\}$,
$\left.\{(1,1)\}\right)$ is core determining.
\end{continued}

\section{Illustration: oligopoly entry with two types of players}
\label{section:BR44}
\begin{figure}[htbp]
\begin{center}
\includegraphics[width=15cm]{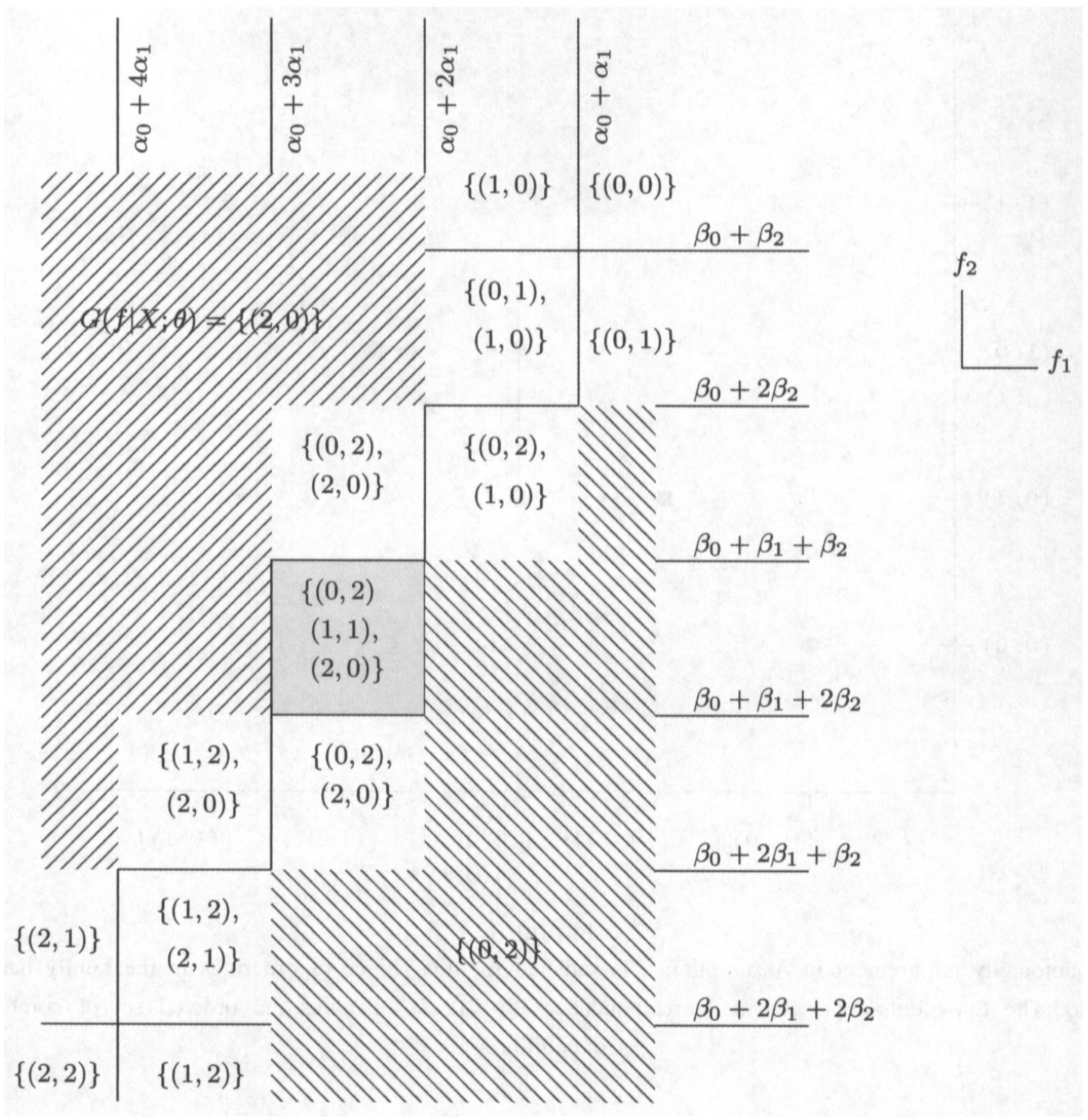}
\caption{{\scriptsize Model correspondence in the oligopoly entry
game with two types of firms, two of each type.}}
\label{figure:BR44}
\end{center}
\end{figure}

We now turn to a more substantive illustration of our methods to compute the identified set,
first, to show the operational usefulness of corollary~\ref{theorem:cd}, and second, to illustrate
the power of the combinatorial approach.
To do so, we consider the oligopoly entry game with two types of players presented in appendix~A
of \cite{BT:2006}.
The profit function of type 1 firms depends on the total number of
firms in the market, but not on the type of those firms, whereas
profits of type 2 firms depend both on the number and on the type of
firms present in the market. The latent variable is the fixed cost
$f_1$ for firms of type 1 and $f_2$ for firms of type 2. $(f_1,f_2)$
is uniformly distributed over $[0,1]^2$. The model is simplified by
assuming linearity of profits in firm number as follows.
\begin{eqnarray*}
&&\pi_1(Y_1,Y_2,X,f;\theta)=\alpha_0+\alpha_1(Y_1+Y_2)+\alpha_2 X-f_1\\
&&\pi_2(Y_1,Y_2,X,f;\theta)=\beta_0+\beta_1Y_1+\beta_2Y_2+\beta_3 X-f_2,
\end{eqnarray*}
with $\alpha_1,\beta_1,\beta_2$ strictly negative and
$\beta_2>\beta_1$ to fix ideas (profit of type 2 firms will decrease
by a larger amount if a type 1 firm enters the market than if a type
2 firm does). The set of observable outcomes is
$\mathcal{Y}=\{(i,j):\;i,j=0,1,2\}$, where $i$ denotes the
number of type 1 firms and $j$ the number of type 2 firms present in
the market. $\mathcal{Y}$ can be ordered lexicographically, where
the number of firms present in the market is considered first, and
then the identity of firms (type 1 dominating type 2)\footnote{The
order could be rationalized by total profit in the industry, but it
is not necessary for the construction of a core determining class
nor the computation of the identified set.}. Hence
$(0,0)\precsim_\mathcal{Y}$ $(0,1)\precsim_\mathcal{Y}$
$(1,0)\precsim_\mathcal{Y}$ $(0,2)\precsim_\mathcal{Y}$
$(1,1)\precsim_\mathcal{Y}$ $(2,0)\precsim_\mathcal{Y}$
$(1,2)\precsim_\mathcal{Y}$ $(2,1)\precsim_\mathcal{Y}$ $(2,2)$. The
model correspondence is represented in figure~\ref{figure:BR44},
which is taken from \cite{BT:2006}.

\subsection{Submodular optimization approach}
The implementation of the submodular optimization method is extremely simple. Once the values of the likelihood have been derived (analytically in the present case, but more often by simulation following the method proposed in \cite{BHR:2005}), all that remains to do to check whether a value $\theta$ belongs to the identified set, is to minimize the function defined for all $A\subseteq\mathcal{Y}$ by $A\mapsto\mathcal{L}(A\vert X;\theta)-P(A\vert X)$ using for instance the Matlab SFO toolbox routine $SFO-min-norm-point.m$ implementing Fujishige's algorithm (see page 293 of \cite{Fujishige:2005}).

\subsection{Optimal transportation approach}
\begin{figure}[htbp]
\begin{center}
\includegraphics[width=8cm]{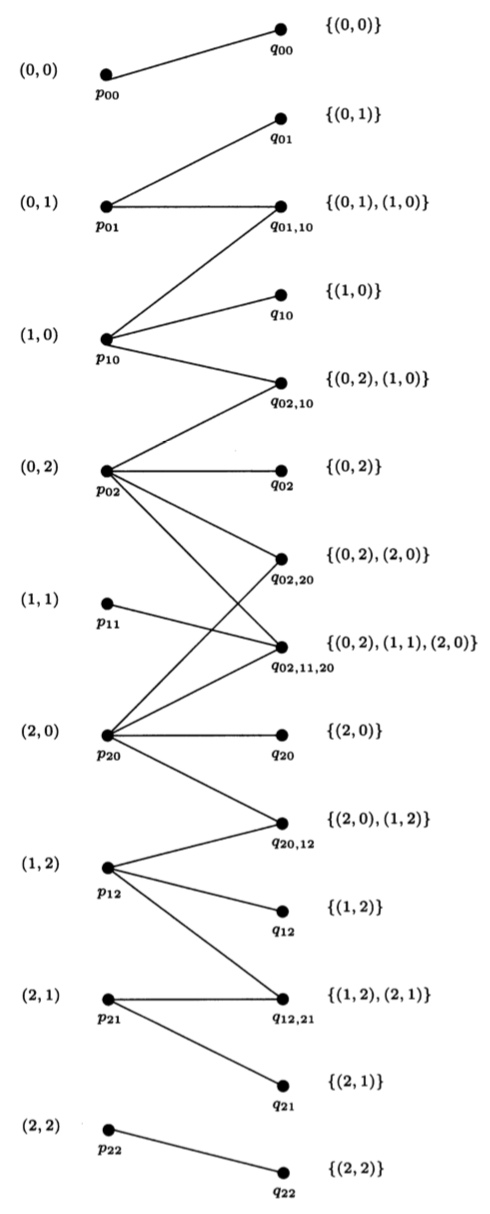}
\caption{{\scriptsize Bi-partite graph representing the admissible
connections between observable outcomes and combinations of
equilibria in the two-type oligopoly entry model.}}
\label{figure:BR44graph}
\end{center}
\end{figure}

Consider now the linear programming strategy for computing the identified set. The bipartite graph
corresponding to this example is represented in figure~\ref{figure:BR44graph}.

As shown in theorem~\ref{theorem:matching}, a value of the
parameter vector is in the identified set if and only if there
exists a zero cost transportation plan for the transfer of masses
$p_y$ on the elements of $\mathcal{Y}$ to masses $q_u$ on the
elements of $\mathcal{U}^\ast$. A transportation plan is a set of
nonnegative numbers attached to all pairs
$(y,u)\in\mathcal{Y}\times\mathcal{U}^\ast$ (which represents the
amount of mass from $y$ that is transferred to $u$ via the edge
$(y,u)$). In our application, the transportation cost from $y$ to
$u$ is zero if $y$ and $u$ are connected by an edge in the graph of
figure~\ref{figure:BR44graph}, and 1 otherwise. If the algorithm
returns a zero cost transportation plan, it means that mass is
transferred through edges of the graph only, and for instance the
pair $\left((1,1),\{(0,2),(1,1),(2,0)\}\right)$ is assigned a non
negative number (i.e. some mass is transported there), but the pair
$((1,1),\{(2,0),(0,2)\})$ is assigned zero (i.e. no mass is
transported there). The existence of a zero cost transportation plan
is equivalent to the existence of a joint distribution on
$\mathcal{Y}\times\mathcal{U}^\ast$ which is supported on the
graph of figure~\ref{figure:BR44graph} and has the correct marginal
distributions, hence, it is equivalent to the fact that $\theta$ is
in the identified set, as we showed in
theorem~\ref{theorem:matching}.

The minimum cost transportation problem is equivalent to the dual
maximum flow problem, as described in the previous section. Mass
flows through the network with $25$ nodes, which include the
\emph{source}, the $9$ elements of $\mathcal{Y}$, the $14$ elements
of $\mathcal{U}^\ast$ and the \emph{sink} (mass flows in the
direction
$\mathrm{Source}\rightarrow\mathcal{Y}\rightarrow\mathcal{U}^\ast\rightarrow\mathrm{Sink}$).
A network is characterized by its \emph{adjacency matrix}, which
gives all the links between nodes with their capacity. In the case
of interest here, the adjacency matrix is given in
table~\ref{table:adjacency}. Maximum flow programs take this
adjacency matrix as an input, and return the maximum flow through
the network it characterizes. This maximum flow cannot be larger
than $\sum_{y\in\mathcal{Y}}p_y=\sum_{u\in\mathcal{U}^\ast}q_u=1$,
and it is equal to $1$ if and only if $\theta$ is in the identified
set.

\subsection{Core determining class approach}
\begin{figure}[htbp]
\begin{center}
\includegraphics[width=15cm]{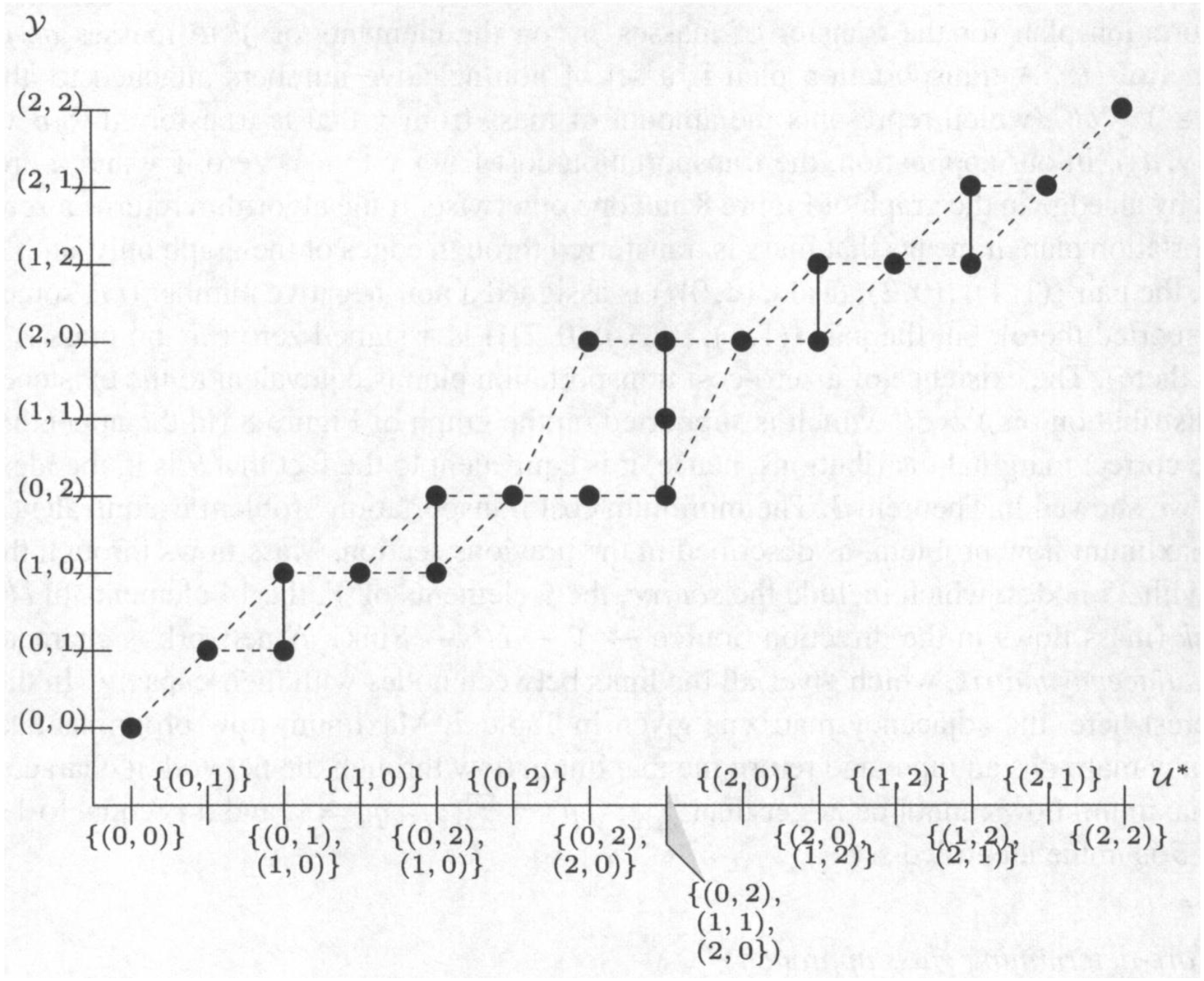}
\caption{{\scriptsize Assumption~\ref{assumption:monotonicity} is
satisfied in the oligopoly entry game with two types of firms, two
of each type, up to the fact that the image of
$G(\{(0,2),(2,0)\}\vert X;\theta)$ is not connected.}}
\label{figure:BR44mon}
\end{center}
\end{figure}

We now illustrate the usefulness of corollary~\ref{theorem:cd} for the determination of a core determining class
in this example. Figure~\ref{figure:BR44mon} graphs the orderings that satisfy assumption~\ref{assumption:monotonicity}
up to the fact that the set of equilibria is not always connected. Indeed, in the ordering of outcomes,
$(1,1)$ comes between $(0,2)$ and $(2,0)$, or more precisely, $(0,2)\precsim_\mathcal{Y}(1,1)
\precsim_\mathcal{Y}(2,0)$. Now $(1,1)$ is not an equilibrium when $\alpha_0+3\alpha_1<
f_1\leq\alpha_0+2\alpha_1$ and $\beta_0+\beta_1+\beta_2<f_2\leq\beta_0+2\beta_2$,
so the set of equilibria $\{(0,2),(2,0)\}$ is disconnected in that case.
However, since $(1,1)$ is observed only when $\epsilon\in\{(0,2),(1,1),(2,0)\}$,
the mass $p_{11}$ can be removed from $q_{02,11,20}$. Indeed, $(1,1)$ is isolated in the sense that all its mass is necessarily transferred to $\{(0,2),(1,1),(2,0)\}$, which is the only predicted combination of equilibria in which it appears. Hence, after checking that $p_{11}\leq q_{02,11,20}$, we can remove $p_{11}$ and $q_{02,11,20}$ and replace $q_{02,20}$ by $q_{02,20}+q_{02,11,20}-p_{11}$. Then, the monotonicity and connectedness conditions hold, and theorem~\ref{theorem:cd} can be applied directly to $\mathcal{Y}\backslash\{(1,1)\}$
and $\mathcal{U}^\ast$, yielding the class $\mathcal{A}$ $=( \{(0,0)\},$ $\{(0,0),$ $(0,1)\},$ $\{(0,0),$ $(0,1),$ $(1,0)\},$
$\{(0,0),$ $(0,1),$ $(1,0),$ $(0,2)\},$
$\{(0,0),$ $(0,1),$ $(1,0),$ $(0,2),$ $(2,0)\},$
$\{(0,0),$ $(0,1),$ $(1,0),$ $(0,2),$ $(2,0),$ $(1,2)\},$
$\{(0,0),$ $(0,1),$ $(1,0),$ $(0,2),$ $(2,0),$ $(1,2),$ $(2,1)\},$
$\{(0,1),$ $(1,0),$ $(0,2),$ $(2,0),$ $(1,2),$ $(2,1),$ $(2,2)\},$
$\{(1,0),$ $(0,2),$ $(2,0),$ $(1,2),$ $(2,1),$ $(2,2)\},$
$\{(0,2),$ $(2,0),$ $(1,2),$ $(2,1),$ $(2,2)\},$
$\{(2,0),$ $(1,2),$ $(2,1),$ $(2,2)\},$
$\{(1,2),$ $(2,1),$ $(2,2)\},$ $\{(2,1),$ $(2,2)\},$ $\{(2,2)\})$.
Note that its cardinality is $2\times7=14$, as opposed to the cardinality of the power set of
$\mathcal{Y}$ which is $2^9=512$.

\begin{table}
\vskip15pt \caption{Adjacency matrix for the two-type oligopoly model.}
\label{table:adjacency}\begin{center}
\begin{tabular}{c|cccccccccc}
&$(0,0)$&$(0,1)$&$(1,0)$&$(0,2)$&$(1,1)$&$(2,0)$&$(1,2)$&$(2,1)$&$(2,2)$&\emph{Sink}\\\hline
\emph{Source}&$p_{00}$&$p_{01}$&$p_{10}$&$p_{02}$&$p_{11}$&$p_{20}$&$p_{12}$&$p_{21}$&$p_{22}$&\\
$\{(0,0)\}$&$\infty$&&&&&&&&&$q_{00}$\\
$\{(0,1)\}$&&$\infty$&&&&&&&&$q_{01}$\\
$\{(0,1),(1,0)\}$&&$\infty$&$\infty$&&&&&&&$q_{01,10}$\\
$\{(1,0)\}$&&&$\infty$&&&&&&&$q_{10}$\\
$\{(1,0),(0,2)\}$&&&$\infty$&$\infty$&&&&&&$q_{02,10}$\\
$\{(0,2)\}$&&&&$\infty$&&&&&&$q_{02}$\\
$\{(0,2),(2,0)\}$&&&&$\infty$&&$\infty$&&&&$q_{02,20}$\\
$\{(0,2),(1,1),(2,0)\}$&&&&$\infty$&$\infty$&$\infty$&&&&$q_{02,11,20}$\\
$\{(2,0)\}$&&&&&&$\infty$&&&&$q_{20}$\\
$\{(2,0),(1,2)\}$&&&&&&$\infty$&$\infty$&&&$q_{20,12}$\\
$\{1,2\}$&&&&&&&$\infty$&&&$q_{12}$\\
$\{(1,2),(2,1)\}$&&&&&&&$\infty$&$\infty$&&$q_{12,21}$\\
$\{(2,1)\}$&&&&&&&&$\infty$&&$q_{21}$\\
$\{(2,2)\}$&&&&&&&&&$\infty$&$q_{22}$
\end{tabular}
\end{center}
\end{table}

\subsection{Efficiency comparison of the combinatorial methods}
As an illustration of the procedure, we compute the identified set
for the two-type oligopoly model with the following distributional
hypotheses and normalization restrictions. The fixed cost vector
$(f_1,f_2)$ is assumed to be uniformly distributed on $[0,1]^2$.
$\alpha_0$ and $\beta_0$ are set equal to $1$. As previously noted,
we assume that monopoly profits are larger than oligopoly profits,
and that a type two firm's profit decreases more if a type one firm
enters than a type two firm, hence $0>\alpha_0$ and
$0>\beta_2>\beta_1$. We can therefore calculate the probabilities of
each combination of equilibria $u\in\mathcal{U}^\ast$. These
probabilities are computed in a Matlab program file available on
request.

The values of the model likelihood are derived from these
probabilities, and the function $\mathcal{L}(.\vert
X;\theta)-P(.\vert X)$ is then minimized over all subsets of
$\mathcal{Y}$ using the Matlab routine SFO-min-norm-point.m from the
SFO toolbox. An idea of the computational efficiency of this
procedure is given by the fact that an order of $10^3$ values of the
parameter can be tested in one second on a standard laptop computer.

To implement the minimum cost of transportation/maximum flow method, the probabilities of each value of the equilibrium correspondence are entered together with the
true frequencies of observable outcomes into the adjacency matrix of
table~\ref{table:adjacency} and the Matlab routine maxflow.m of the BGL library returns a
flow of $1$ if the value of $\theta=(\alpha_1,\beta_1,\beta_2)'$
(used to compute the predicted probabilities) belongs to the
identified set, and a flow strictly smaller than $1$ if it doesn't.
To give an idea of the efficiency of the method, we can test $10^5$
values of $\theta=(\alpha_1,\beta_1,\beta_2)'$ in less than a second
on a standard portable computer. Hence this method is faster than the general submodular minimization method, but only applies to pure strategy equilibria.

Finally, the additional information yielded by the existence of a core determining class for this problem allows us to test a value of the parameter by a constrained submodular optimization, where the search is limited to the sets in the core determining class.

\begin{figure}[htbp]
\begin{center}
\includegraphics[width=15cm]{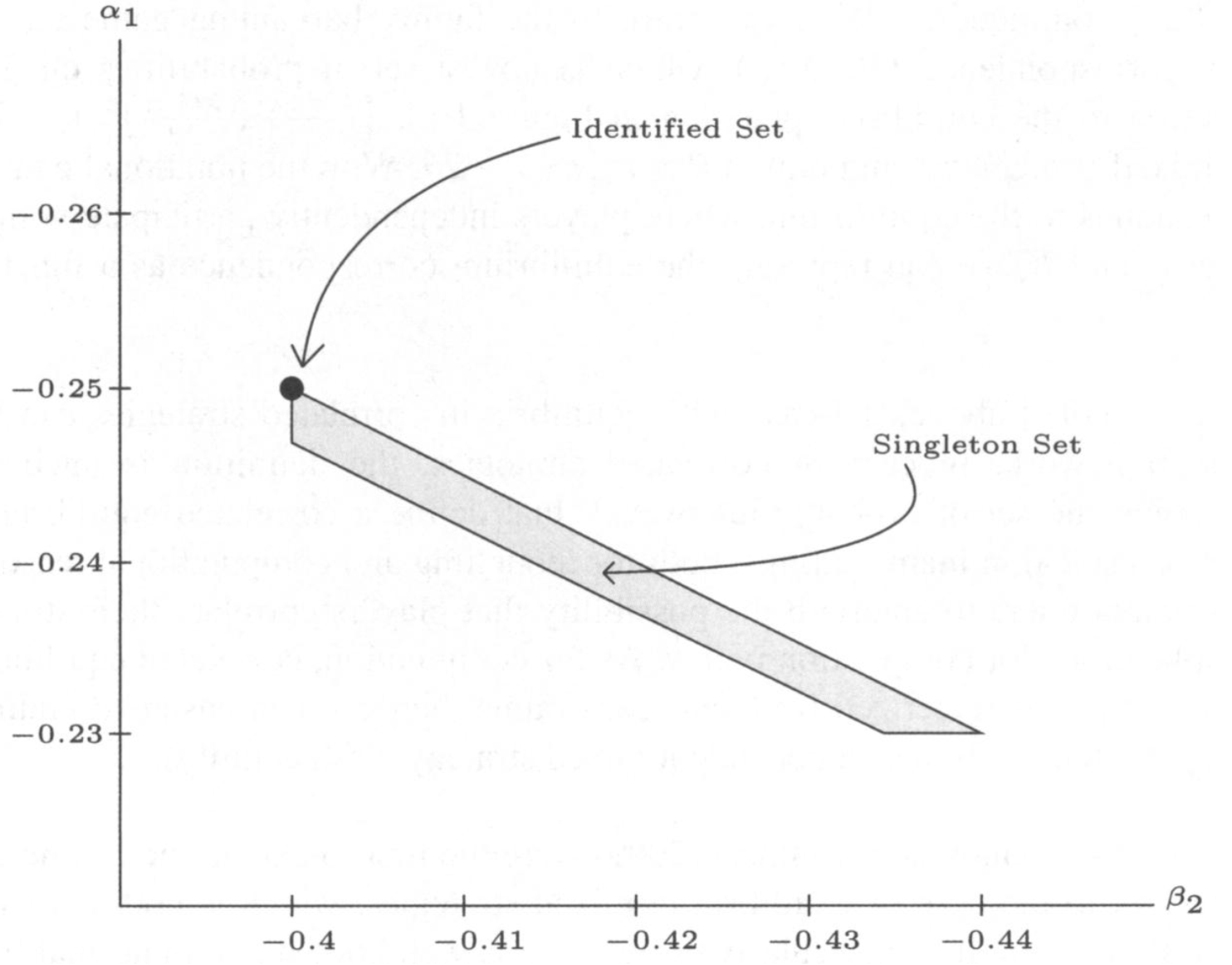}
\caption{{\scriptsize Projection of the identified set, and of the
set characterized by the Singleton class on the $(\beta_2,\alpha_1)$
space. The projection of the identified set is a single point
$(\beta_2,\alpha_1)=(-0.4,-0.25)$. The frequencies of observable
variables are
$(p_{00},p_{01},p_{10},p_{02},p_{11},p_{20},p_{12},p_{21},p_{22})=
(0.1,0.15,0.15,0.1,0,0.5,0,0,0)$.
}} \label{figure:abj}
\end{center}
\end{figure}

For illustration purposes, we compute the identified set for a given choice of the observable frequencies,
namely $(p_{00},p_{01},p_{10},p_{02},p_{11},p_{20},p_{12},p_{21},p_{22})=
(0.1,0.15,0.15,0.1,0,0.5,0,0,0)$, and compare it to the set obtained
by imposing the inequality restrictions on the \emph{Singleton class} only. The latter corresponds to
the set of values of the parameters such that $p_{00}\leq q_{00}$, $p_{01}\leq q_{01}+q_{01,10}$,
$p_{10}\leq q_{01,10}+q_{10}+q_{02,10}$, $p_{02}\leq q_{02,10}+q_{02}+q_{02,20}+q_{02,11,20}$, $p_{11}\leq q_{02,11,20}$,
$p_{20}\leq q_{02,20}+q_{02,11,20}+q_{20}+q_{20,12}$, $p_{12}\leq q_{20,12}+q_{12}+q_{12,21}$, $p_{21}\leq q_{12,21}+q_{21}$ and $p_{22}\leq q_{22}$.
It turns out the values of $\alpha_1$ and $\beta_2$ are identified (under these specific
values for the true probabilities of observable variables, which were chosen for the simulation purposes from the parameter values), and all values of $\beta_1<\beta_2$
are compatible with the given frequencies. The set defined by the \emph{Singleton class} restrictions, however, is much larger, as shown by its projection on the $(\beta_2,\alpha_1)$ space in figure~\ref{figure:abj}.
There are also many values of the observed frequencies, for which the identified set is empty, so that the model is rejected,
but the set defined by the \emph{Singleton class} restrictions is non-empty, so that it fails to reject the model.

\section{Extension to equilibria in mixed strategies}
\label{section:mixed strategies}

We now show how our results extend to the case where mixed strategy
equilibria are allowed. As before, we consider a game parameterized
by the deterministic parameter vector $\theta$, a vector of
covariates $X$ and unobserved heterogeneity parameter $\epsilon$
with distribution $\nu(.\vert X;\theta)$. The observable outcomes of
the game are equilibrium actions profiles $Y$ whose realizations
belong to the finite set $\mathcal{Y}$. Call $P(.\vert X)$ the true
distribution of observable outcomes $Y$ conditionally on $X$. The
equilibrium correspondence $G(\varepsilon\vert X;\theta)$ is now the
set of Nash equilibria of the game in mixed strategies (including
pure strategies as degenerate mixed strategies).
Call $\sigma$ the equilibrium strategy profile that is actually
selected within the set $G(\epsilon\vert X;\theta)$, i.e. the model
predicts $\sigma\in G(\epsilon\vert X;\theta)$. The outcome $Y$ is a
random variable with distribution $\sigma$, so it can be written
$Y=Q_\sigma(V)$, where $Q_\sigma$ is the quantile transform
associated with $\sigma$, and $V$ is a random variable with uniform
distribution, independent of $\epsilon$ and $\sigma$, conditionally
on $X$. The variable $V$ is traditionally interpreted as an
independent signal that agents receive and on whose realization they
base their action. The identified set is therefore defined as the
set $\Theta_I$ of values of the parameter such that $Y=Q_\sigma(V)$
and $\sigma\in G(\epsilon\vert X;\theta)$ as specified above.

\begin{definition}[Identified set]\label{definition:mixed identified set}
The identified set is equal to \begin{eqnarray*}&&\Theta_I=
\left\{\theta\in\Theta:\;X-\mbox{a.s.},\;\exists
(Y,\epsilon,\sigma,V):\;\right.\\&&\left.\hskip20pt Y\sim
P_{|X},\;\epsilon\sim\nu(.|X;\theta),\;V\sim U[0,1],\;
Y=Q_\sigma(V),\;\sigma\in G(\epsilon\vert
X;\theta),\;V\independent\epsilon,\sigma\right\}\end{eqnarray*}\end{definition}

\begin{figure}[htbp]
\begin{center}
\includegraphics[width=15cm]{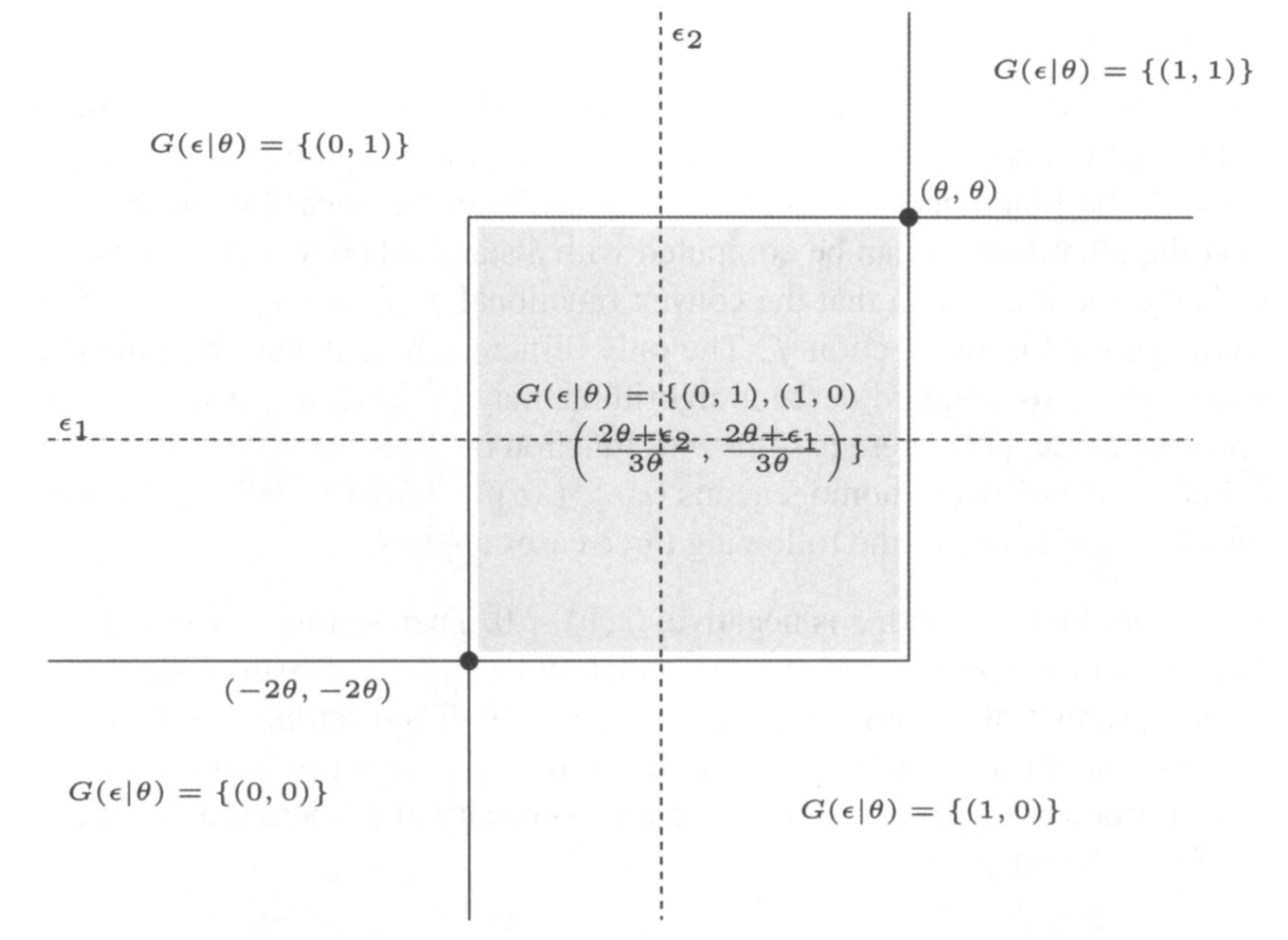}
\caption{{\scriptsize Model correspondence for the family bargaining case with mixed strategy equilibria.}} \label{figure:FBmixed}
\end{center}
\end{figure}

\begin{continued}[Example \protect\ref{subsubsection:family bargaining} continued]
We now return to the family bargaining game and derive the
equilibrium correspondence $G(\epsilon\vert X;\theta)$, which is now
a set of probabilities on $\mathcal{Y}$. We note that in addition to
the equilibria appearing in figure~\ref{figure:FB}(a),
$\{(\frac{2\theta+\epsilon_2}{3\theta},\frac{2\theta+\epsilon_1}{3\theta})\}$
is a Nash equilibrium in mixed strategies if and only if
$\theta>\epsilon_1,\epsilon_2>-2\theta$. With the notational
convention that $(\alpha,\alpha)$ corresponds to the equilibrium,
where players independently participate with probability $\alpha$,
we can represent the equilibrium correspondence as a function of
$\epsilon$ in figure~\ref{figure:FBmixed}.
\end{continued}

\cite{BMM:2008} (hereafter BMM) were the first to extend the
characterization of the identified set to the case of equilibria in
mixed strategies, which is both a very important and nontrivial
problem. It is particularly useful, as it is well known that in
normal form games, existence of equilibria is guaranteed in mixed
strategies, but not necessarily in pure strategies, except in
special classes of games such as supermodular games. It is a
nontrivial problem, since the likelihood of the model is no longer
the distribution function of a random set. Nonetheless, we go beyond
the results of BMM and show that in certain classes of games the
identified set can be characterized by a finite set of inequalities.
We also show that our efficient submodular optimization method is
still applicable to this case, as the identified set continues to be
characterized by the fact that true data distribution is contained
in the core of the model likelihood as in
theorem~\ref{theorem:sharp}. In the general case, where the
likelihood is not necessarily submodular, we show that the
identified set can be computed efficiently through a convex
optimization problem, which we describe.

We first show now that the identified set $\Theta_I$ defined in
definition~\ref{definition:mixed identified set} is indeed equal to
$\Theta_I^\ast$ defined in BMM. Indeed: suppose $\theta\in\Theta_I$.
Call $\mu(.\vert\epsilon, X;\theta)$ the distribution of the random
element $\sigma$ (hence a distribution on the simplex). The latter
has support $G(\epsilon\vert X;\theta)$. Then, for any value
$y\in\mathcal{Y}$, we have \begin{eqnarray}P(Y=y\vert X)&=&
  \int \int P(Y=y\vert \epsilon,\sigma,X;\theta)d\mu(\sigma\vert\epsilon,X;\theta)d\nu(\epsilon\vert X;\theta)\nonumber\\&=&\int \int P(Q_\sigma(V)=y\vert \epsilon,\sigma,X;\theta)d\mu(\sigma\vert \epsilon,X;\theta)d\nu(\epsilon\vert X;\theta)\nonumber\\&=&
 \int \left(\int P_\sigma(Y=y\vert \epsilon,X;\theta)d\mu(\sigma\vert \epsilon,X;\theta)\right)d\nu(\epsilon\vert X;\theta)\label{equation:mixed}\end{eqnarray}
 by independence.
 Conversely, if (\ref{equation:mixed}) holds, we can construct the random
 quadruplet $(Y,V,\epsilon,\sigma)$ with the required properties.

The fundamental work by BMM was the first to derive a
characterization of the identified set in the case of multiple mixed
strategy equilibria, which we restate here as the following
proposition:

\begin{proposition}[BMM 2009]\label{proposition:BMM}\begin{equation*}
{ \Theta _{I}=\left\{ \theta \in \Theta :\;\forall f\;\mbox{
function on }\;\mathcal{Y}:E_{P}(f(Y)|X)\leq \tilde{\mathcal{L}}
\left( f\vert X;\theta\right) \right\} ,}
\end{equation*}%
where%
\begin{equation*}
\tilde{\mathcal{L}}\left( f\vert X;\theta\right) { =}\int \left(
\max_{\sigma \in G(\epsilon |X;\theta )}{ E}_{\sigma }%
{ (f(Y))}\right) { d\nu (\epsilon |X;\theta ),~f\;%
\mbox{ function on }\;\mathcal{Y}.}
\end{equation*}
\end{proposition}

For completeness, in the appendix we give a different proof of this
result, as it fits more naturally with our presentation than the
original random set theoretic proof given in \cite{BMM:2008}.
We now show that this characterization allows efficient computation
of the identified set as the solution of a convex optimization
problem. In addition, under regularity conditions on the equilibrium
correspondence of the game, we show that we can derive an exact
characterization of the identified set with a finite collection of
inequalities, as was the case when considering equilibria in pure
strategies only.

\subsection{Efficient computation}
In this section, we give a simple procedure, which is valid for any
normal form game, and is recommended to practitioners who do not
wish to restrict attention to pure strategy Nash equilibria.  In
proposition~\ref{proposition:BMM}, the functional
$\tilde{\mathcal{L}}$ is convex, as a maximum of linear functionals.
As a result, we show that the identified set can be computed with a
standard convex optimization program. Indeed, $\Theta_I$ is the set
of $\theta$'s such that the convex functional
$\phi(f)=\tilde{\mathcal{L}}(f\vert X;\theta)-E_P(f(Y)\vert X)$
remains non negative for all function $f$. Suppose there is some $v$
(a function on $\mathcal{Y}$, hence a vector in $\mathbb{R}^{d_y}$)
such that $\phi(v)<0$. Since $\phi$ is positively homogeneous (i.e.
$\phi(\lambda v)=\lambda\phi(v)$ for all $\lambda\geq0$), then for
each $z\in\mathbb{R}^{d_y}$, such that $z\ne0$, one of the following
three cases applies:
\begin{itemize}\item The inner product of $v$ with $z$ is negative, $\langle z,v\rangle<0$. Then setting $\lambda=-\langle z,v\rangle^{-1}$, we have $\langle z,\lambda v\rangle=-1$ and $\phi(\lambda v)=\lambda\phi(v)<0$, so that $\min_{\{w: \langle z, w\rangle=-1\}}\phi(w)<0$. \item The inner product of $v$ with $z$ is positive, $\langle z,v\rangle>0$. Then setting $\lambda=\langle z,v\rangle^{-1}$, we have $\langle z,\lambda v\rangle=1$ and $\phi(\lambda v)=\lambda\phi(v)<0$, so that $\min_{\{w: \langle z, w\rangle=1\}}\phi(w)<0$. \item The inner product is zero, $\langle z,v\rangle=0$. Then by convexity of $\phi$, there is a $\tilde{v}$ in the neighborhood of $v$ such that $\phi(\tilde{v})<0$ and $\langle z,\tilde{v}\rangle>0$.\end{itemize}

Hence, if we take any $z$, for instance $z=(1,0,\ldots,0)$, and
define $H=\{w: \langle z,w\rangle=0\}$, $\phi$ takes negative values
if and only if the following statement is true
\[\min\left(\min_{w\in H}\phi(w+z),\hskip5pt\min_{w\in
H}\phi(w-z)\right)<0\] Since $H$ is a convex set, the program above
is a completely standard convex optimization program (we implement
the procedure with the matlab optimization toolbox).

\subsection{Cases with a submodular likelihood}
 \label{subsection:mixed submodular} We now turn to the efficient
 computation of the identified set in the case where the likelihood is
 submodular. This occurs when the game satisfies the following assumption,
 taken from \cite{Shapley:71}. Recall that the upper envelope
 $\phi$ of a family of probability functions $A\mapsto\mu_i(A)$, $i=1,\ldots,I$ is
 defined as $A\mapsto\phi(A)=\max_{i\in I}\mu_i(A)$.

\begin{definition}[Regular core (Shapley)]
A normal form game is said to have a \emph{regular core} if the
upper envelope $A\mapsto\sup_{\sigma\in G}\sigma(A)$ of the
equilibrium correspondence $G$ is submodular.
\label{definition:regular core}\end{definition}

We now show a very simple and easy to check sufficient condition for
a regular core.

\begin{lemma}[Sufficient condition for a regular core]
A normal form game with multiple equilibria including only one
equilibrium in proper mixed strategies and any number of equilibria
in pure strategies has a regular core.\label{lemma:regular
core}\end{lemma}

\begin{remark} Note that this sufficient condition is very easy to check,
and it is satisfied for large classes of games (see
\cite{EE:2004} for games of strategic complementarities, and
in particular $2\times 2$ games, when unstable equilibria are
removed).\end{remark}

\begin{theorem}[Characterization of the identified set]\label{theorem:submodular mixed identified set} If the game has a regular core for all $\theta$, $(X,\epsilon)$-almost surely, then
\begin{equation*}
\Theta_I=\left\{\theta\in\Theta:\;\left(\forall B\subseteq\mathcal{Y}:
P(B\vert X)\leq \mathcal{L}(B\vert X;\theta)\right);\;X-\mathrm{a.s.}\right\}
\end{equation*}
 where \[\mathcal{L}(B\vert X;\theta)=
\int\left(\max_{\sigma\in G(\epsilon\vert X;\theta)}\sigma(B)\right)d\nu(\epsilon\vert X;\theta).\]
\end{theorem}

In this case, we give an exact characterization of the identified
set with only a finite number of inequalities, unlike BMM's
characterization. One implication is trivial:
it is obvious to see that $E_{P}{ (f(Y)|X)\leq \tilde{\mathcal{L}}
}\left( f\vert X;\theta \right)$ implies $ P(B|X)\leq
\mathcal{L}(B|X;\theta )$ (it suffices to take $f\left( {y}\right)
=1_{B}\left(y\right) $). The converse is far from trivial, and
follows from the characterization of Choquet integrals (see
\cite{Schmeidler:86}).

This implies that the problem of checking whether a value of the
parameter is in the identified set is equivalent to the problem of
checking whether the distribution of $Y$ is in the core of a Choquet
capacity. This was already the case in \cite{GH:2006a} when we
restricted ourselves to pure strategy equilibria. However, in the
latter case, the Choquet capacity was infinitely alternating, as it
was defined as the distribution of a random set. With mixed strategy
equilibria, it no longer has this property, hence it is no longer
the distribution of a random set.


\begin{continued}[Example \protect\ref{subsubsection:family bargaining} continued]
In the family bargaining example, the model likelihood can be
derived in the following way: If the support of $\epsilon$ belongs
to $[-2\theta,\theta]^2$, then
\begin{itemize}\item $\mathcal{L}(\{(0,1)\}\vert \theta)=\mathcal{L}(\{(1,0)\}\vert \theta)=1$
\item $\mathcal{L}(\{(0,0)\}\vert \theta)=\mathbb{E}_\nu[(\theta-\epsilon_2)
(\theta-\epsilon_1)/(9\theta^2)]$
\item $\mathcal{L}(\{(1,1)\}\vert \theta)=\mathbb{E}_\nu[(2\theta+\epsilon_2)
(2\theta+\epsilon_1)/(9\theta^2)]$
\item $\mathcal{L}(\{(1,1),(0,0)\}\vert \theta)=\mathcal{L}(\{(1,1)\}\vert \theta)+\mathcal{L}(\{(0,0)\}\vert \theta)$.
\end{itemize}

The general case, where the domain of $\epsilon$ is unrestricted, can be deduced very simply from the above (see details in appendix~\ref{section:likelihood derivation}).
\end{continued}

As explained in section~\ref{subsection:submodular}, the strategy to compute the identified set efficiently in the case, where the likelihood is submodular is based on the following observation.

\begin{corollary}[Computation of the identified set]\label{corollary:submodular function}
The identified set is obtained by minimization of a submodular set function: $\Theta_I=\left\{\theta\in\Theta:\;\min_{B\subseteq\mathcal{Y}}
\left(\mathcal{L}(B\vert X;\theta)-P(B\vert X)\right)=0,\;X-\mbox{a.s.}\right\}$.
\end{corollary}

As stated in definition~\ref{definition:submodular}, a set function $\varphi$ is called submodular if $\varphi(A\cup
B)+\varphi(A\cap B)\leq \varphi(A)+\varphi(B)$, and $\varphi(B) =
\mathcal{L}(B\vert X;\theta)-P(B\vert X)$ is indeed submodular, as shown in the proof of theorem~\ref{theorem:submodular mixed identified set}.
The notion of submodularity is the analog of convexity for functions defined on lattices, such as set functions. Hence, minimizing a submodular set
function is akin to minimizing a convex function. As noted in section~\ref{subsection:submodular}, it is a classical
problem in combinatorial optimization (see for instance
\cite{Topkis:98} chapter~2), and several polynomial-time
algorithms exist (see for instance \cite{Fujishige:2005}). We
implement this using Andreas Krause's Matlab SFO toolbox. Note that, as explained in section~\ref{subsection:transportation}, the more efficient optimal transportation method does not apply when mixed strategy equilibria are considered.


\subsection{Monte Carlo simulations}
In order to illustrate our fundamental characterization of the
identified set, we propose a series of Monte Carlo simulations based
on our example~\ref{subsubsection:family bargaining}. For three
different values of the interaction parameter
$\theta=0.25,0.5,0.75$, we represent the core of the likelihood in
the four dimensional simplex. We simulate observed strategy profiles
according to probability distributions $P$ in the core of the model.
We consider three cases. First, the case where the probability
distribution $P$ of the simulated data is the barycenter of the core
(this case is called ``central DGP''). Second a case where $P$ is an
extreme point of the core (this case is called ``extreme DGP''), and
finally an intermediate case (called ``intermediate DGP''). In each
case, and for each value of $\theta$, we simulated $10000$ samples
of size $n=100,1000$, excluded the $5\%$ most distant (case with
``confidence=0.95'') or the $10\%$ most distant (case with
``confidence=0.9'') and showed graphically how the set of remaining
empirical distributions $P_n$ intersects with the core. The $36$
graphics pertaining to all cases described are given in the
appendix. In all graphs on figures
\ref{figure:L-L-L-L} to \ref{figure:L-M-H-M}, the red tetrahedron is
the four dimensional simplex, whose extreme points correspond to the
dirac masses on each of the equilibrium profiles $00,01,10,11$. All
points inside the tetrahedron have barycentric coordinates
corresponding to the vector of probabilities attached to each
profile. The blue diamond-shaped polyhedron is the core computed for
each of the values of the parameter $\theta=0.25,0.5,0.75$, i.e. the
set of distributions of equilibrium profiles that are compatible
with the model for that particular value of $\theta$. The true
distribution is a point in the simplex. If the true distribution is
a point in the core, then the value of $\theta$ is in the identified
set.

\begin{figure}[htbp]
\centering
\subfigure[{\scriptsize $\theta=0.25$, Central DGP, $n=100$, $90\%$ confidence}]{
\includegraphics[width = 2in]{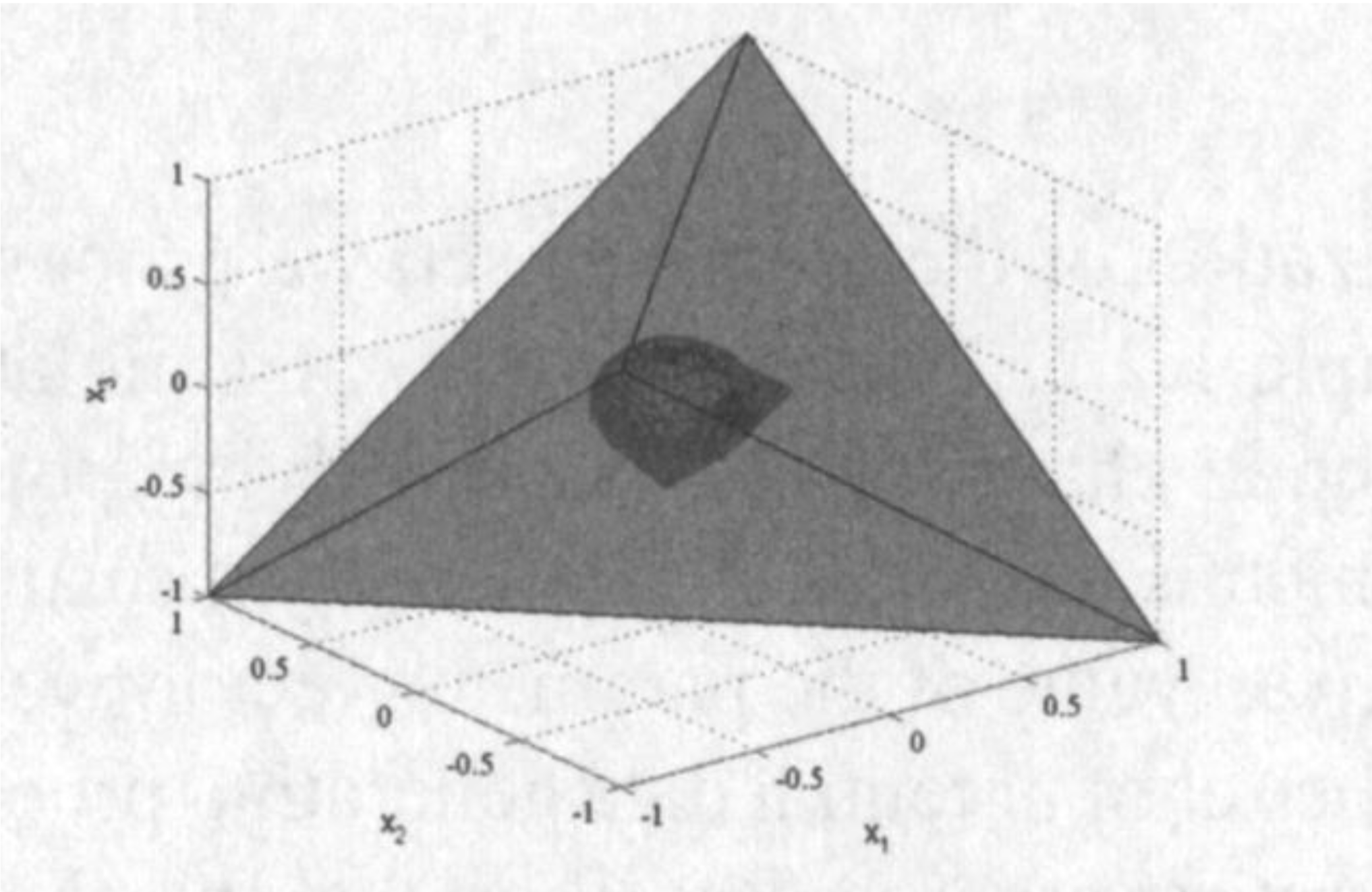}
\label{figure:L-L-L-L}
}
\subfigure[{\scriptsize $\theta=0.25$, Central DGP, $n=100$, $95\%$ confidence}]{
\includegraphics[width = 2in]{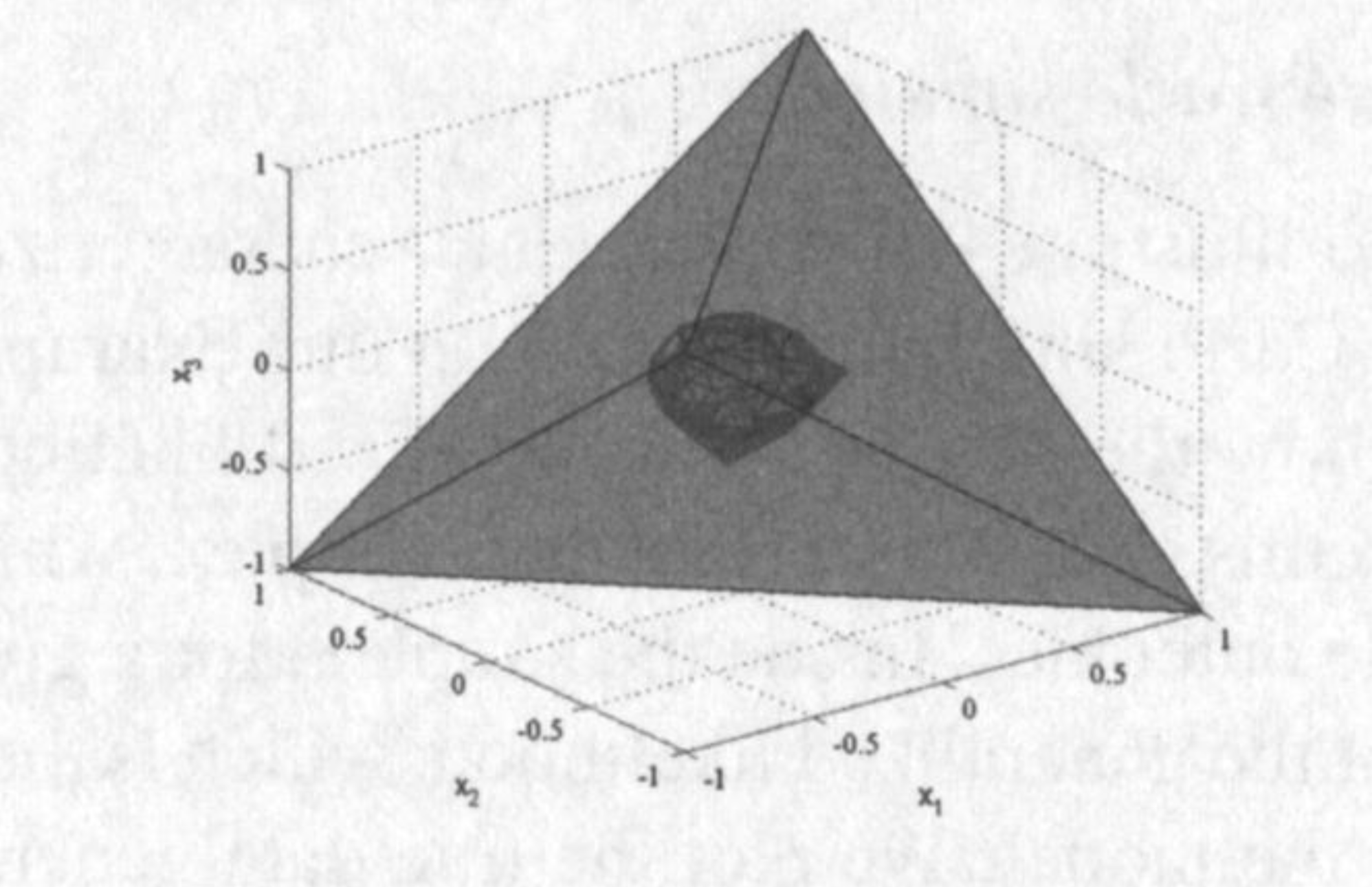}
\label{figure:L-L-L-M}
}
\subfigure[{\scriptsize $\theta=0.25$, Central DGP, $n=1000$, $90\%$ confidence}]{
\includegraphics[width = 2in]{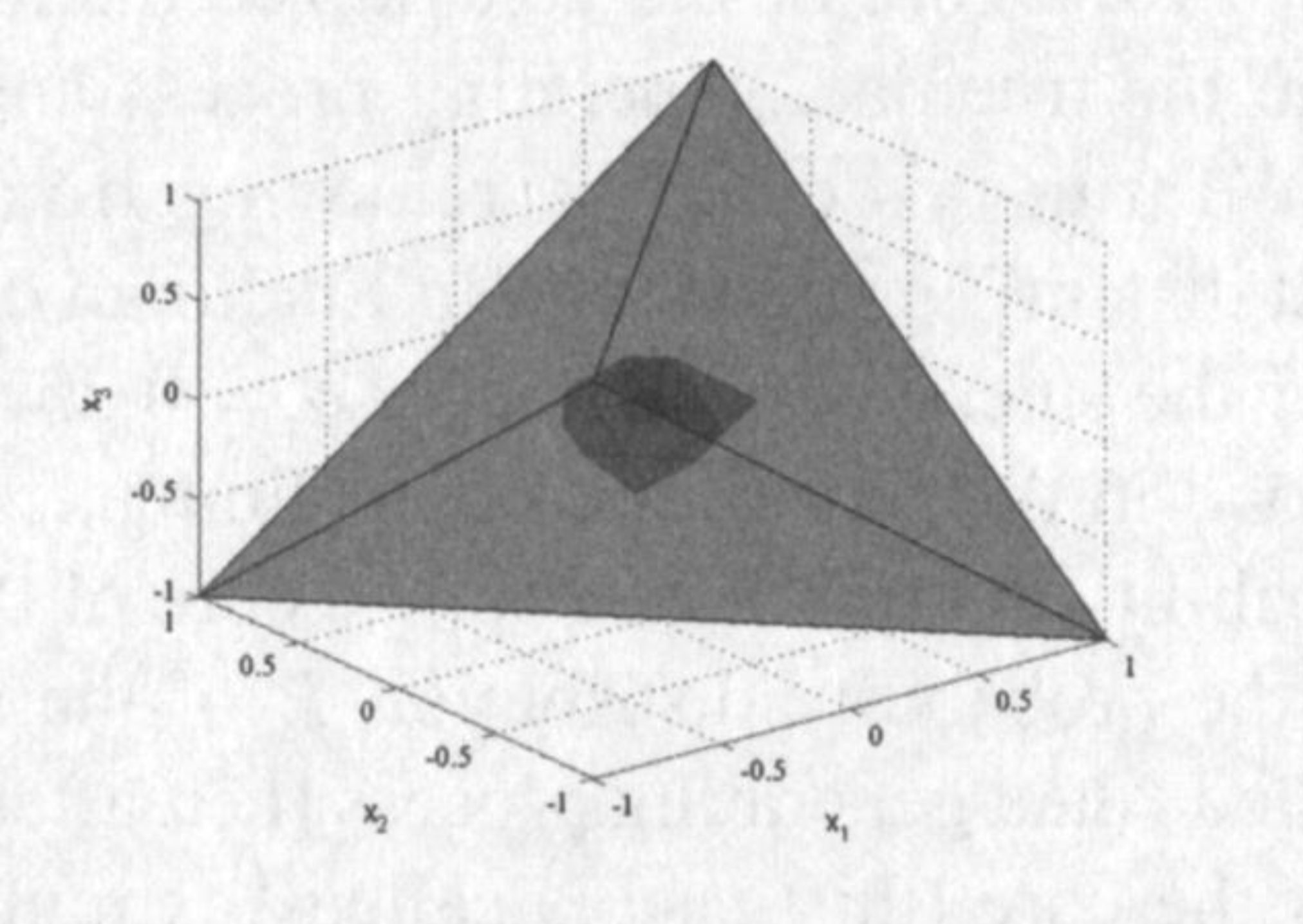}
\label{figure:L-L-H-L}
}
\subfigure[{\scriptsize $\theta=0.25$, Central DGP, $n=1000$, $95\%$ confidence}]{
\includegraphics[width = 2in]{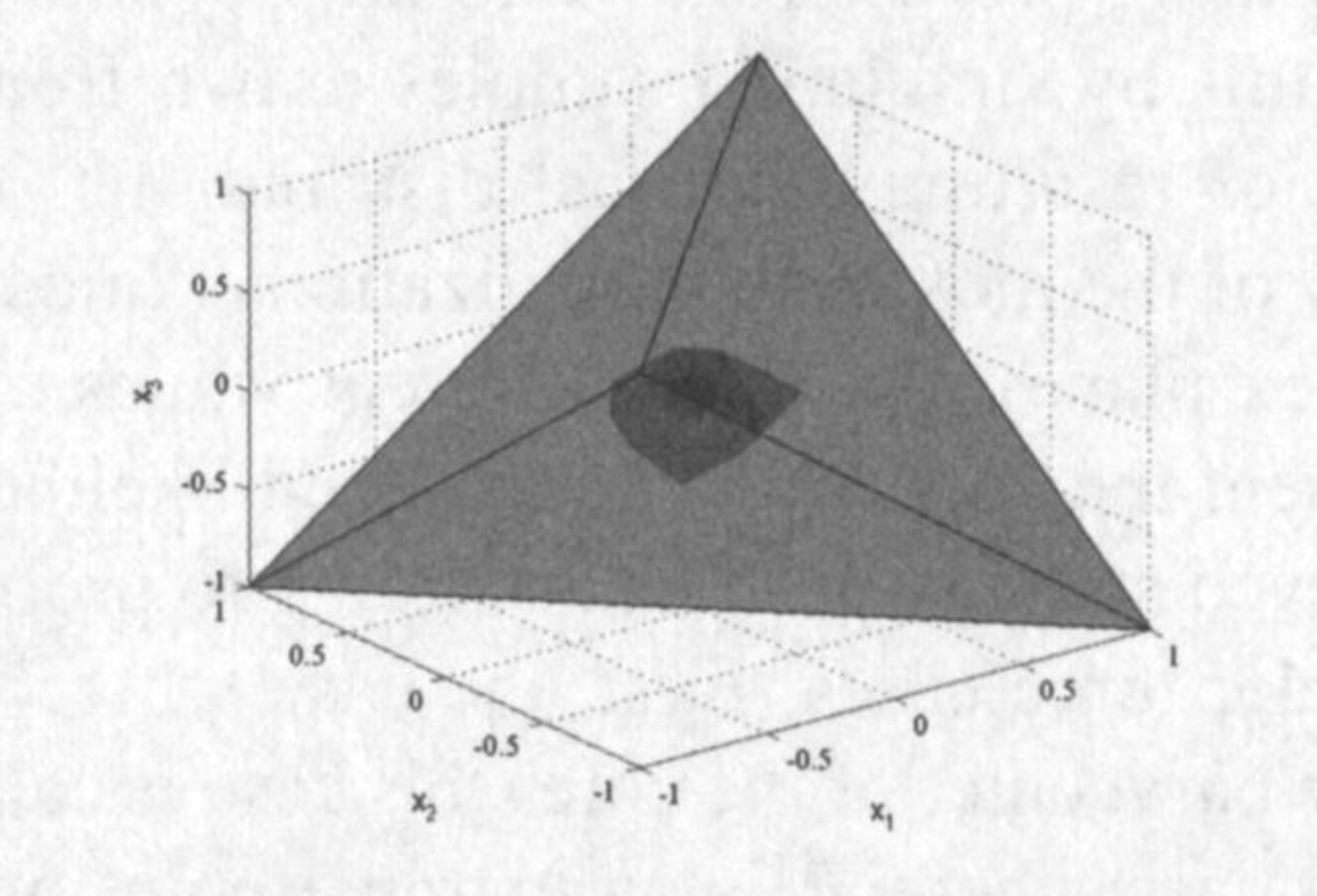}
\label{figure:L-L-H-M}
}
\subfigure[{\scriptsize $\theta=0.25$, Intermediate DGP, $n=100$, $90\%$ confidence}]{
\includegraphics[width = 2in]{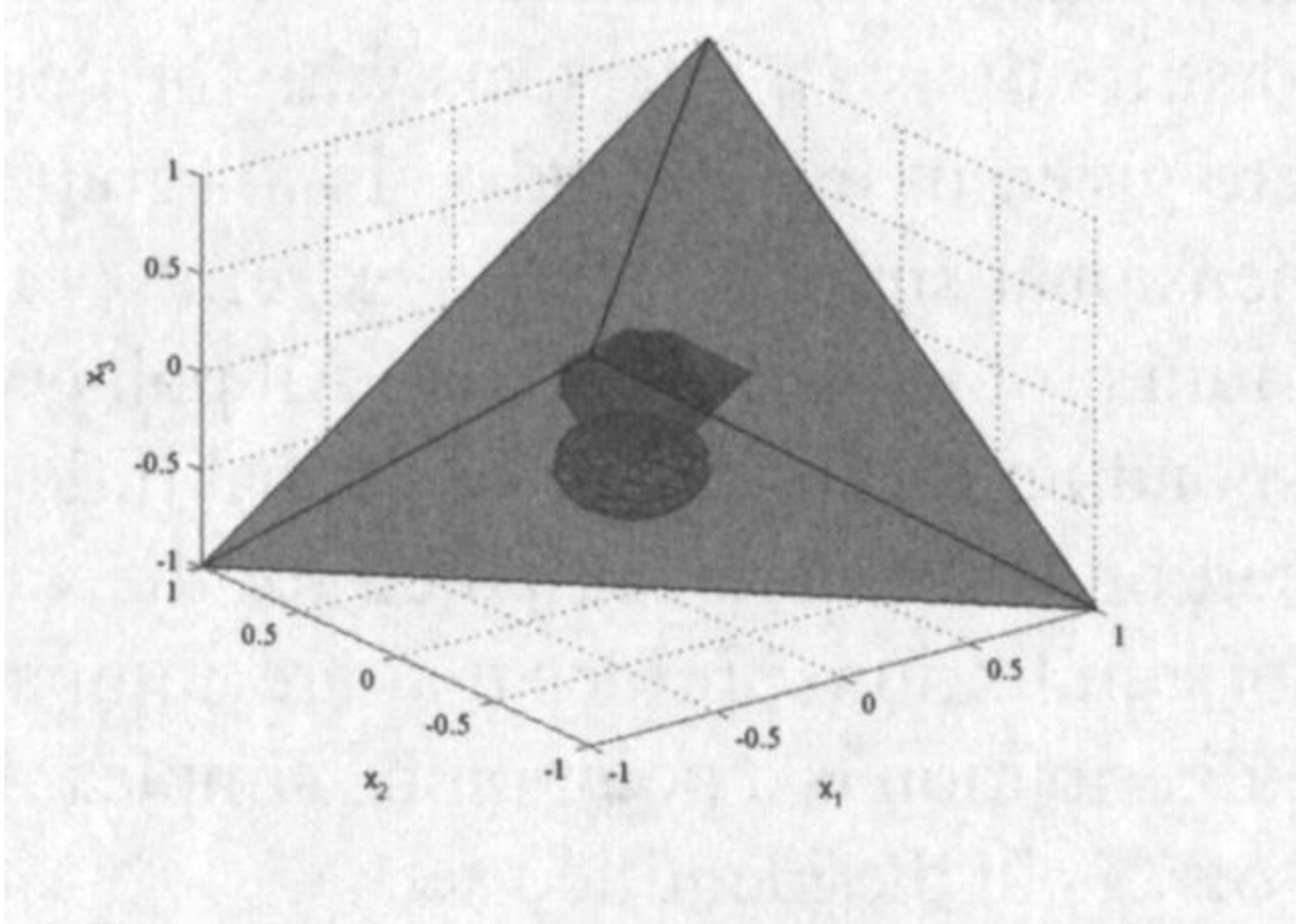}
\label{figure:L-M-L-L}
}
\subfigure[{\scriptsize $\theta=0.25$, Intermediate DGP, $n=100$, $95\%$ confidence}]{
\includegraphics[width = 2in]{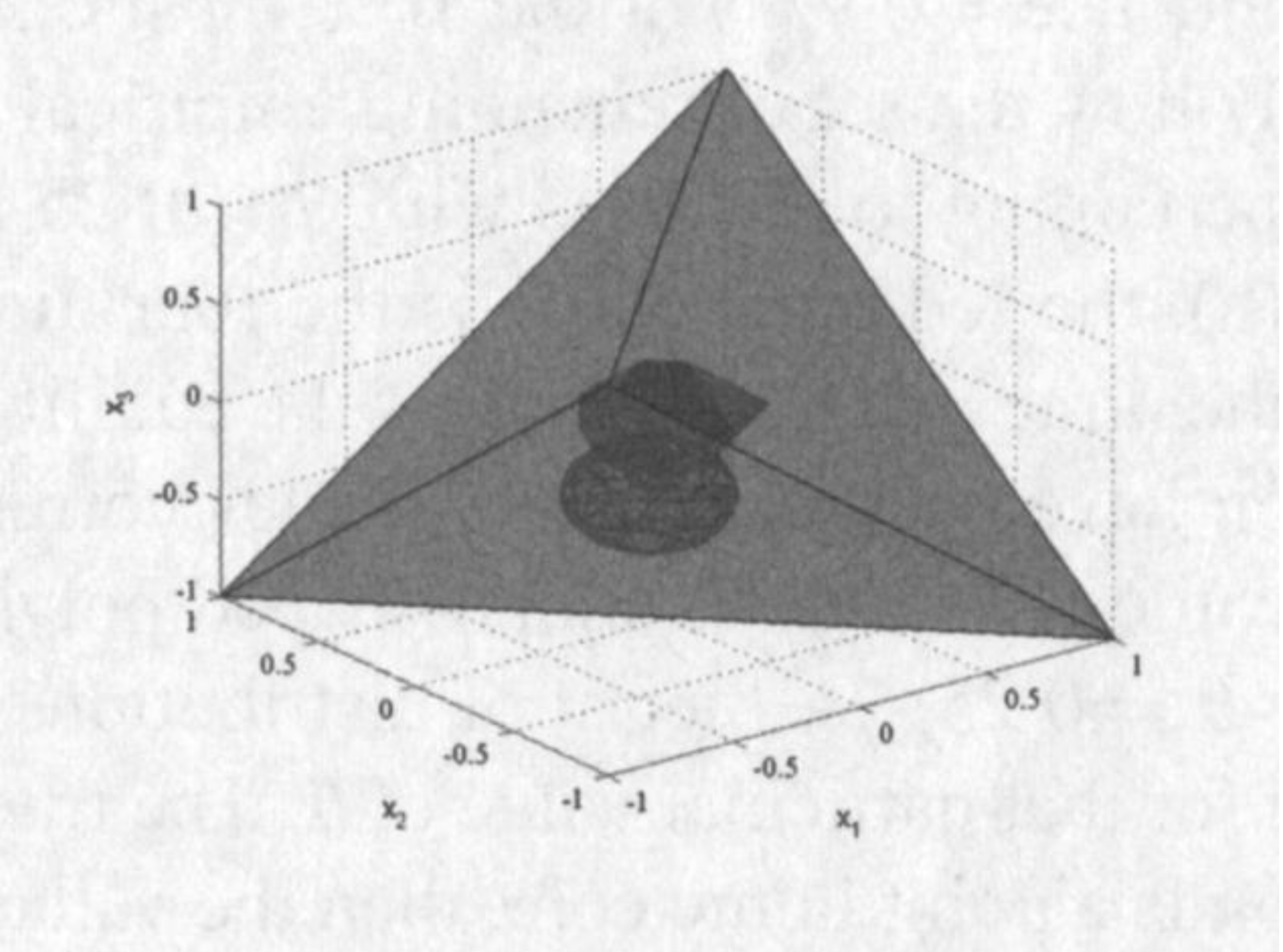}
\label{figure:L-M-L-M}
}
\subfigure[{\scriptsize $\theta=0.25$, Intermediate DGP, $n=1000$, $90\%$ confidence}]{
\includegraphics[width = 2in]{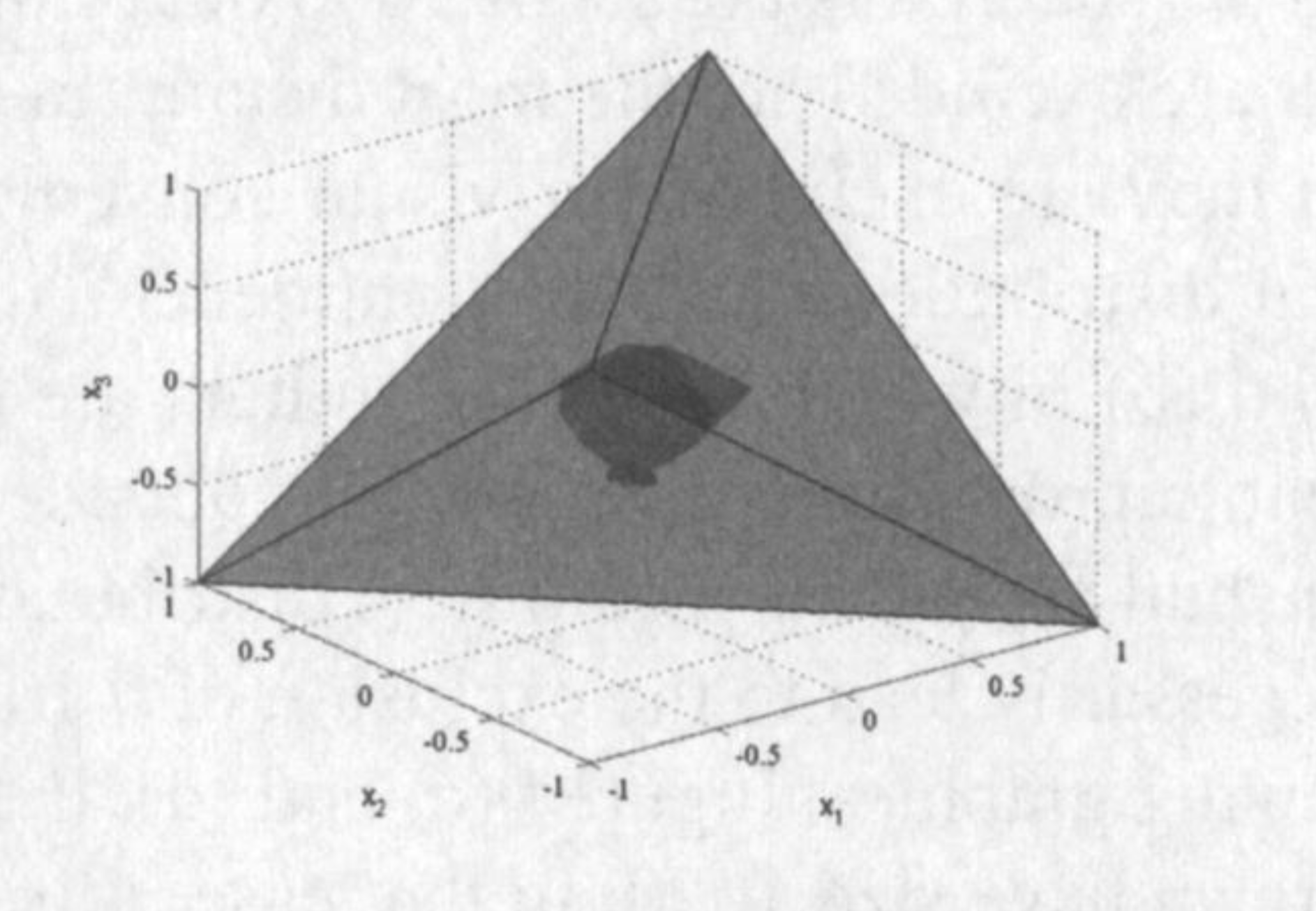}
\label{figure:L-M-H-L}
}
\subfigure[{\scriptsize $\theta=0.25$, Intermediate DGP, $n=1000$, $95\%$ confidence}]{
\includegraphics[width = 2in]{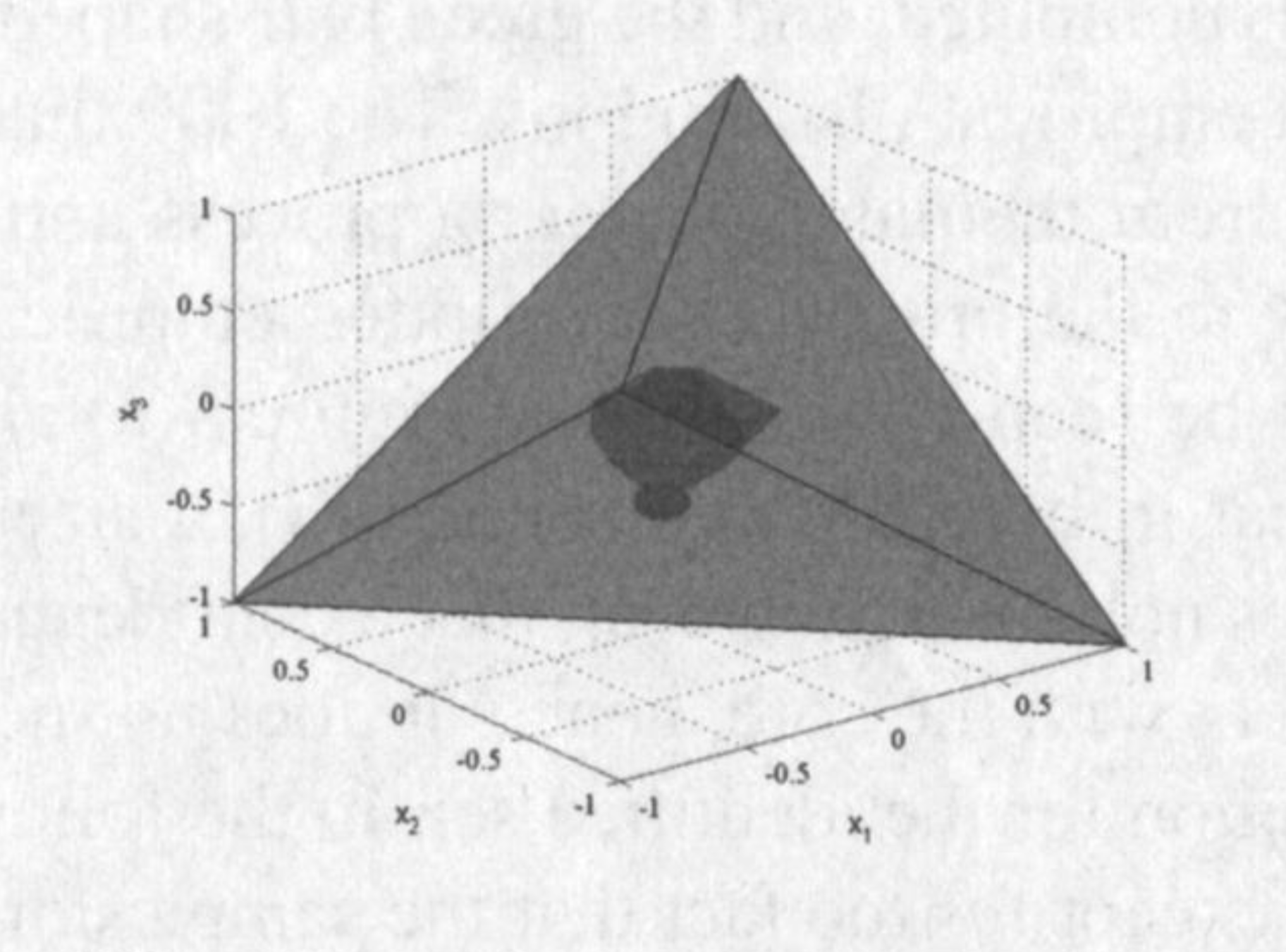}
\label{figure:L-M-H-M}
}
\caption[]{}
\label{figure:FB1}
\end{figure}

In figure~\ref{figure:L-L-L-L}, the core is plotted for the case $\theta=0.25$ and $10,000$ data series of size $n=100$ are drawn from the analytical center of the core. For each simulated sample, the empirical distribution is computed, and the green ball-shaped polyhedron is the set of $9,000$ out of the $10,000$ computed empirical distributions. The $1,000$ that are excluded are the most distant (in Euclidian distance) from the data generating process, and they are excluded for visual convenience. The sensitivity to the proportion of plotted empirical distributions (called ``confidence'' in the captions) can be seen by comparing figures~\ref{figure:L-L-L-L} and~\ref{figure:L-L-L-M}, in which only $500$ outliers are excluded. We see that in both figures~\ref{figure:L-L-L-L} and~\ref{figure:L-L-L-M}, a sizeable proportion of draws falls outside the core, which does not mean, however, that a confidence ball around such empirical distributions would not intersect with the core, hence it does not necessarily lead to the exclusion of $\theta$ from a confidence region for the identified set. In the following graphics, figures~\ref{figure:L-L-H-L} and \ref{figure:L-L-H-M}, the setup is identical, except for the fact that the samples drawn have size $1,000$. In that case, it is clear that all drawn empirical distributions are well within the core, which would lead to include $\theta$ in any reasonably constructed confidence region for the identified set. The observations are similar for figures~\ref{figure:L-M-L-L} to~\ref{figure:L-M-H-M}, where the data are generated from the mid-point between the analytical center and an extreme point of the core.
As a result, a larger proportion of simulated draws falls outside the core when $n=100$, but again, all draws a well within the core for $n=1,000$, so that $\theta$ would always be contained in a reasonably constructed confidence region for the identified set. In figures~\ref{figure:L-M-L-L} to~\ref{figure:L-M-H-M}, the data is generated from an extreme point of the core. In such a case, a point in the neighbourhood of the data generating process is more likely to fall outside the core than inside it. Hence, for all sample sizes, the proportion of empirical distributions that fall outside the core is relatively stable, and larger than the proportion that falls inside the core. However, for larger sample sizes, we see that simulated empirical distributions are more likely to be within a predetermined neighbourhood of the core, so that the value $\theta$ is less likely to be excluded from a confidence region for the identified set. The cases where $\theta= 0.5$ and $\theta= 0.75$ are very similar and we do not report them here.

\section{Empirical illustration: determinants of long term care for elderly parents}\label{section:application}
The applicability of the methodology proposed is best illustrated on
data pertaining to our example~\ref{subsubsection:family
bargaining}. We estimate the determinants of long term care option
choices for elderly parents in American families. The data consists
of a sample of 948 families with two children drawn from the
National Long Term Care Survey, sponsored by the National Institute
of Aging and conducted by the Duke University Center for Demographic
Studies under Grant number U01-AG007198, \cite{LTCS:82}.
Elderly people were interviewed in 1984 about their living and care
arrangements. The survey questions include gender and age of the
children, the distance between homes of parents and each of the
children, the disability status of the elderly parent (where
disability is referred to as problems with ``Activities of Daily
Living or Instrumental Activities of Daily Living (ADL)'') and the number of hours per week each of the
children devote to the care of the elderly parent. The dependent
variable is the care provision for the parent. The parent is asked
to list children (either at home or away from home) and how much
each provides help. If only one child is listed as providing
significant help, that child is designated the primary care giver.
If more than one child is listed, the one providing the most hours
is designated the primary care giver. If no child is listed or if
the parent lives in a nursing home, then the parent is designated as
``living alone.''

The observable choice of care option is modeled as in
\cite{ES:2002} as the outcome of a family bargaining game. The
payoff to family member $i$, $i=1,2$ is represented as the sum of
three terms. The first term $V_{ij}$ represents the value to child
$i$ of care option $j$, where $j>0$ means child $j$ becomes the
primary care giver and $j=0$ means the parent remains alone or is
moved to a nursing home. The matrix $V=(V_{ij})_{ij}$ is known to
both children. We suppose it takes the form
\[V=\left(\begin{array}{lll}\hskip5pt0\hskip10pt&
\beta_0+ADL\beta&\beta_0+ADL\beta\\
\hskip5pt0\hskip10pt&\beta_0+ADL\beta+D_1\psi+F\alpha&
\beta_0+ADL\beta\\\hskip5pt0\hskip10pt& \beta_0+ADL\beta
&\beta_0+ADL\beta+D_2\psi\end{array}\right)\] where $ADL$ is a dummy
variable that takes the value 1 if the parent has problems with
activities of daily living or instrumental activities of daily living, $D_i$ is the distance between parent and
child $i$, $F=1$ if child $1$ (who is first-born) is female, and
zero otherwise and $\theta=(\beta_0,\beta,\psi,\alpha)'$ is a four
dimensional parameter vector, unknown to the analyst. The value a family attaches to the fact that a non disabled parent is taken care of by one of the children is measured by $\beta_0$, whereas for a disabled parent, it is $\beta_0+\beta$. $\psi$ measures the disutility to the caregiver of living more than an hour away from the parent, and $\alpha$ measures the incremental utility of taking care of the parent for the firstborn daughter, as compared with a firstborn son.

In the data, $85\%$ of interviewed parents have problems with
activities of daily living. $70\%$ of parents live alone. $46\%$ of
first-born children are female, whereas $56\%$ of first-born who are
primary care-takers are female. The female first-born effect was
identified in \cite{ES:2002}, and we wish to see whether it is
robust to our analysis that takes multiple equilibria in the game
seriously.

Both children simultaneously decide whether or not to take part in
the long term care decision. Suppose $M$ is the set of children who
participate. The option chosen is option $j$ which maximizes the sum
$\sum_{i\in M}V_{ij}$ among the available options (only
participating children can become primary care givers). It is
assumed that participants abide with the decision and that benefits
are then shared equally among children participating in the decision
through a monetary transfer $s_i$, which is the second term in the
children's payoff.
The third term $\epsilon_i$ in the payoff is a random benefit from
participation, which is $0$ for children who decide not to
participate and distributed according to $\nu(.\vert \theta)$ for children
who participate. All children observe the realizations of
$\epsilon$, whereas the analyst only knows its distribution (the negative shock on participation is calibrated with a normal distribution $N(-3,1)$ for $\epsilon$. An alternative would have been a negative exponential distribution for $\epsilon$).

The payoff matrix of the participation game can be determined in the following way. \begin{itemize}
\item If both children decide to participate,
child i's payoff (for $i=1,2$) is \[
X_i=\max\left(0,\beta_0+ADL\beta+\frac{1}{2}(D_1\psi+F\alpha),
\beta_0+ADL\beta+\frac{1}{2}D_2\psi\right)+\epsilon_i.\]
\item If neither child participates, their payoffs are 0.
\item If only child $1$ participates, child $1$'s payoff is \[
X_1=\max\left(0,\beta_0+ADL\beta+D_1\psi+F\alpha\right)+\epsilon_1\]
and child $2$'s payoff is
\[X_2=
\left(\beta_0+ADL\beta\right)1_{\{\beta_0+ADL\beta+D_1\psi+F\alpha\geq0\}}.\]
\item If only child $2$ participates, child $2$'s payoff is \[
X_2=\max\left(0,\beta_0+ADL\beta+D_2\psi\right)+\epsilon_2\] and
child $1$'s payoff is
\[X_1=
\left(\beta_0+ADL\beta\right)1_{\{\beta_0+ADL\beta+D_2\psi\geq0\}}.\]
\end{itemize} We do not observe participation, but only the chosen
care option. To each equilibrium strategy profile corresponds a
(almost surely) unique care option choice, hence for each
participation shock $\epsilon$, we can derive the $G(\epsilon\vert
F,D,ADL;\theta)$ as the set of probability measures on the set of
care options $\{0,1,2\}$ induced by mixed strategy profiles, which
are probabilities on the set of participation profiles $\{(N,N),
(N,P), (P,N), (P,P)\}$. The Gambit software is a good option for the
computation of $G(\epsilon\vert F,D,ADL;\theta)$.

The methodology proposed in the paper allows the construction of the
identified set based on the hypothetical knowledge of the true
distribution of the data. In order to account for sampling
uncertainty, we appeal to the bootstrap procedure proposed in
\cite{GHQ:2008} to construct confidence regions for the
identified set in such situations. The method relies on a bootstrap
determination of a set function $A\mapsto\underline{P}(A\vert
F,D,ADL)$ which is dominated by $P(A\vert F,D,ADL)$ (uniformly over
$A\subseteq\{0,1,2\}$, $X$, $D$ and $ADL$) with probability
$1-\alpha$ (the chosen level of significance, here $0.95$). Once
$\underline{P}$ is determined, one proceeds as recommended in
section~\ref{subsection:mixed submodular}. That is to say, we keep
in the identified set only values of $\theta$ such that for all
observed values of the explanatory variables, the minimum over
$A\subseteq\{0,1,2\}$ of the function $\mathcal{L}(A\vert
F,D,ADL;\theta)-\underline{P}(A\vert F,D,ADL;\theta)$ is non
negative. The search over the 4-dimensional parameter space is
conducted in the following way. First one finds a value of the
parameter which lies within the confidence region (for instance the
corresponding estimates in the analysis of \cite{ES:2002}
which achieves identification by removing multiplicity of
equilibria), then one chooses a coarse discretization of
$[-\pi,\pi]^3$ to search in all directions for the frontier point of
the confidence region, which is assumed to be arc-connected (on each
arc, we use a dichotomic search). The resulting frontier points are
the extreme points of the confidence polytope. Each value of the
parameter can be tested in a fraction of a second on a standard
laptop, and the region can be constructed in a few hours, again on a
standard laptop without parallel processing. The confidence region
is a four dimensional polytope and 3, 2 or 1-dimensional confidence
regions for subsets or linear combinations of parameters can be
easily visualized as cuts from the confidence region using the
matlab multiparametric MPT toolbox.

\begin{figure}[htbp]
\centering
\subfigure[{\scriptsize }]{
\includegraphics[width = 2.5in]{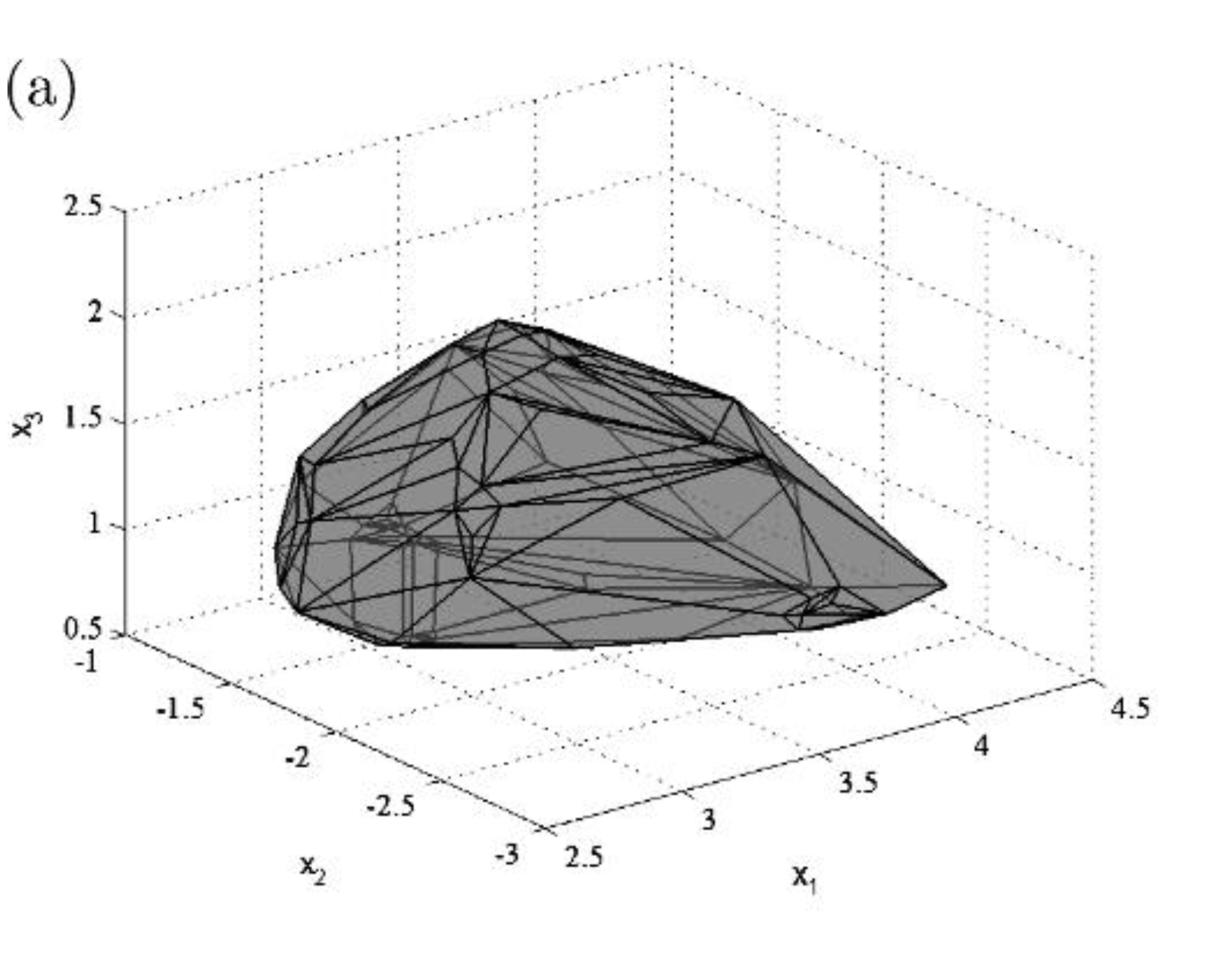}
\label{figure:}
}
\subfigure[{\scriptsize }]{
\includegraphics[width = 2.5in]{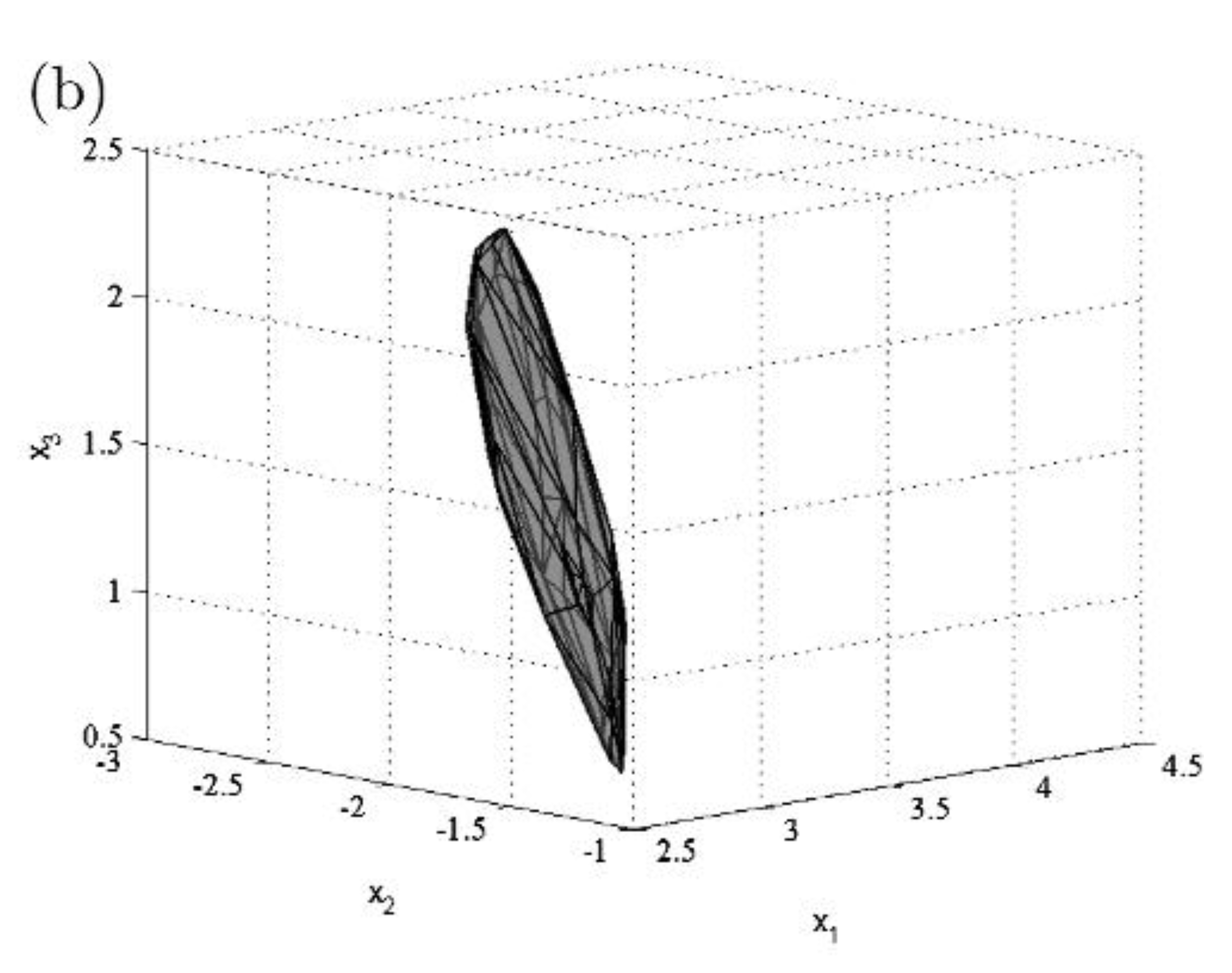}
\label{figure:}
}
\subfigure[{\scriptsize }]{
\includegraphics[width = 2.5in]{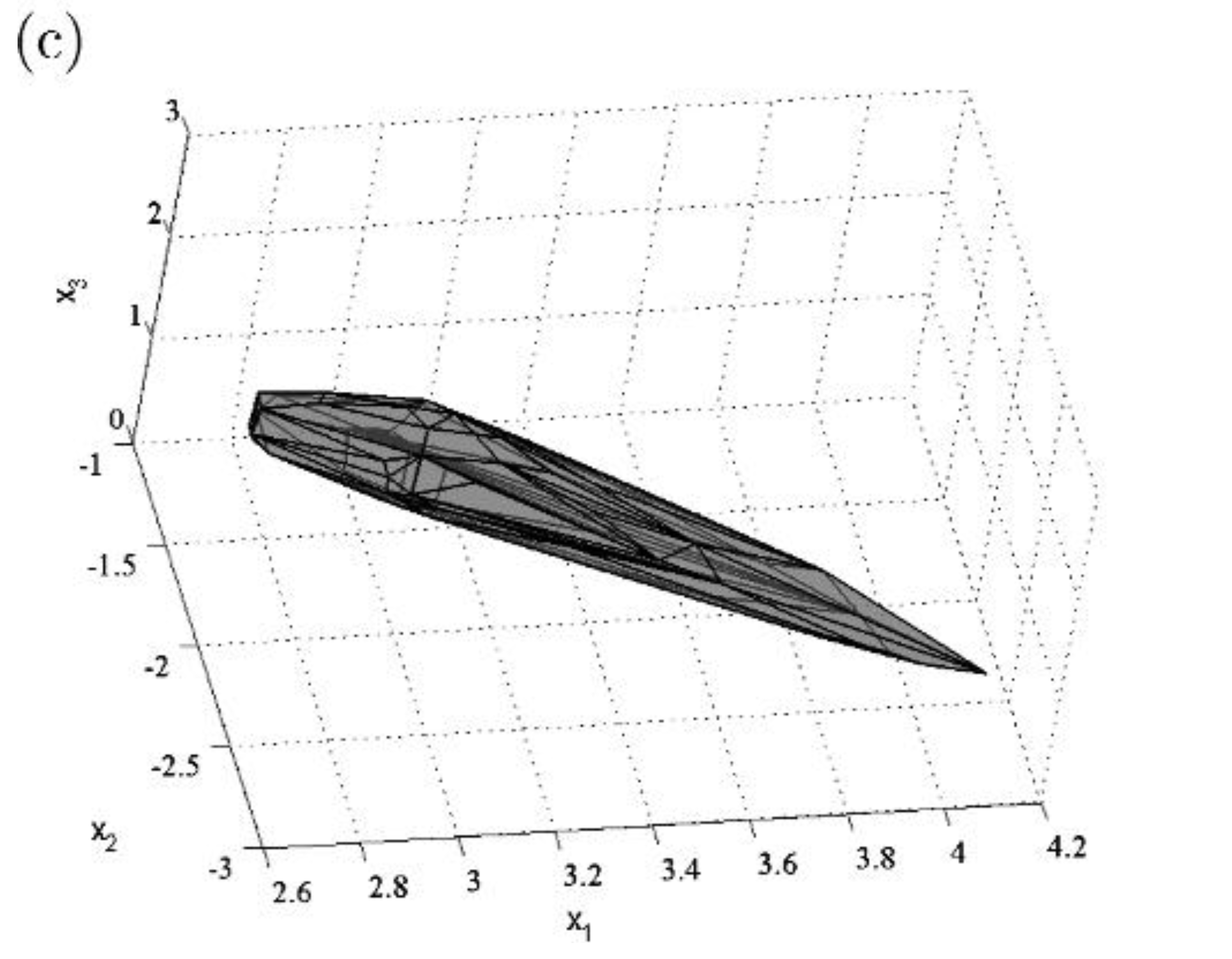}
\label{figure:}
}
\subfigure[{\scriptsize }]{
\includegraphics[width = 2.5in]{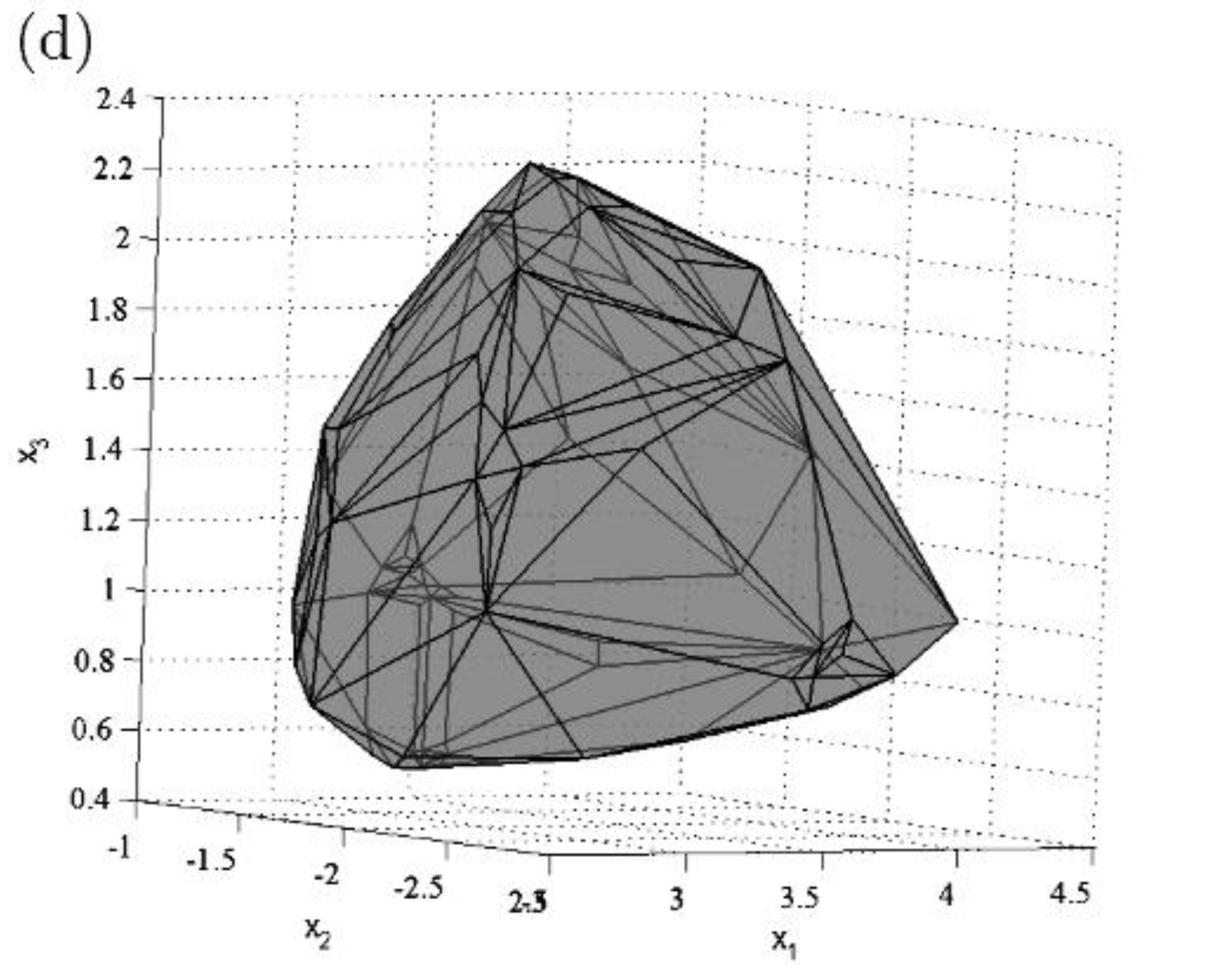}
\label{figure:}
}
\caption[]{$(\beta,\psi,\alpha)$ region at $\beta_0=0$.}
\label{figure:constrainedCR}
\end{figure}

\begin{figure}[htbp]
\centering
\subfigure[{\scriptsize $(\psi,\alpha)$ region at $(\beta_0,\beta)=(0,3)$.}]{
\includegraphics[width = 2.5in]{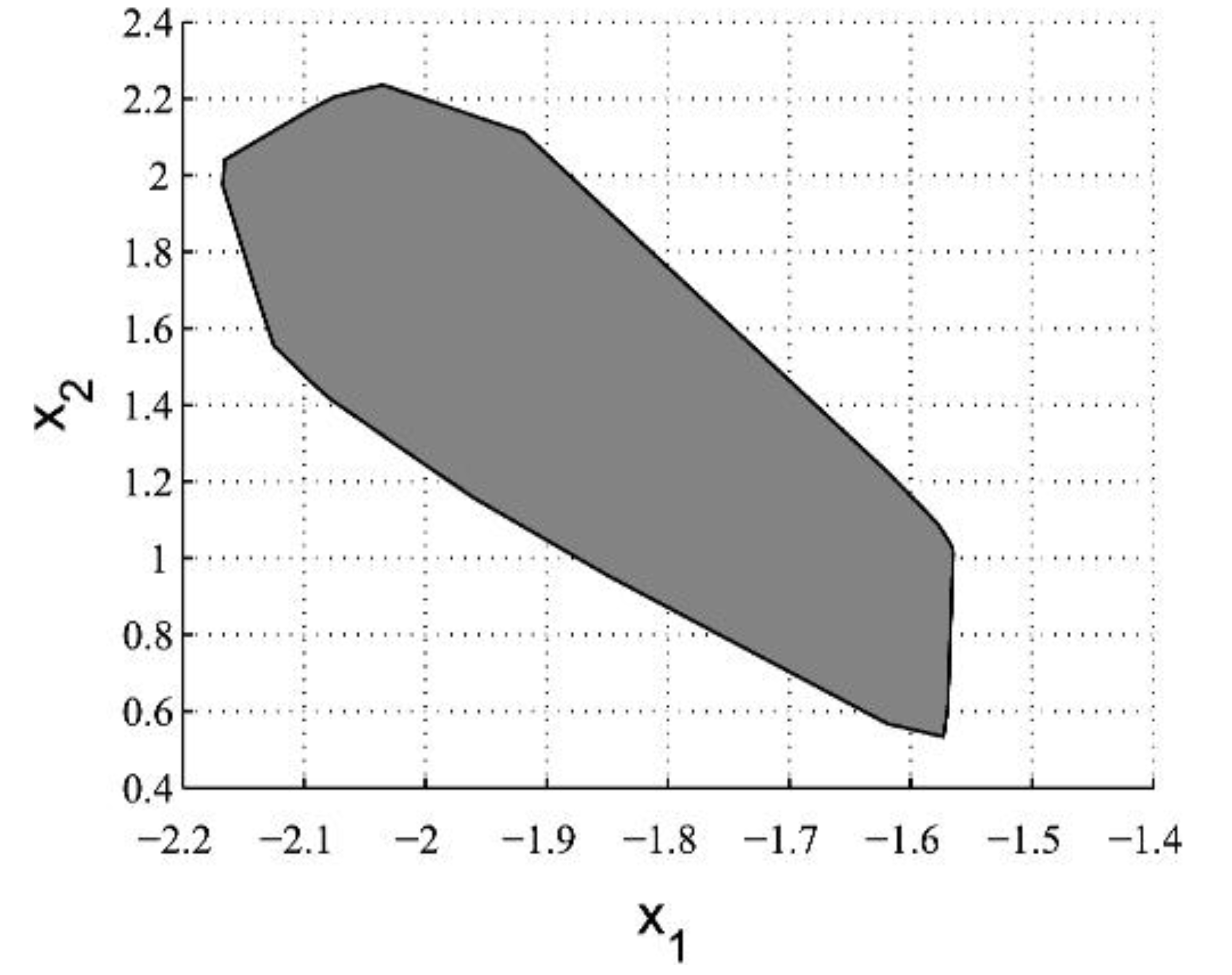}
\label{figure:slicedCRbeta=3}
}
\subfigure[{\scriptsize $(\beta,\psi)$ region at $(\beta_0,\alpha)=(0,2)$.}]{
\includegraphics[width = 2.5in]{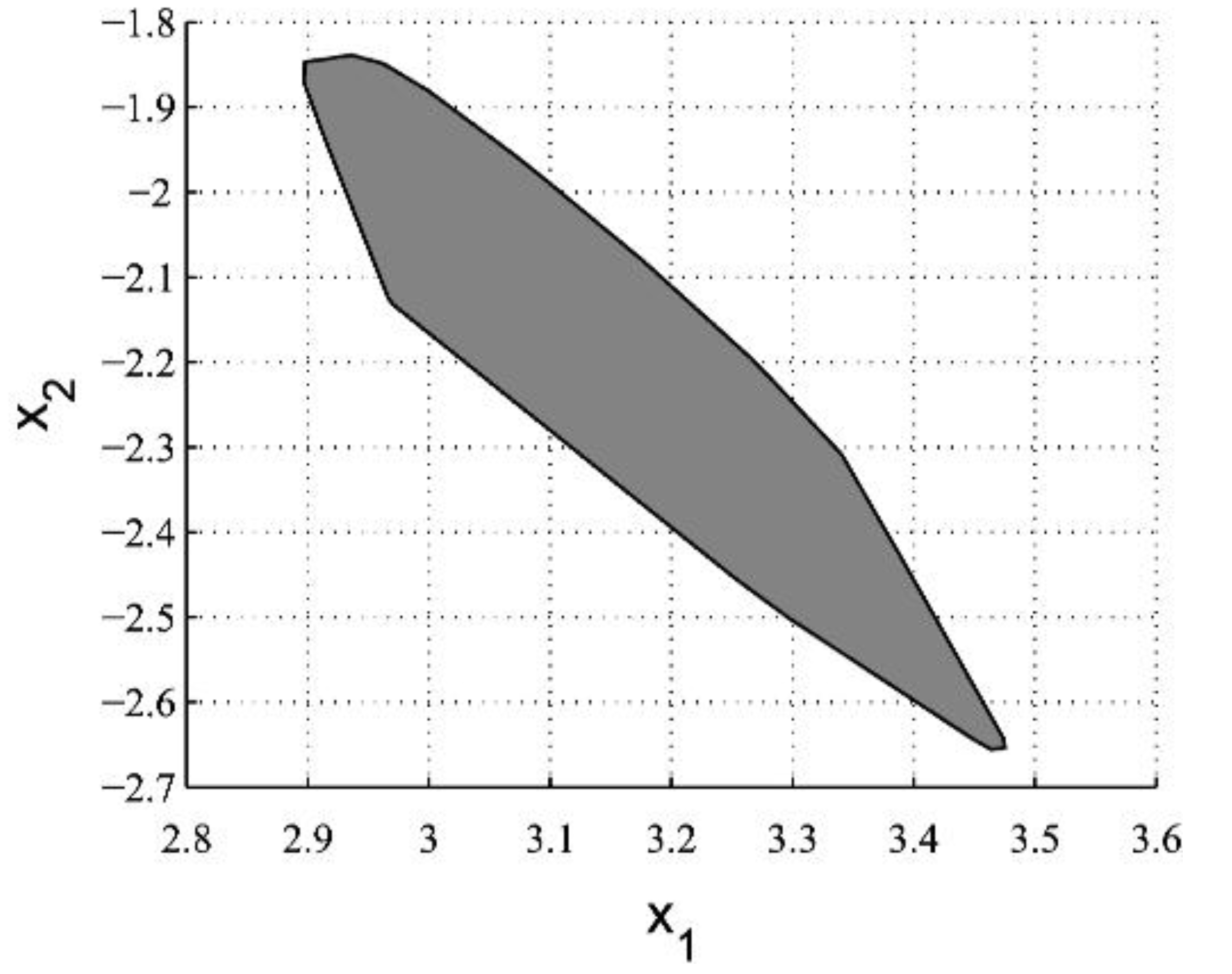}
\label{figure:slicedCRalpha=2}
}
\caption[]{Bivariate confidence regions}
\label{figure:slices}
\end{figure}

The variables chosen were those that were significant in
\cite{ES:2002}. We test the significance of each of the
individual parameters by checking whether the hyperplanes defined by
$\beta_0=0$, $\beta=0$, $\psi=0$ and $\alpha=0$ intersect with our
$95\%$ confidence region. The range of values of $\beta_0$ is
$\beta_0\in[-4.297,1.426]$. Hence, the hyperplane $\beta_0=0$ has
non empty intersection with the confidence region, which means we
fail to reject (at the $5\%$ level) the null hypothesis that the
family values identically the fact that a non disabled parent lives
alone or is taken care of by a family member. On the other hand, the
range of values for $\beta$ is $\beta\in[1.976,7.297]$, so the half
space $\beta\leq0$ has empty intersection with the confidence
region, and we reject the null hypothesis that a parent's disability
does not increase the value of care provided by the family. We also
reject the hypothesis that distance between parent and caretaker is
insignificant, as the range for $\psi$ is $\psi\in[-2.981,-1.306]$.
Finally, we reject the hypothesis that gender of the firstborn child
does not affect the chosen care option, as the range for $\alpha$ is
$\alpha\in[0.409,2.405]$. Indeed, our results are still consistent
with \cite{ES:2002} in the finding that families value care
provided by a firstborn daughter more than care provided by a
firstborn son. Since the hypothesis that $\beta_0=0$ is not
rejected, we slice the confidence region at $\beta_0=0$ to obtain
the constrained confidence region, which is three dimensional and is
represented (from four different angles) in
figure~\ref{figure:constrainedCR}. In the constrained region, the
ranges for the three remaining parameters are considerably tighter,
as $\beta\in[2.717,4.062]$, $\psi\in[-2.867,-1.365]$ and
$\alpha\in[0.532,2.290]$. Two dimensional regions are also given for
$(\alpha,\psi)$, when $\beta=3$, for $(\beta,\psi)$ when $\alpha=2$
and $(\beta,\alpha)$ when $\psi=-2$. These regions are plotted in
figure~\ref{figure:slices}. Notice the thin diagonal shapes of the
regions, which makes them rather informative, as for instance
simultaneous large values of $\psi$ and $\alpha$ are rejected, which
roughly means that firstborn daughters living close by are not the
only possible caregivers. Similarly, $\beta$ and $\psi$ are not
simultaneously very large, so that either the disability effect is
very large or the distance effect is very large, but not both.

\section*{Conclusion}
In the context of models with multiple equilibria in pure and mixed strategies, we have proposed
an equivalence result between the existence of an equilibrium
selection mechanism compatible with the data and a finite set of
inequalities characterizing the core of the model likelihood, and
provided methods to reduce this number of inequalities to be checked
with an appeal to the notion of core determining families and to
efficient easily implementable combinatorial methods. The issue of statistical inference on the identified feature thus characterized is taken up
in \cite{GH:2006d} and \cite{GH:2006c}, which complement
the seminal work of \cite{CHT:2007}.

\newpage

\begin{appendix}

\section{Proofs of results in the main text}
\subsection{Proof of theorem~\ref{theorem:sharp}}
It suffices to show that for all $\theta\in\Theta$, statement 1 and statement 2 are equivalent,
where statement 1 and statement 2 are defined as the following.
Statement 1: $P(A\vert X)
\leq\nu(\epsilon:\;G(\epsilon\vert X;\theta)\cap A\ne\varnothing\vert X;\theta)$ for all $A$ measurable subset of
$\mathcal{Y}$, $X$-almost surely. Statement 2: For almost all $\epsilon$, there exists a probability
measure $\pi(.\vert\epsilon,X)$ with support $G(\epsilon\vert X;\theta)$ such that $P(A\vert X)
=\int_\mathcal{U}\pi(A\vert X;\theta)\nu(d\epsilon\vert X;\theta)$
for all $A$ measurable subset of $\mathcal{Y}$, $X$-almost surely. We proceed in six steps.
\emph{Step 1:}
Since $G(\epsilon\vert X;\theta)$ is nonempty and closed by assumption, the set $\Delta(G(\epsilon\vert X;\theta))$ of probability measures on
$\mathcal{Y}$ with support $G(\epsilon\vert X;\theta)$ is convex and closed in the
topology of convergence in distribution.
\emph{Step 2:}
Since $G(.\vert X;\theta)$ is a measurable correspondence, for any $f\in\mathcal{C}_b(\mathcal{Y})$, the set of all continuous and bounded real functions on $\mathcal{Y}$,
the map $\epsilon\mapsto\sup\{\int fd\mu:\mu\in\Delta(G(\epsilon\vert X;\theta))\}$ is measurable.
\emph{Step 3:}
By step 1 and step 2, we can apply theorem~3 of \cite{Strassen:65} to conclude
that statement 2 is equivalent to
$\int_\mathcal{Y}f(y)P(dy)\leq\int_\mathcal{U}\sup\{\int fd\mu:\mu\in\Delta(G(\epsilon\vert X;\theta))\}
\nu(d\epsilon\vert X;\theta)$ for all $f\in\mathcal{C}_b(\mathcal{Y})$.
\emph{Step 4:} For any bounded continuous function $f$, we have
$\sup\{\int fd\mu:\mu\in\Delta(G(\epsilon\vert X;\theta))\}=\max\{f(y):y\in G(\epsilon\vert X;\theta)\}$.
\emph{Step 5:}
Call $\rho$ the Choquet capacity defined for all measurable subset $A$ of $\mathcal{Y}$
by $\rho(A)=\nu(\epsilon:\;G(\epsilon\vert X;\theta)\cap A\ne\varnothing\vert X;\theta)=\nu(G^{-1}(A\vert X;\theta)\vert X;\theta)$.
We show that $\int_\mathcal{U}\max\{f(y):y\in G(\epsilon\vert X;\theta)\}\nu(d\epsilon\vert X;\theta)=\int_\mathrm{Choquet}
fd\rho$, where the latter is the Choquet integral with respect to the Choquet capacity functional~$\rho$,
which is defined by $\int_\mathrm{Choquet}fd\rho=
\int_0^{\infty}\rho(\{f\geq v\})\,\mbox{d}v + \int_{-\infty}^0\,(\rho(\{f\geq
v\})-1)\,\mbox{d}v$. The latter can be rewritten $\int_0^{\infty}\nu\bigl(\epsilon:\;
 G(\epsilon\vert X;\theta)\cap\left\{f\geq v\right\}\ne\varnothing\vert X;\theta\bigr)\,\mbox{d}v +
\int_{-\infty}^0\,(\nu\bigl(\epsilon:\;
 G(\epsilon\vert X;\theta)\cap\left\{f\geq v\right\}\ne\varnothing\vert X;\theta\bigr)-1)\,\mbox{d}v$, which is equal
 to $\int_0^{\infty}\nu\bigl(\epsilon:\;\max_{y\in
G(\epsilon\vert X;\theta)}f(y)\geq v\vert X;\theta\bigr)\,\mbox{d}v +
\int_{-\infty}^0\,(\nu\bigl(\epsilon:\;\max_{y\in G(\epsilon\vert
X;\theta)} f(y)\geq v\vert X;\theta\bigr)-1)\,\mbox{d}v=
\int_\mathcal{U}\max\{f(y):y\in G(\epsilon\vert
X;\theta)\}\nu(d\epsilon\vert X;\theta)$, as we set out to show. \emph{Step
6:} Finally, by monotone continuity, we have that
$\int_\mathcal{Y}f(y)P(dy\vert X)\leq\int_\mathrm{Choquet}fd\rho$
for all $f\in\mathcal{C}_b(\mathcal{Y})$ is equivalent to the fact
that $P(.\vert X)$ is in the core of the Choquet capacity functional
$\rho$, which is statement~1.

\subsection{Proof of proposition~\ref{proposition:cd}}
We first state a general criterion for the core determining property. \begin{proposition}
A class $\mathcal{A}(\theta)$ of subsets of $\mathcal{Y}$ is core determining for
the Choquet capacity $A\mapsto\mathcal{L}(A\vert X;\theta)=\nu(\epsilon:\;G(\epsilon\vert X;\theta)\cap A\ne\varnothing\vert X;\theta)$
if for every measurable subset $A$ of $\mathcal{Y}$,
there exists nonnegative integers $K,L,N$, $\alpha_k$, $k=1,\ldots,K$, and elements $A_1,\ldots,A_K$ of $\mathcal{A}(\theta)$ such that for almost all $y\in\mathcal{Y}$ and almost all $\epsilon\in\mathcal{U}$,
\begin{eqnarray}1_A(y)\leq\frac{1}{N}\left(\sum_{k=1}^K\alpha_k1_{A_k}(y)-L\right) \hskip10pt \mathrm{ and }
\hskip10pt 1_{\{G(\epsilon\vert X;\theta)\cap A\ne\varnothing\}}(\epsilon)
\geq\frac{1}{N}\left(\sum_{k=1}^K\alpha_k1_{\{G(\epsilon\vert X;\theta)\cap
A_k\ne\varnothing\}}(\epsilon)-L\right). \label{equation:core
determining}\end{eqnarray} \label{proposition:cd}\end{proposition}

\begin{proof}[Proof of proposition~\ref{proposition:cd}] Subtracting the second inequality in
equation~\ref{equation:core determining} yields: $$1_A-1_{\{\epsilon:\,G(\epsilon\vert X;\theta)\cap A\ne\varnothing\}}\leq\frac{1}{N}\left(\sum_{k=1}^K\alpha_k(1_{A_k}-1_{\{\epsilon:\,G(\epsilon\vert X;\theta)\cap A_k\ne\varnothing\}})\right)$$
Taking expectations (conditionally on $X$) on both sides of the previous equation yields
\[P(A\vert X)-\nu(\epsilon:\;G(\epsilon\vert X;\theta)\cap A\ne\varnothing\vert X;\theta)\leq\frac{1}{N}\left(\sum_{k=1}^K\alpha_k(P(A_k\vert X)-\nu(\epsilon:\;G(\epsilon\vert X;\theta)\cap A_k\ne\varnothing\vert X;\theta))\right).\] This in turn implies that $P(A\vert X)\leq\nu(\epsilon:\;G(\epsilon\vert X;\theta)\cap A\ne\varnothing\vert X;\theta)$ if $P(A_k\vert X)\leq\nu(\epsilon:\;G(\epsilon\vert X;\theta)\cap A_k\ne\varnothing\vert X;\theta)$
for each $k$, which means that $\mathcal{A}(\theta)$ is indeed core determining, which completes the proof.\end{proof}

We now turn to the proof of theorem~\ref{theorem:cd}. We consider the equivalent problem
where the set of latent variables $\mathcal{U}$ is replaced by the set of predicted combinations
of equilibria $\mathcal{U}^\ast$. We keep the same notation for the equilibrium correspondence
$G$, and call $u$ the elements of $\mathcal{U}^\ast$. For simplicity, we also
drop the dependence on $X$ and $\theta$ in the notation, so that $G:u\mapsto G(u)$ is a correspondence between $\mathcal{U}^\ast$ and $\mathcal{Y}$. Note that by construction $G(u)=u\in2^\mathcal{Y}$, but it is not the identity
when considered as a correspondence.
Let $A$ be a subset of $\mathcal{Y}$. Call $K_y$ the cardinality of $\mathcal{Y}$ and $K_u$
the cardinality of $\mathcal{U}^\ast$. List all elements of $\mathcal{Y}$ as $y_k$, $k=1\ldots,K_y$
and all elements of $\mathcal{U}^\ast$ as $u_k$, $k=1,\ldots,K_u$. For any $u\in\mathcal{U}^\ast$, define $k_u$ by $u=u_{k_u}$
and for any $y\in\mathcal{Y}$, define $k_y$ by $y=y_{k_y}$. As usual, denote
$G^{-1}(A)$ the set $\{u\in\mathcal{U}^\ast:G(u)\cap A\ne\varnothing\}$. Call $\Delta1_{G^{-1}(A)}(u_1)=
1_{G^{-1}(A)}(u_1)$ and for $k\geq2$, $\Delta1_{G^{-1}(A)}(u_k)=1_{G^{-1}(A)}(u_k)-1_{G^{-1}(A)}(u_{k-1})$.
Call $\Delta1^+_{G^{-1}(A)}$ and $\Delta1^-_{G^{-1}(A)}$ the positive and negative parts of
$\Delta1_{G^{-1}(A)}$.  By construction, we have for any $u\in\mathcal{U}^\ast$,
\begin{eqnarray*}1_{G^{-1}(A)}(u)&=&\sum_{k=1}^{k_u}\Delta1_{G^{-1}(A)}(u_k)\\&=&
\sum_{k=1}^{K_u}1_{\{u_k,\ldots,u_{K_u}\}}(u)\Delta1_{G^{-1}(A)}(u_k)\\&=&
\sum_{k=1}^{K_u}1_{\{u_k,\ldots,u_{K_u}\}}(u)\Delta1^+_{G^{-1}(A)}(u_k)-
\sum_{k=1}^{K_u}1_{\{u_k,\ldots,u_{K_u}\}}(u)\Delta1^-_{G^{-1}(A)}(u_k)\\&=&
\sum_{k=1}^{K_u}1_{\{u_k,\ldots,u_{K_u}\}}(u)\Delta1^+_{G^{-1}(A)}(u_k)+
\sum_{k=1}^{K_u}1_{\{u_1,\ldots,u_{k-1}\}}(u)\Delta1^-_{G^{-1}(A)}(u_k)-
\sum_{k=1}^{K_u}\Delta1^-_{G^{-1}(A)}(u_k).\end{eqnarray*} We then apply
proposition~\ref{proposition:cd} with the
following choice of parameters. $K=2K_u$, $\alpha_j/N=
\Delta1^+_{G^{-1}(A)}(u_j)$, $j=1\ldots,K_u$,
$\alpha_j/N=\Delta1^-_{G^{-1}(A)}(u_{j-K_u})$,
$j=K_u+1,\ldots,2K_u$, and
$L/N=\sum_{k=1}^{K_u}\Delta1^-_{G^{-1}(A)}(u_k)$. There remains to
show that \begin{eqnarray*}1_A(y)\leq\sum_{k=1}^{K_u}1_{\{\inf
G(u_k),\ldots,y_{K_y}\}}(y)\Delta1^+_{G^{-1}(A)}(u_k)+
\sum_{k=1}^{K_u}1_{\{y_1,\ldots,\sup
G(u_{k-1})\}}(y)\Delta1^-_{G^{-1}(A)}(u_k)-
\sum_{k=1}^{K_u}\Delta1^-_{G^{-1}(A)}(u_k)\end{eqnarray*} to
complete the proof. The latter follows from
\begin{eqnarray}&&\sum_{k=1}^{K_u}1_{\{\inf
G(u_k),\ldots,y_{K_y}\}}(y)\Delta1^+_{G^{-1}(A)}(u_k)+
\sum_{k=1}^{K_u}1_{\{y_1,\ldots,\sup
G(u_{k-1})\}}(y)\Delta1^-_{G^{-1}(A)}(u_k)-
\sum_{k=1}^{K_u}\Delta1^-_{G^{-1}(A)}(u_k)\nonumber\\&=&
\sum_{k=1}^{K_u}1_{\{\inf
G(u_k),\ldots,y_{K_y}\}}(y)\Delta1^+_{G^{-1}(A)}(u_k)-
\sum_{k=1}^{K_u}1_{\{y_{k_{\sup
G(u_{k-1})}+1},\ldots,y_{K_u}\}}(y)\Delta1^-_{G^{-1}(A)}(u_k).
\label{equation:Poincare}\end{eqnarray}
We have $k_{\inf G(u_k)}< k_{\sup G(u_{k-1})}+1$ (otherwise, there would be a $y$
that belongs to none of the $u$'s, i.e. that is never an equilibrium outcome,
and it could be eliminated from the analysis). Hence (\ref{equation:Poincare})
is equal to
\begin{eqnarray*}
&&\sum_{k=1}^{K_u}1_{\{\inf G(u_k),\ldots,y_{k_{\sup
G(u_{k-1})}+1}\}}(y)\Delta1^+_{G^{-1}(A)}(u_k)\\&&\hskip100pt+
\sum_{k=1}^{K_u}1_{\{y_{k_{\sup
G(u_{k-1})}+1},\ldots,y_{K_u}\}}(y)(\Delta1^+_{G^{-1}(A)}(u_k)-\Delta1^-_{G^{-1}(A)}(u_k))\\&=&
\sum_{k=1}^{K_u}1_{\{\inf G(u_k),\ldots,y_{k_{\sup
G(u_{k-1})}+1}\}}(y)\Delta1^+_{G^{-1}(A)}(u_k)+
\sum_{k=1}^{K_u}1_{\{y_{k_{\sup
G(u_{k-1})}+1},\ldots,y_{K_u}\}}(y)\Delta1_{G^{-1}(A)}(u_k)\\&\geq&
\sum_{k=1}^{K_u}1_{\{\inf G(u_k),\ldots,y_{k_{\sup
G(u_{k-1})}+1}\}}(y)\Delta1_{G^{-1}(A)}(u_k)+
\sum_{k=1}^{K_u}1_{\{y_{k_{\sup
G(u_{k-1})}+1},\ldots,y_{K_u}\}}(y)\Delta1_{G^{-1}(A)}(u_k)\\&=&
\sum_{k=1}^{K_u}1_{\{\inf
G(u_{k}),\ldots,y_{K_u}\}}(y)\Delta1_{G^{-1}(A)}(u_k)=1_A(y),\end{eqnarray*}
Which completes the proof.

\subsection{Proof of Theorem~\ref{theorem:matching}}
First note that any specification of the latent variable $\epsilon$
that produces the same combinations of equilibria listed in $\mathcal{U}^\ast$
with the same probabilities are observationally equivalent. We can therefore replace
$\mathcal{U}$ by $\mathcal{U}^\ast$, where each $u\in\mathcal{U}^\ast$
has probability $Q(u\vert X;\theta)=\nu(\epsilon:\;G(\epsilon\vert X;\theta)=u\vert X;\theta)$,
and redefine $G$ as the correspondence from $\mathcal{U}^\ast$ to $\mathcal{Y}$
defined by $G(u)=u$.
By theorem~\ref{theorem:sharp}, $\theta$ belongs to the identified set
if and only if for any subset $A$ of $\mathcal{Y}$, $P(A\vert X)
\leq\nu(\epsilon:\;G(\epsilon\vert X;\theta)\cap A\ne\varnothing\vert X;\theta)$ or equivalently
$P(A\vert X)\leq Q(G^{-1}(A)\vert X;\theta)$. By proposition~1 of \cite{GH:2006d},
this is equivalent to the existence of a probability $\pi$ on $\mathcal{Y}\times\mathcal{U}^\ast$
with marginal distributions $P(.\vert X)$ and $Q(.\vert X;\theta)$ and such that
$\pi\{(y,u)\in\mathcal{Y}\times\mathcal{U}^\ast: y\in u\}=1$, in other words
such that it is supported on the subset of pairs $(y,u)$ such that $y\in u$. This completes the proof.

\subsection{Proof of Proposition~\ref{proposition:BMM}}
The term in parenthesis in (\ref{equation:mixed}) is a convex combination of elements of $G(\epsilon\vert X;\theta)$, hence the existence of an equilibrium mechanism with the required property is equivalent to the existence of a family $\Pi(.,\epsilon\vert X;\theta)$ of probability measures on $\mathcal{Y}$ such that $\Pi(.,\epsilon\vert X;\theta)\in\mathrm{co}(G(\epsilon\vert X;\theta))$, $\epsilon$-almost surely. By Choquet's Theorem (see \cite{Choquet:53}), this implies
that for any function $f$ on $\mathcal{Y}$, \[\sum_{\mathcal{Y}} f(y)\Pi(y,\epsilon\vert X;\theta)\leq\max\left\{\sum_{\mathcal{Y}}f(y)s(y); s\in G(\epsilon\vert X;\theta)\right\},\] which, by integration over $\epsilon$, implies \[\sum_\mathcal{Y}f(y)\mathbb{P}(Y=y\vert X)\leq
\int \left\{\max_{s\in G(\epsilon\vert X;\theta)}\sum_\mathcal{Y}f(y)s(y)\right\}d\nu(\epsilon\vert X).\]
The latter yields the desired representation after dividing both sides by the norm of $f$.

Conversely, suppose that for all functions $f$ on $\mathcal{Y}$, we have \[\sum_\mathcal{Y}f(y)\mathbb{P}(y\vert X)\leq \int L_{\epsilon\vert X;\theta}(f)d\nu(\epsilon\vert X;\theta),\] where \[L_{\epsilon\vert X;\theta}(f)=\max\left\{\sum_{\mathcal{Y}}f(y)s(y); s\in G(\epsilon\vert X;\theta)\right\}.\] By the Strassen Disintegration Theorem (see \cite{Strassen:65}), there exists a map $\epsilon\mapsto\Pi(.,\epsilon\vert X;\theta)$ such that for all $y\in\mathcal{Y}$, $\mathbb{P}(Y=y\vert X)=\int\Pi(y,\epsilon\vert X;\theta)d\nu(\epsilon\vert X)$ and $\Pi(.,\epsilon\vert X;\theta)\leq L_{\epsilon\vert X;\theta}(.)$, $\epsilon$-almost surely. Again by the Choquet Theorem, the latter condition is equivalent to $\Pi(.,\epsilon\vert X;\theta)\in\mathrm{co}(G(\epsilon\vert X;\theta))$, $\epsilon$-almost surely, and the result follows.

\subsection{Proof of Lemma~\ref{lemma:regular core}}
Consider a game with equilibrium correspondence $G$ and call $\mu$ the upper envelope of the equilibrium correspondence, i.e. for all $A\subset\mathcal{Y}$, $\mu(A)=\sup_{\sigma\in G}\sigma(A)$. Let $Core(\mu)$ be the core of $\mu$ (probabilities set-wise dominated by $\mu$) and for a given subset $A$ of $\mathcal{Y}$, call $C_A$ the set of probabilities in $Core(\mu)$ that attain the supremum at $A$, i.e. such that $P(A)=\mu(A)$. Section~3.1 of \cite{Shapley:71} shows that the game has a regular core if and only if for each $P\in Core(\mu)$, the class of subsets $\{A\subset\mathcal{Y}\}$ such that $P\in C_A$ is closed under union and intersection. Consider a pure strategy equilibrium in $G$, denoted $\delta_y$. Then $\delta_y\in C_A$ if and only if $y\in A$, and the class of sets that contains $y$ is closed under unions and intersections. Consider now the unique proper mixed strategy equilibrium $\sigma$ in $G$. Then, $\sigma\in C_A$ is and only if $A$ does not meet the domain of any pure strategy equilibrium in $G$, a property which is also preserved under unions and intersections. The result follows.

\subsection{Proof of Theorem~\ref{theorem:submodular mixed identified set}} By proposition~\ref{proposition:BMM} , the identified set is characterized as the set of parameter values $\theta$ such that for all functions $f$ on $\mathcal{Y}$,
$E_P(f(Y)\vert X)\leq \int\left(\max_{\sigma\in G(\epsilon\vert X;\theta)}E_\sigma(f(Y))\right)d\nu(\epsilon\vert X;\theta)$.
By \cite{Schmeidler:86}, the functional \[\tilde{\mathcal{L}}: f\mapsto \int\left(\max_{\sigma\in G(\epsilon\vert X;\theta)}E_\sigma(f(Y))\right)d\nu(\epsilon\vert X;\theta)\]
is the Choquet integral associated with the Choquet capacity $\mathcal{L}:B\mapsto \mathcal{L}(B\vert X;\theta)=\tilde{\mathcal{L}}(1_B\vert X;\theta)$, because the latter is submodular, as a mixture of submodular capacities, and (dropping the dependence on $X$ and $\theta$ from notation) the Choquet integral satisfies the following two properties:
\begin{itemize}
\item Monotonicity: If $f(y)\leq t(y)$ for all $y\in\mathcal{Y}$, then $\tilde{\mathcal{L}}(f)\leq\tilde{\mathcal{L}}(t)$,
\item Comonotonic additivity: If $f$ and $t$ are comonotonic, i.e. if $(f(y_1)-f(y_2))(t(y_1)-t(y_2))\geq0$ for all $y_1,y_2\in\mathcal{Y}$, then $\tilde{\mathcal{L}}(f+t)=\tilde{\mathcal{L}}(f)+\tilde{\mathcal{L}}(t)$.
\end{itemize}
Note that submodularity of $\mathcal{L}:B\mapsto \mathcal{L}(B)=\tilde{\mathcal{L}}(1_B)$ (i.e. $\mathcal{L}(A\cap B)+\mathcal{L}(A\cup B)\leq \mathcal{L}(A)+\mathcal{L}(B)$ for all $A,B\subseteq\mathcal{Y}$) is equivalent to convexity of the Choquet integral $\tilde{\mathcal{L}}$ (see proposition~3 of \cite{Schmeidler:86} or theorem~6.13 page 212 of \cite{Fujishige:2005}). By definition of the Choquet integral, $\tilde{\mathcal{L}}$ can be written \[\tilde{\mathcal{L}}(f)=
\int_0^\infty \mathcal{L}(f\geq x)dx+\int_{-\infty}^0(\mathcal{L}(f\geq x)-1)dx.\]
Whenever a probability $P$ on $\mathcal{Y}$ satisfies $P(B)\leq \mathcal{L}(B)$
for all $B\in\subseteq\mathcal{Y}$, it follows that \begin{eqnarray*}E_P(f(Y))&=&\sum_{y\in\mathcal{Y}}f(y)P(y)\\&=&
\int_0^\infty P(f\geq x)dx+\int_{-\infty}^0(P(f\geq x)-1)dx\\&\leq&
\int_0^\infty \mathcal{L}(f\geq x)dx+\int_{-\infty}^0(\mathcal{L}(f\geq x)-1)dx\\&=&
\tilde{\mathcal{L}}(f).\end{eqnarray*} The converse is immediately seen by taking the indicator functions for $f$, and the result follows.

\section{Complements}
\subsection{Likelihood in the example~\ref{subsubsection:family bargaining}}
\label{section:likelihood derivation}
\begin{itemize}\item $\mathcal{L}(\{(0,1)\}\vert \theta)=
\nu(\epsilon_1<\theta,\epsilon_2>-2\theta)$
\item $\mathcal{L}(\{(1,0)\}\vert \theta)=
\nu(\epsilon_2<\theta,\epsilon_1>-2\theta)$
\item $\mathcal{L}(\{(0,1),(1,0)\}\vert\theta)=1-
\nu(\epsilon_1<-2\theta,\epsilon_2<-2\theta)
-\nu(\epsilon_1>\theta,\epsilon_2>\theta)$
\item $\mathcal{L}(\{(0,1),(1,0),(1,1)\}\vert\theta)=1-
\nu(\epsilon_1<-2\theta,\epsilon_2<-2\theta)$
\item $\mathcal{L}(\{(0,1),(1,0),(0,0)\}\vert\theta)=1
-\nu(\epsilon_1>\theta,\epsilon_2>\theta)$
\item $\mathcal{L}(\{(0,0)\}\vert \theta)=\mathbb{E}_\nu[(\theta-\epsilon_2)
(\theta-\epsilon_1)/(9\theta^2)1{\{\epsilon\in[-2\theta,\theta]^2\}}]
+\nu(\epsilon_1<-2\theta,\epsilon_2<-2\theta)$
\item $\mathcal{L}(\{(1,1)\}\vert \theta)=\mathbb{E}_\nu[(2\theta+\epsilon_2)
(2\theta+\epsilon_1)/(9\theta^2)1{\{\epsilon\in[-2\theta,\theta]^2\}}]
+\nu(\epsilon_1>\theta,\epsilon_2>\theta)$
\item $\mathcal{L}(\{(1,1),(0,0)\}\vert\theta)=
\mathcal{L}(\{(1,1)\}\vert\theta)+\mathcal{L}(\{(0,0)\}\vert \theta)$.
\item $\mathcal{L}(\{(0,0),(0,1)\}\vert\theta)=\mathcal{L}(\{(0,1)\}\vert\theta)+
\nu(\epsilon_1<-2\theta,\epsilon_2<-2\theta)$
\item $\mathcal{L}(\{(0,0),(1,0)\}\vert\theta)=\mathcal{L}(\{(1,0)\}\vert\theta)+
\nu(\epsilon_1<-2\theta,\epsilon_2<-2\theta)$
\item $\mathcal{L}(\{(1,1),(0,1)\}\vert\theta)=\mathcal{L}(\{(0,1)\}\vert\theta)+
\nu(\epsilon_1>\theta,\epsilon_2>\theta)$
\item $\mathcal{L}(\{(1,1),(1,0)\}\vert\theta)=\mathcal{L}(\{(1,0)\}\vert\theta)+
\nu(\epsilon_1>\theta,\epsilon_2>\theta)$
\item $\mathcal{L}(\{(1,1),(0,0),(0,1)\}\vert\theta)=\mathcal{L}(\{(0,1)\}\vert\theta)+
\nu(\epsilon_1>\theta,\epsilon_2>\theta)+
\nu(\epsilon_1<-2\theta,\epsilon_2<-2\theta)$
\item $\mathcal{L}(\{(1,1),(0,0),(1,0)\}\vert\theta)=\mathcal{L}(\{(1,0)\}\vert\theta)+
\nu(\epsilon_1>\theta,\epsilon_2>\theta)+
\nu(\epsilon_1<-2\theta,\epsilon_2<-2\theta)$\end{itemize}

\end{appendix}

\pagebreak
\markboth{References}{References}
\printbibliography

@unpublished{ABBP:2007,
        author="D. Ackerberg and L. Benkard and S. Berry and A. Pakes",
        title="Econometric tools for analyzing market outcomes",
        year= 2007,
        note= "{\em Handbook of Econometrics}, Volume 6A",
}

@unpublished{ABJ:2003,
        author="Donald Andrews and Steven Berry and Panle Jia",
        title="Placing bounds on parameters of entry games in the presence of multiple equilibria",
        year= 2003,
        note= "unpublished manuscript",
}

@unpublished{AS:2007,
        author="Donald Andrews and Gustavo Soares",
        title="Inference for Parameters Defined by Moment Inequalities Using Generalized Moment Selection ",
        year= 2007,
        note={unpublished manuscript},
}

@article{Artstein:83,
        author="Zvi Artstein",
        title="Distributions of random sets and random selections",
        journal="Israel Journal of Mathematics",
        year= 1983,
        volume=46,
        pages=313-324,
}

@unpublished{BM:2007,
        author="Ari Beresteanu and Francesca Molinari",
        title="Asymptotic properties for a class of partially identified models",
        year= 2007,
        note={forthcoming in Econometrica},
}

@article{Berry:92,
        author="Steve Berry",
        title="Estimation of a model of entry in the airline industry",
        journal={Econometrica},
        year=1992,
        volume=60,
        pages="889--917",
}

@inproceedings{BT:2006,
        author="Steve Berry and Elie Tamer",
        title="Identification in models of oligopoly entry",
        booktitle="Advances in Economics and Econometrics",
        year= 2006,
        publisher= "Cambridge University Press",
        pages="46--85"
}

@article{BR:90,
        author="Tim Bresnahan and P. Reiss",
        title="Entry in monopoly markets",
        journal={Review of Economic Studies},
        year=1990,
        volume=57,
        pages="531--553",
}

@unpublished{Canay:2007,
        author="Ivan Canay",
        title="EL Inference for Partially Identified Models: Large Deviation Optimality
        and Bootstrap Validity",
        year= 2007,
        note= "unpublished manuscript",
}

@article{CHT:2007,
        author="Victor Chernozhukov and Han Hong and Elie Tamer",
        title=" Estimation and Confidence Regions for Parameter Sets in Econometric Models",
        journal={Econometrica},
        year= 2007,
        volume=75,
        pages="1243--1285"
}

@article{Choquet:53,
        author="Gustave Choquet",
        title="Theory of capacities",
        journal={Annales de l'Institut Fourier},
        year= 1954,
        volume=5,
        pages="131--295",
}

@unpublished{CT:2006,
        author="Federico Ciliberto and Elie Tamer",
        title="Market structure and multiple equilibria in airline markets",
        year= 2006,
        note= "unpublished manuscript",
}

@article{FF:57,
        author="L. Ford and D. Fulkerson",
        title="A simple algorithm for finding maximal network flows and an application to the Hitchcock problem",
        journal={Canadian Journal of Mathematics},
        year= 1957,
        volume=9,
        pages="210--218",
}

@unpublished{GH:2006a,
        author="Alfred Galichon and Marc Henry",
        title="Inference in incomplete models",
        year= 2006,
        note="Columbia University Discussion Paper 0506-28 available at
        http://www.columbia.edu/cu/economics/discpapr/DP0506-28.pdf",
}

@unpublished{GH:2006c,
        author="Alfred Galichon and Marc Henry",
        title="Dilation Bootstrap. A methodology for constructing confidence
        regions with partially identified models",
        year= 2006,
        note="unpublished manuscript",
}

@unpublished{GH:2006d,
        author="Alfred Galichon and Marc Henry",
        title="A test of non-identifying restrictions and confidence regions for
        partially identified parameters",
        year= 2007,
        note="unpublished manuscript",
}

@unpublished{Gillies:53,
        author="D. Gillies",
        title="Some theorems on n-person games",
        year= 1953,
        note={Princeton Ph.D.},
}

@article{Heckman:78,
        author="James Heckman",
        title="Dummy endogenous variables in a simultaneous equations system",
        journal={Econometrica},
        year=1978,
        volume=46,
        pages="931--960",
}

@article{HSC:97,
        author="James Heckman and Jeffrey Smith and Nancy Clements",
        title="Making the most out of programme evaluation and social experiments: accounting for heterogeneity
        in programme impacts",
        journal={Review of Economic Studies},
        year=1997,
        volume=64,
        pages="487--535",
}

@article{IM:2004,
        author="Guido Imbens and Charles Manski",
        title="Confidence intervals for partially identified parameters",
        journal={Econometrica},
        year=2004,
        volume=72,
        pages="1845--1859",
}

@article{Jovanovic:89,
        author="Boyan Jovanovic",
        title="Observable implications of models with multiple equilibria",
        journal={Econometrica},
        year= 1989,
        volume=57,
        pages="1431--1437",
}

@article{Koopmans:49,
        author="Tjallin Koopmans",
        title="Optimum utilization of the transportation system",
        journal={Econometrica},
        year=1949,
        volume=17,
        pages="136--146",
}

@article{KR:50,
        author="Tjallin Koopmans and O. Reiersol",
        title="The identification of structural characteristics",
        journal={Annals of Mathematical Statistics},
        year= 1950,
        volume=21,
        pages="165--181",
}

@article{Manski:90,
        author="Charles Manski",
        title="Nonparametric bounds on treatment effects",
        journal={American Economic Review},
        year=1990,
        volume=80,
        pages="319--323",
}

@article{MT:2002,
        author="Charles Manski and Elie Tamer",
        title="Inference on Regressions with Interval Data on a Regressor or Outcome",
        journal={Econometrica},
        year=2002,
        volume=70,
        pages="519--546",
}

@unpublished{PPHI:2004,
        author="Ariel Pakes and Jack Porter and Kate Ho and Joy Ishii",
        title="Moment inequalities and their application",
        year= 2004,
        note= "unpublished manuscript",
}

@book{PS:98,
        author={Christos Papadimitriou and Kenneth Steiglitz},
        title={Combinatorial Optimization: Algorithms and Complexity},
        year=1998,
        publisher={Dover},
}

@unpublished{RS:2006a,
        author="Joe Romano and Azeem Shaikh",
        title="Inference for identifiable parameters in partially identified econometric models",
        year= 2006,
        note="forthcoming in the {\em Journal of Statistical Planning and Inference}",
}

@unpublished{Rosen:2006,
        author="Adam Rosen",
        title="Confidence sets for partially identified parameters that satisfy a
        finite number of moment inequalities",
        year= 2006,
        note={unpublished manuscript},
}

@article{Schmeidler:86,
        author="David Schmeidler",
        title="Integral representation without additivity",
        journal={Proceedings of the American Mathematical Society},
        year= 1986,
        volume=97,
        pages="255--261",
}

@article{Strassen:65,
        author="Victor Strassen",
        title="The existence of probability measures with given marginals",
        journal={Journal of Mathematical Statistics},
        year= 1965,
        volume=36,
        pages="423--439",
}

@article{Tamer:2003,
        author="Elie Tamer",
        title="Incomplete simultaneous discrete response model with multiple equilibria",
        journal={Review of Economic Studies},
        year= 2003,
        volume=70,
        pages="147--165"
}

@Article{GLM:80,
  author={Gourieroux, C and Laffont, J J and Monfort, A},
  title={{Coherency Conditions in Simultaneous Linear Equation Models with Endogenous Switching Regimes}},
  journal={Econometrica},
  year=1980,
  volume={48},
  number={3},
  pages={675-695},
  month={April},
  doi={},
  url={https://ideas.repec.org/a/ecm/emetrp/v48y1980i3p675-95.html}
}

@Article{DJ:94,
  author={Dagsvik, John and Jovanovic, Boyan},
  title={{Was the Great Depression a low-level equilibrium?}},
  journal={European Economic Review},
  year=1994,
  volume={38},
  number={9},
  pages={1711-1729},
  month={December},
  doi={},
  url={https://ideas.repec.org/a/eee/eecrev/v38y1994i9p1711-1729.html}
}

@Article{BMT:2002,
  author={Bisin, Alberto and Moro, Andrea and Topa, Giorgio},
  title={Empirical Content of Models with Multiple Equilibria},
  journal={unpublished manuscript},
  year=2002}

@article{EGH:2008,
 ISSN = {09382259, 14320479},
 URL = {http://www.jstor.org/stable/25619992},
 author = {Ivar Ekeland and Alfred Galichon and Marc Henry},
 journal = {Economic Theory},
 number = {2},
 pages = {355--374},
 publisher = {Springer},
 title = {Optimal Transportation and the Falsifiability of Incompletely Specified Economic Models},
 volume = {42},
 year = {2010}
}

@article{ES:2002,
 author = {Maxim Engers and Steven Stern},
 journal = {International Economic Review},
 pages = {73–114},
 title = {Family Bargaining and Long Term Care},
 volume = {43},
 year = {2002}
}

@article{BMM:2008,
author = {Beresteanu, Arie and Molchanov, Ilya and Molinari, Francesca},
year = {2009},
month = {01},
pages = {},
title = {Sharp Identification Regions in Models with Convex Predictions: Games, Individual Choice, and Incomplete Data},
volume = {79},
journal = {Centre for Microdata Methods and Practice, Institute for Fiscal Studies, CeMMAP working papers},
doi = {10.3982/ECTA8680}
}

@article{Shapley:71,
	title = {Cores of convex games},
	volume = {1},
	issn = {1432-1270},
	url = {https://doi.org/10.1007/BF01753431},
	doi = {10.1007/BF01753431},
	pages = {11--26},
	number = {1},
	journaltitle = {International Journal of Game Theory},
	shortjournal = {International Journal of Game Theory},
	author = {Shapley, Lloyd S.},
	date = {1971-12-01}
}

@article{BV:85,
author = {Bjorn, Paul and Vuong, Quang},
year = {1984},
month = {08},
pages = {},
title = {Simultaneous Equations Models for Dummy Endogenous Variables: A Game Theoretic Formulation with an Application to Labor Force Participation}
}

@article{BD:2001,
 ISSN = {00346527, 1467937X},
 URL = {http://www.jstor.org/stable/2695928},
 author = {William A. Brock and Steven N. Durlauf},
 journal = {The Review of Economic Studies},
 number = {2},
 pages = {235--260},
 publisher = {[Oxford University Press, Review of Economic Studies, Ltd.]},
 title = {Discrete Choice with Social Interactions},
 volume = {68},
 year = {2001}
}

@article{BHR:2005,
author = {Bajari, Patrick and Hong, Han and Ryan, Stephen P.},
title = {Identification and Estimation of a Discrete Game of Complete Information},
journal = {Econometrica},
volume = {78},
number = {5},
pages = {1529-1568},
doi = {https://doi.org/10.3982/ECTA5434},
url = {https://onlinelibrary.wiley.com/doi/abs/10.3982/ECTA5434},
eprint = {https://onlinelibrary.wiley.com/doi/pdf/10.3982/ECTA5434},
year = {2010}
}

@article{Sweeting:2004,
author = {Sweeting, Andrew},
year = {2005},
month = {02},
pages = {},
title = {Coordination Games, Multiple Equilibria and the Timing of Radio Commercials}
}

@book{Moulin:95,
        author={Hervé Moulin},
        title={Cooperative Microeconomics},
        year=1995,
        publisher={Princeton: Princeton University Press},}

@book{Topkis:98,
        author={Donald M. Topkis},
        title={Supermodularity and Complementarity},
        year=1998,
        publisher={Princeton: Princeton University Press},}

@book{Fujishige:2005,
        author={Satoru Fujishige},
        title={Submodular Functions and Optimization},
        year=2005,
        publisher={Amsterdam, The Netherlands: Elsevier},}

@article{Vives:90,
title = {Nash equilibrium with strategic complementarities},
author = {Vives, Xavier},
year = {1990},
journal = {Journal of Mathematical Economics},
volume = {19},
number = {3},
pages = {305-321},
url = {https://EconPapers.repec.org/RePEc:eee:mateco:v:19:y:1990:i:3:p:305-321}
}

@article{MR:90,
 ISSN = {00129682, 14680262},
 URL = {http://www.jstor.org/stable/2938316},
 author = {Paul Milgrom and John Roberts},
 journal = {Econometrica},
 number = {6},
 pages = {1255--1277},
 publisher = {[Wiley, Econometric Society]},
 title = {Rationalizability, Learning, and Equilibrium in Games with Strategic Complementarities},
 volume = {58},
 year = {1990}
}

@article{EE:2004,
author = {Echenique, Federico},
year = {2003},
month = {02},
pages = {33-44},
title = {Mixed equilibria in games of strategic complementarities},
volume = {22},
journal = {Economic Theory},
doi = {10.1007/s00199-002-0277-8}
}

@article{LTCS:82,
author = {Duke},
year = {1999},
title = {National Long Term Care Survey},
journal = {Public use data set produced and distributed by the Duke University Center for Demographic Studies with funding from the National Institute on Aging under Grant No. U01-AG007198},
}

@article{GHQ:2008,
author = {Galichon, Alfred and Henry, Marc},
year = {2011},
month = {10},
pages = {1264-1298},
title = {Set Identification in Models with Multiple Equilibria},
volume = {78},
journal = {The Review of Economic Studies},
doi = {10.1093/restud/rdr008}
}
\pagebreak

\end{document}